\documentclass{aa}  
\usepackage{graphicx}

\usepackage{txfonts}
\usepackage{caption}
\usepackage{hyperref}
\usepackage{xcolor}
\newcommand{\iraf}{{\fontfamily{qcr}\selectfont IRAF}}
\newcommand{\daophot}{{\fontfamily{qcr}\selectfont DAOPHOT II}}

\begin{document}

   \title{Investigating the origin of  optical flares from the TeV blazar \\S4 0954$+$65}

   \subtitle{}

    \author{Ashwani Pandey\inst{1}
        \and
        Rumen Bachev\inst{2}
         \and
        Bo\.zena Czerny\inst{1}
        \and
         Paul J. Wiita\inst{3}
         \and
          Alok C. Gupta\inst{4}    
          \and
          Anton Strigachev\inst{2}
          \and
          Adam Popowicz\inst{5}
             }

   \institute{Center for Theoretical Physics, Polish Academy of Sciences, Al.Lotnikov 32/46, PL-02-668 Warsaw, Poland\\
              \email{ashwanitapan@gmail.com}
         \and
         Institute of Astronomy and NAO, Bulgarian Academy of Sciences, 1784 Sofia, Bulgaria
         \and
            Department of Physics, The College of New Jersey, 2000 Pennington Road, Ewing, NJ 08628-0718, USA 
        \and  
        Aryabhatta Research Institute of Observational Sciences (ARIES), Manora Peak, Nainital 263001, India
        \and 
        Faculty of Automatic Control, Electronics and Computer Science,
Akademicka 16, 44-100 Gliwice, Poland
                     }

   \date{Received xxx yy, 2023; accepted XXX YY, 2023}

 
  \abstract
   {}
   {To investigate the extreme variability properties of the TeV blazar S4 0954$+$65 using optical photometric and polarization observations carried out between 2017--2023 using 3 ground-based telescopes.}
   {We examined an extensive dataset comprised of 138 intraday (observing duration shorter than a day) light curves (LCs) of S4 0954$+$65 for flux, spectral, and polarization variations on diverse timescales. For the variable LCs, we computed the minimum variability timescales. We investigated flux-flux correlations and colour variations to look for spectral variations on long-term (several weeks to years) timescales. Additionally, we looked for connections between optical R-band flux and polarization degree.}
   {We found significant variations in 59 out of 138 intraday LCs. We detected a maximum change of 0.58$\pm$0.11 in V-band magnitude within $\sim$2.64 hr and a corresponding minimum variability timescale of 18.21$\pm$4.87 minutes on 2017 March 25. During the course of our observing campaign, the source brightness changed by $\sim$4 magnitudes in V and R bands; however, we did not find any strong spectral variations.  The slope of the relative spectral energy distribution was 1.37$\pm$0.04. The degree of polarization varied from $\sim$ 3\% to 39\% during our monitoring. We observed a change of $\sim$120 degrees in polarization angle within $\sim$3 hr on 2022 April 13. No clear correlation was found between optical flux and the degree of polarization. }
   {The results of our optical flux, colour, and polarization study provide hints that turbulence in the relativistic jet could be responsible for the intraday optical variations in the blazar S4 0954$+$65. However, the long-term flux variations may be caused by changes in the Doppler factor. }


   \maketitle
%
\section{Introduction}
According to the traditional orientation-based classification scheme of active galactic nuclei (AGN), blazars are radio-loud sources with relativistic jets pointing very close to our line of sight \citep{1995PASP..107..803U}. 
Depending on the strength of their optical/ultraviolet (UV) emission lines, blazars are further classified as BL Lacertae objects (BLLs; EW\footnote{equivalent width of emission lines in rest frame}$_{rest}$ $< 5$\AA) and flat-spectrum radio quasars (FSRQs; EW$_{rest}$ $> 5$\AA)\citep[][]{1991ApJS...76..813S, 1996MNRAS.281..425M}. The primary characteristics of blazars are high amplitude flux variations throughout the whole electromagnetic spectrum, significant polarizations in all bands in which it can be measured,  and the double-humped shape of their broad-band spectral energy distributions (SEDs) \citep{1995ARA&A..33..163W,1998MNRAS.299..433F,2022MNRAS.510.1809P,2022Natur.611..677L}. The low-energy hump of the SED is attributed to the synchrotron emission from relativistic electrons within the jet, while the high-energy component is usually explained by inverse Compton emission \citep{1994ApJ...421..153S,1996ApJ...461..657B}. However, models dominated by hadronic processes have also been proposed to explain the high-energy hump \citep[e.g.][]{2013ApJ...768...54B}.  

The blazar S4 0954$+$65 was discovered as a radio source and its optical counterpart was identified by \cite{1977MmRAS..84....1C}. It was classified as a BL Lac object by \cite{1984MNRAS.211..105W} and its redshift was first measured to be $z$ = 0.368 \citep{1986AJ.....91..494L,1993A&AS...98..393S}. \cite{2015AJ....150..181L} challenged this value of the redshift and suggested a lower limit of $z \geq$ 0.45. However, \cite{2021MNRAS.504.5258B} recently ruled out this lower limit and determined the redshift of S4 0954$+$65 to be $z$ = 0.3694$\pm$0.0011 using the Mg II line during its low flux state. 

S4 0954$+$65 has been studied several times for flux variations on diverse timescales \citep{1993A&A...271..344W,1999A&A...352...19R,2014AJ....148...42M}. The source exhibited extreme optical intraday variability (IDV) of $\sim$0.7 mag within 7 hr and $\sim$1.0 mag within 5 hr on 2011 March 9 and April 24, respectively, accompanied by changes in the fractional polarization \citep{2014AJ....148...42M}. During its 2015 February outburst, \cite{2015MNRAS.451L..21B} observed rapid intranight flux variability with a change of $\sim$0.7 mag in optical brightness within about 5 h. 
During the same epoch, the source was detected, for the first time, at very high energies ($E \geq$ 100 GeV) by the MAGIC telescopes \citep{2018A&A...617A..30M}. Recently, \cite{2021MNRAS.504.5629R} investigated the nature of the complex variability of S4 0954$+$65 using data from the Whole Earth Blazar Telescope (WEBT) Collaboration and the Transiting Exoplanet Survey Satellite (TESS). They observed extreme flux variability with an increase of 1 mag in brightness in 24 h followed by a decrease of 0.8 mag in brightness in 23 h. They also found strong variations in optical polarization degree (PD) and electric vector polarization angle (PA). However, they did not find any correlation between optical PD and flux.

Variability timescales in blazars span from years to minutes, indicating a variety of underlying physical processes are present \citep[e.g.][and references therein]{1995ARA&A..33..163W,2020ApJ...890...72P,2022A&A...668A.152P,2023MNRAS.522..102R}. These emission mechanisms can be intrinsic e.g. interaction of shocks with turbulent plasma, magnetic reconnection in localized jet regions \citep[e.g.][]{1985ApJ...298..114M,2014ApJ...780...87M,2016ApJ...820...12P} and/or extrinsic e.g. geometrical effects such as a change in viewing angle, and hence in the Doppler boosting \citep[e.g.][]{1992A&A...255...59C,2017Natur.552..374R} in nature. Polarization variations are also often observed in blazars on a variety of timescales. Observed variable polarization provides crucial information about the magnitude and direction of the magnetic field inside the jets. A number of investigations have been done on the connections between optical flux and polarization variations. People have observed correlations, anticorrelations, and no correlations, between optical flux and PD \citep[e.g.][]{2002A&A...385...55H,2006ChJAS...6a.247J,2022MNRAS.510.1809P,2022MNRAS.517.3236R}. It is crucial to investigate the relationship between optical flux and PD variations in order to comprehend how the magnetic field affects blazar jet emission processes. 

With this motivation, we investigate the multiband optical variability properties of TeV blazar S4 0954$+$65 on diverse timescales during 2017--2023. We also examine the optical polarization variability and the correlation between optical flux and PD to probe the origin of low-energy emissions from S4 0954$+$65.

The paper is structured as follows. Details of the observations and description of the data reduction are given in Section \ref{sec:data}. The  results of our optical photometry and polarimetry study are presented in Section \ref{sec:results}. A discussion of our results and conclusions are given in Section \ref{sec:diss_con}. Finally, we summarize our findings in Section \ref{sec:sum}.

\begin{table*}
\caption{\label{tab:telescopes}List of telescopes used for observations. }
    \centering
    \begin{tabular}{cccccc} \hline
  Code & Observatory & Country & Aperture  & Instrument   & No. of data points \\ \hline
   A   & Astronomical Observatory Belogradchik &  Bulgaria & 60 cm &  FLI PL 16803 & 164 B, 1317 V, 1763 R, 1310 I\\
   B   & National Astronomical Observatory Rozhen & Bulgaria & 2 m &  ANDOR iKON-L  & 174 B, 445 R\\
   C   & S.U.T.O. Otivar                          & Spain    & 30 cm & ASI ZWO 1600MM & 93 V, 99 R, 85 I \\ \hline
    \end{tabular}
\end{table*}
\section{Observations and data reduction}\label{sec:data}
We carried out optical photometric monitoring of the TeV blazar S4 0954$+$65 from 2017 March 21 to 2023 April 28 using three ground-based optical telescopes in Bulgaria and Spain listed in Table \ref{tab:telescopes}. We spent a total of 89 nights observing the blazar, gathering a total of 5005 image frames in the $B$, $V$, $R$, and $I$ optical bands. A detailed log of our optical photometric monitoring is given in Table \ref{tab:opt_log}.

For data reduction, we performed the conventional steps, which include cleaning (bias-subtraction, flat-fielding, and cosmic ray removal) of raw images in \iraf, followed by the aperture photometry of cleaned images in \daophot \ to get the instrumental magnitudes. 
 Detailed descriptions of the data reduction procedure are given in \citet[][and references therein]{2019ApJ...871..192P,2020MNRAS.496.1430P,2020ApJ...890...72P}. In addition to the source, each image frame also contains three comparison stars \citep[S2, S3, and S4 from Figure (1) of][] {1999A&A...352...19R}. We generated the differential light curves (DLCs) of the blazar S4 0954$+$65 relative to the comparison stars as well as the DLCs of the comparison stars. The standard deviation of DLCs between comparison stars indicates the observational uncertainties on that particular night, whereas the DLC of the blazar with respect to the comparison stars shows the blazar’s intrinsic variability. First, we selected a steady (having minimum standard deviation) pair of comparison stars. Then we used the comparison star (S4) from that pair, which had a magnitude and colour comparable to those of the blazar, to obtain the calibrated magnitudes. 
 The calibrated magnitudes are dereddened by subtracting the Galactic extinction, A$_{\lambda}$\footnote{taken from \url{https://ned.ipac.caltech.edu/}.}, and converted into flux densities. The observations in different bands (BVRI) on a particular night were carried out quasi-simultaneously (within 20 minutes) by the same telescope.

In addition, we also performed optical (R-band) polarimetry observations using the 60 cm telescope of the Belogradchik observatory between 2022 March 15 and 2023 April 28. To obtain the polarimetric parameters, the polarization degree (PD) and the electric vector polarization angle (PA), we used photometric measurements of the blazar S4 0954$+$65 with respect to the ambient field stars through three polarizing filters (in addition to the R-band filter). The polarizing filters are oriented at 0$-$180, 60$-$240, and 120$-$300 degrees with respect to the North. This approach cannot employ the standard Stokes parameters and requires solving 3 equations for 3 unknowns instead; details are given in \citet{2023MNRAS.522.3018B}.
The location of S4 0954$+$65 in the sky implies the presence of interstellar absorption in that direction of $A_{v} \simeq 0.33$ mag \citep{2011ApJ...737..103S}. The dichroic polarization due to the interstellar dust then can be estimated, following \cite{1992dge..book.....W} as PD$_{dust, max} < 3 A_{v}$ \%, i.e. PD$_{dust,
max} < 1 \%$. Therefore, for the purposes of our study, the ISM dichroic polarization can be neglected.

To resolve the $180^{\circ}$ ambiguity in the PA measurements, we employed the standard procedure wherein the value $\Delta\theta~=~|\theta_{n}-\theta_{n-1}|- \sqrt{\sigma(\theta_{n})^2+\sigma(\theta_{n-1})^2}$ is minimized for consecutive measurements \citep[e.g.][]{2019Galax...7...46B}. Here, $\theta_{n}$ and $\sigma(\theta_{n})$ are respectively the $n^{th}$ measurement of PA and its uncertainty. For $\Delta\theta~ >~90^{\circ}$, $\theta_{n}$ is shifted by $\pm~n~\times~180^{\circ}$, where the integer $n$ is selected to minimize the value of $\Delta\theta$. For $\Delta\theta~ \leq~90^{\circ}$, $\theta_{n}$ remains the same. 

\begin{figure*}
\includegraphics[height=6cm,width=6cm]{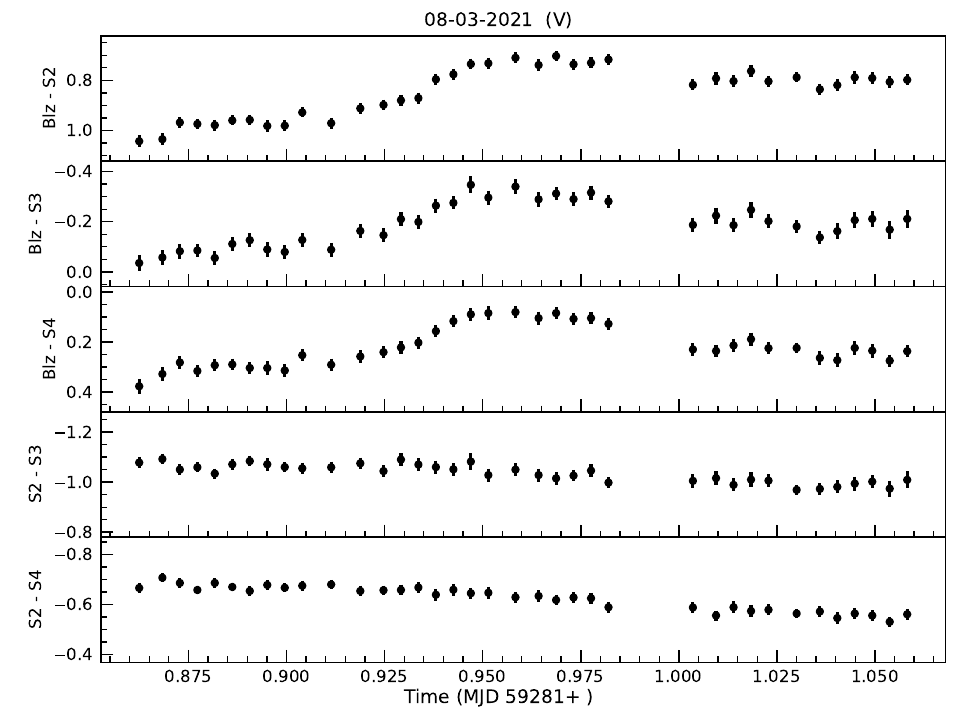}
\includegraphics[height=6cm,width=6cm]{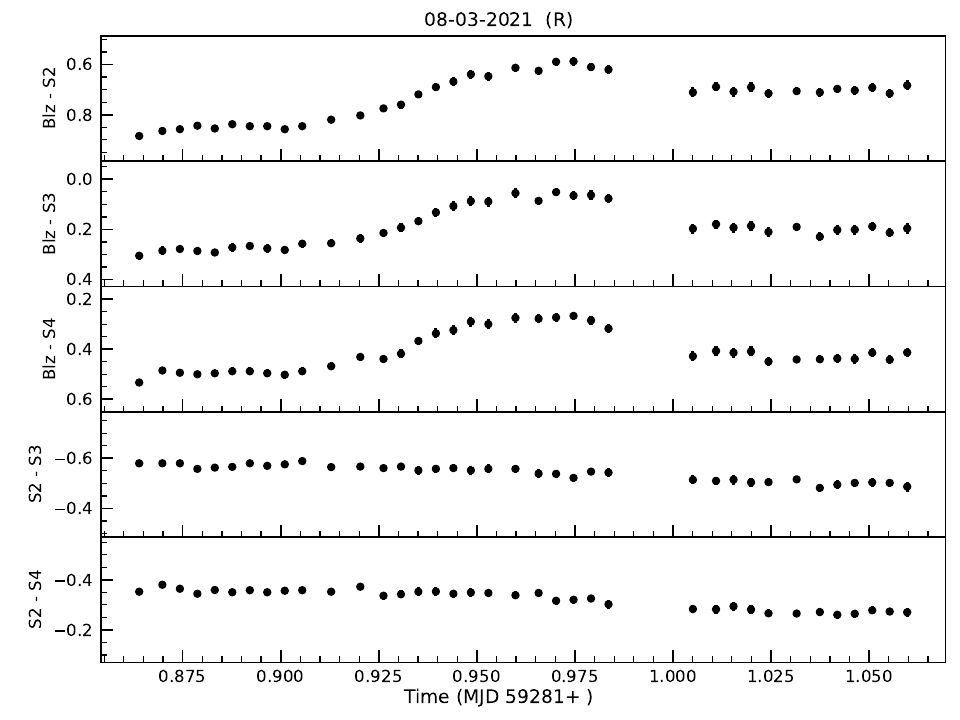}
\includegraphics[height=6cm,width=6cm]{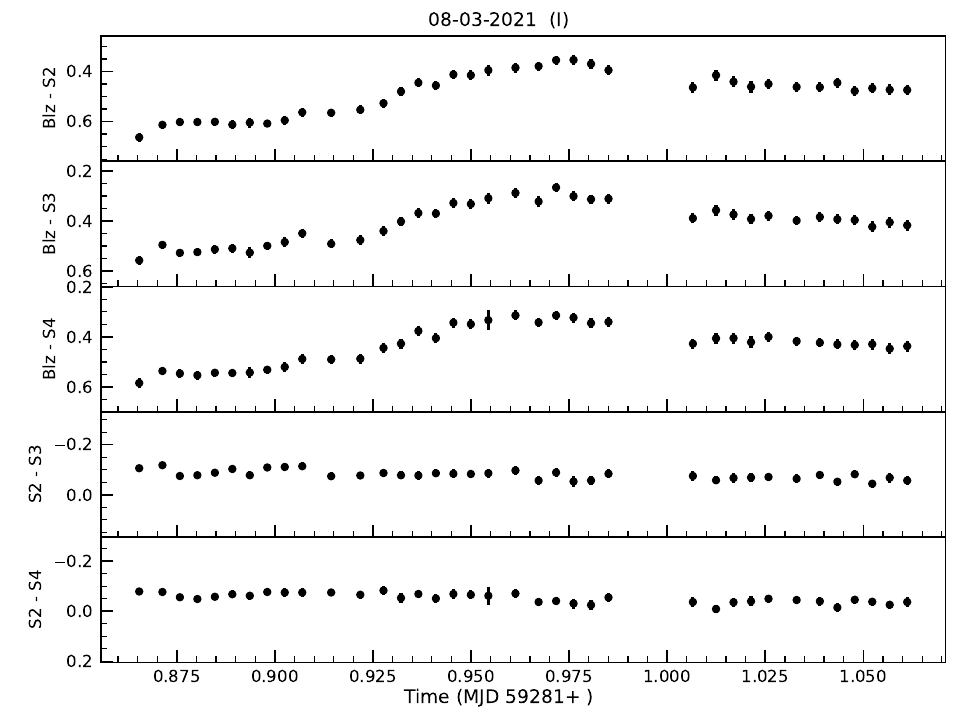}
\caption{\label{fig:idv_lc}Sample DLCs of the TeV blazar S4 0954$+$65. Here, S2, S3 and S4 refer to the comparison stars and Blz refers to the blazar. The observation date and the filter name are mentioned at the top of each plot. All 138 DLCs are shown in Figures \ref{fig:idv_lc1}-\ref{fig:idv_lc12} in the appendix. }
\end{figure*}

\begin{figure}
\includegraphics[height=8cm,width=8cm]{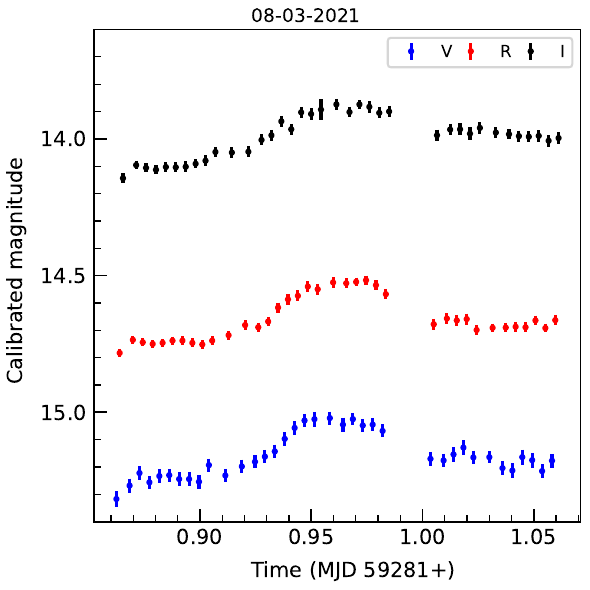}
\caption{\label{fig:idv_cal_lc}The calibrated light curves for the DLCs shown in Figure \ref{fig:idv_lc}.}
\end{figure}

\section{Results}\label{sec:results}
\subsection{Optical flux variability}\label{subsec:flux_var}
\subsubsection{Intraday flux variability}\label{subsubsec:IDV}
In order to ensure that there were enough photometric points available to characterize IDV, we chose DLCs with at least ten data points in a given filter in each night. By applying this criterion, we were able to consider 138 IDV DLCs, a sample of which is shown in Figure \ref{fig:idv_lc}. The calibrated light curves for the DLCs shown in Figure \ref{fig:idv_lc} are plotted in Figure \ref{fig:idv_cal_lc}. 

We examined the DLCs of TeV blazar S4 0954$+$65 for IDV using the power-enhanced $F-$test, which is one of the most powerful and reliable statistical tests to detect microvariability in blazars \citep{2014AJ....148...93D}. 
In the power-enhanced $F-$test, the variance of the DLC of blazar is compared with the combined variance of more than one comparison star to detect flux variations. A detailed description of the power-enhanced $F-$test is provided in our previous papers \citep{2019ApJ...871..192P,2020MNRAS.496.1430P}. Here, we briefly discuss its main steps. First, we estimate the power-enhanced {\it F}-test statistics, which is given by
\begin{equation} \label{eq:Fenh}
F_{\rm enh} = \frac{s_{\rm blz}^2}{s_c^2},
\end{equation}  
where $s_{\rm blz}^2$ is the variance of the source DLC and $s_c^2$ is the combined variance of $k$ comparison stars DLCs. The value of $s_c^2$ is calculated using Equation (2) of \cite{2019ApJ...871..192P}.
The number of degrees of freedom in the numerator and denominator of the F-statistics are $\nu_1$=$N-1$ and $\nu_2$=$k(N-1)$, respectively, where $N$ is the number of data points.
We then compare the value of $F_{\rm enh}$ with the critical value ($F_{\rm c}$) at $\alpha$ = 0.01 (99\% confidence level). If $F_{\rm enh} \geq  F_{\rm c}$, we refer to the light curve as  variable (V); otherwise, we refer to it as non-variable (NV). The $F_{\rm enh}$ test results are given in Table \ref{tab:var_res}. Using this criterion we find that S4 0954+65 displays IDV variability on 59 of the 138 nights of our observations.

\subsubsection{Flux variability amplitude}\label{subsubsec:var_amp}
To quantify the amplitude of variations in the variable IDV light curves, we estimated the variability amplitude, $Amp$, which is defined as \citep{1996A&A...305...42H}
\begin{equation}
    Amp (\%) = 100 \sqrt{(A_{max} - A_{min})^2 - 2 \sigma^2},
\end{equation}
where A$_{max}$ and A$_{min}$ are the maximum and minimum magnitudes of the calibrated light curve, respectively, while $\sigma$ is the measurement error. The error in the variability amplitude is calculated using the error propagation 
as 
\begin{equation}
    \sigma_{Amp} (\%) = 100 \times  \left( \frac{A_{max} - A_{min}}{Amp} \right ) \times   \sqrt{\sigma^2_{max} + \sigma^2_{min}} ,
\end{equation}
where $\sigma_{min}$ and $\sigma_{max}$ are the uncertainties in the minimum (A$_{min}$) and maximum (A$_{max}$) calibrated magnitudes, respectively. 
The values of the variability amplitude and its error are given in Table \ref{tab:var_res} for the variable light curves. For a non-variable light curve, we put a `--'. 

\subsubsection{Variability timescale}
For each variable light curve, we also determined the variability timescale following \citet{1974ApJ...193...43B},
\begin{equation}
    \tau_{ij} = \frac{dt}{ln(F_i/F_j)},
\end{equation}
where $dt$ represents the time interval between two separate flux measurements $F_i$ and $F_j$ such that $|F_i - F_j| > \sigma_{F_i} + \sigma_{F_j}$. The minimum flux variability timescale is estimated as $\tau_{var}$ = min($\tau_{ij}$). The uncertainties in $\tau_{ij}$ were obtained by standard error propagation \citep{1992drea.book.....B}. The value of $\tau_{var}$ and its uncertainty $\sigma$($\tau_{var}$) are given for the variable light curves in Table \ref{tab:var_res} whenever $\tau_{var} \geq \sigma$ ($\tau_{var}$), otherwise we put a `--'. We noticed that there were four occasions when the estimated timescale was more than the length of the light curve which we denoted by a `*' in Table \ref{tab:var_res}.

\begin{figure*}
\centering
\includegraphics[height=16cm,width=18cm]{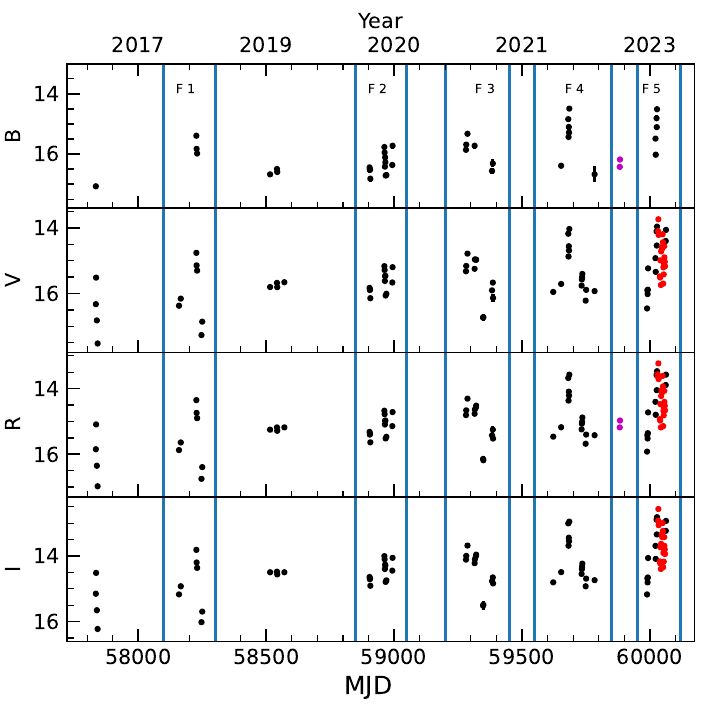}
\caption{\label{fig:ltv_lc}Long-term light curves of S4 0954$+$65 in $B$,$V$, $R$, and $I$ bands. 
Here, the observations performed with telescopes A, B, and C are denoted by black, magenta, and red colours, respectively. The vertical lines indicate the epochs of flaring events F1, F2, F3, F4, and F5. }
\end{figure*}

\begin{table*}
\caption{Results of LTV analysis of S4 0954$+$65.}        
\label{tab:ltv_var_res}                   
\centering 
\resizebox{\textwidth} {!}{                     
\begin{tabular}{ccccccc}           
\hline\hline                		 
Band &  Brightest magnitude/MJD  &  Faintest magnitude/MJD  & Average magnitude &  $\Delta$mag &  $\tau_{\rm var}$ (in hr)\\
\hline
B  &  14.269$\pm$0.186/59684.77789  &  17.191$\pm$0.469/58970.00546  &  16.059$\pm$0.006  &  2.922 &  34.57$\pm$8.50 \\ 
V  &  13.653$\pm$0.023/60032.99529  &  17.621$\pm$0.089/57840.91645  &  15.420$\pm$0.001  &  3.968 &  20.51$\pm$0.19 \\ 
R  &  13.114$\pm$0.022/60032.99905  &  17.082$\pm$0.070/57840.83056  &  14.948$\pm$0.001  &  3.968 &  20.90$\pm$0.20 \\ 
I  &  12.430$\pm$0.029/60033.00280  &  16.308$\pm$0.090/57840.77654  &  14.242$\pm$0.001  &  3.878 &  22.88$\pm$0.36 \\ 
\hline                          
\end{tabular}}
\end{table*}

\subsubsection{Long-term multiband flux variability}\label{subsubsec:LTV}
The daily averaged optical multiband ($BVRI$) LCs of the TeV blazar S4 0954$+$65 for the entire monitoring period are shown in Figure \ref{fig:ltv_lc}. The source showed clear variations in all the bands. The minimum, maximum, and averaged magnitudes of the source, together with the change in magnitude ($\Delta$mag), and the minimum variability timescale in each optical band are listed in Table \ref{tab:ltv_var_res}. During our observing campaign, the brightest state we observed the source to be in was $R_{mag}$ = 13.11 on 2023 March 29, while the faintest state, with $R_{mag}$ = 17.08, was recorded on 2017 March 28. The V and R band light curves showed maximum fluctuations ($\Delta$ mag $\sim$ 4) and shortest variability timescales ($\sim$21 hr). The significantly higher variability timescale and the smaller values of $\Delta$ mag for the $B$ band are almost certainly due to the fewer observations in the $B$ band. 

\begin{figure*}
\includegraphics[height=6cm,width=6cm]{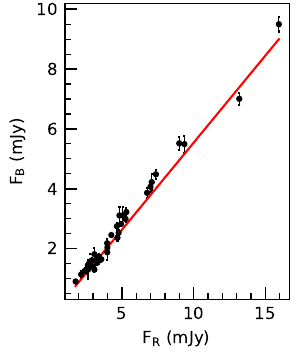}\includegraphics[height=6cm,width=6cm]{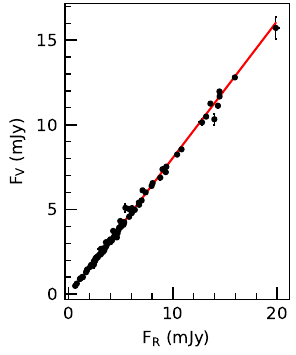}\includegraphics[height=6cm,width=6cm]{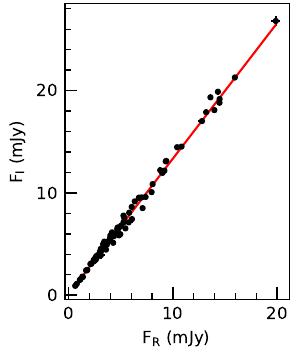}
\caption{\label{fig:flux_flux}Flux-flux diagrams of S4 0954$+$65 for optical wavelengths. The solid red line represents the straight line fit indicating that the fluxes are correlated.}
\end{figure*}

\begin{figure}
\includegraphics[height=7cm,width=8cm]{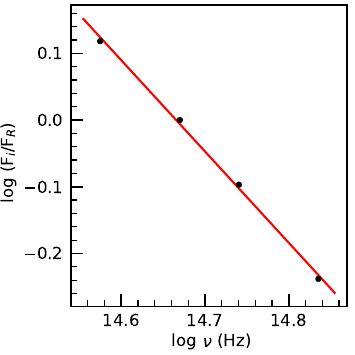}
\caption{\label{fig:sed}Mean observed relative SED of S4 0954$+$65 for the entire monitoring period. }
\end{figure}

\subsection{Optical spectral variability}\label{subsec:op_spec_var}
 To study the spectral variability of the blazar S4 0954+65 for our entire monitoring period, we adopted the technique used by \cite{2008ApJ...672...40H}. It is based on the assumption that the radiation has two components,  one constant and one variable, and that the variable component causes all the changes in the flux. This method involves plotting the flux-flux diagrams for pairs of bands. If the spectral properties of the variable component remain unchanged during a given time interval, the flux-flux diagrams will follow a linear relationship. The slopes of these lines are the flux ratios for the respective pairs of bands. The reverse is also true, with a few limitations: a linear relationship between fluxes at two separate bands during a period of flux variability implies that the slope (flux ratio) remains constant. Such a linear relation for multiple bands would indicate that the mean relative SED of the variable component remains unchanged for the entire period and can be determined from the slopes of these lines. 

 \begin{table}
    \caption{ \label{tab:flare_analysis}Results of the flux and spectral variability analysis of optical flares. }
\centering
\resizebox{0.5\textwidth} {!}{                     
\begin{tabular}{ccccc}           
\hline\hline 
Flare & Duration & $\Delta$mag & $\tau_{\rm var}$ (in hr) &  $\alpha$ \\ \hline
F1 & MJD 58100-58300 & 2.40 & 68.60$\pm$0.86 &  1.04$\pm$0.21 \\
F2 & MJD 58850-59050 & 0.96 & 45.08$\pm$1.28 &  1.37$\pm$0.07 \\
F3 & MJD 59200-59450 & 1.88 & 95.41$\pm$43.08 & 1.13$\pm$0.02 \\
F4 & MJD 59550-59850 & 2.10 & 37.24$\pm$0.15 &  1.37$\pm$0.04 \\
F5 & MJD 59950-60120 & 2.69 & 20.90$\pm$0.20 &  1.25$\pm$0.02 \\
\hline                          
\end{tabular}}
\end{table}

\begin{figure}
\centering
\includegraphics[height=5cm,width=8cm]{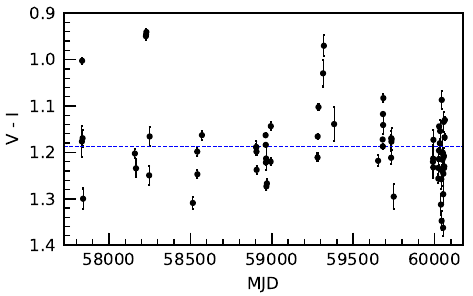}
\includegraphics[height=5cm,width=8cm]{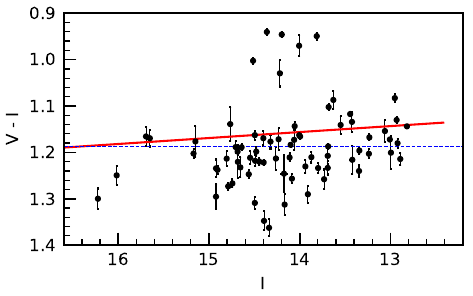}
\caption{\label{fig:color}The variations of V-I colour indices with time (top panel) and I-band magnitude (bottom panel). The horizontal blue line denotes the average value of the V-I colour indices, while the red solid line in the bottom panel represents a linear fit.}
\end{figure}
The flux-flux diagrams for S4 0954$+$65 are plotted in Figure \ref{fig:flux_flux}. As can be seen from the figure, a straight line fits all the flux-flux plots very well. Using the slopes (log($F_{B}/F_{R}$) = $-0.24$, log($F_{V}/F_{R}$) = $-0.10$, log($F_{R}/F_{R}$)  $\equiv$ 0.0, log($F_{I}/F_{R}$) = 0.12) of the lines we constructed the mean relative SED of S4 0954$+$65, shown in Figure \ref{fig:sed}. The SED follows a power-law ($F_{\nu} \propto \nu^{-\alpha}$) with spectral index $\alpha$ = 1.37$\pm$0.04, which is consistent with the value (1.32$\pm$0.05) obtained by \cite{2015ARep...59..551H} during 2008--2012. It is important to note that here we are only using optical data in $BVRI$ bands which cover a relatively narrow spectral range. In a few cases, it has been found that the flux-flux plots deviated from the linear relationship in the infrared (IR) bands \citep{2010A&A...510A..93L,2020ApJ...902...61L}. However, \cite{2015ARep...59..551H} observed that for S4 0954$+$65, the flux-flux plots followed a linear relationship even in the IR bands.

In addition, we also estimated the colour indices for the total observing period to examine colour variations with time and magnitude. We acquired 38 B$-$V, 40 B$-$R, 38 B$-$I, 86 V$-$R, 82 V$-$I, and 86 R$-$I colour indices with average values of 0.646$\pm$0.014, 1.146$\pm$0.014, 1.811$\pm$0.014, 0.503$\pm$0.003, 1.187$\pm$0.002, and  0.679$\pm$0.003, respectively. The V-I colour indices, having the highest average value among the frequently measured indices, are plotted against time and I band magnitude in Figure \ref{fig:color}. The correlation coefficient is just 0.23 ($p-$value $>$ 0.05) and the slope of the linear fit is 0.01, indicating that there was an insignificant relationship between the V$-$I colour indices and I magnitude. 

\begin{figure*}
\includegraphics[height=6cm,width=6cm]{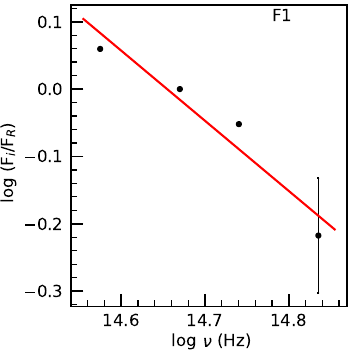}\includegraphics[height=6cm,width=6cm]{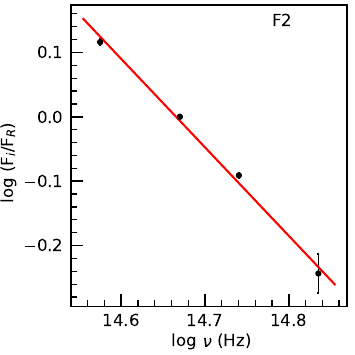}\includegraphics[height=6cm,width=6cm]{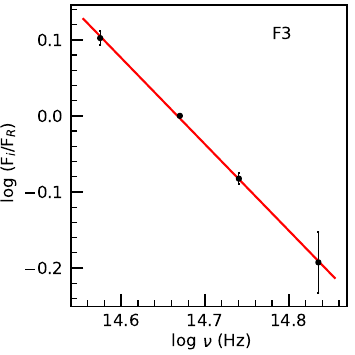} \\ \includegraphics[height=6cm,width=6cm]{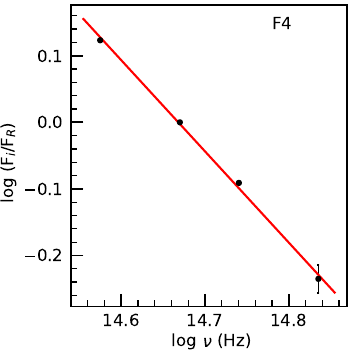}\includegraphics[height=6cm,width=6cm]{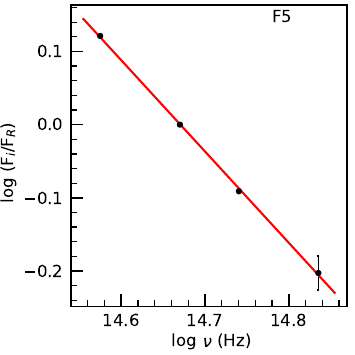}
\caption{\label{fig:flare_sed}The relative SEDs for different flaring epochs.}
\end{figure*}

\begin{table*}
\centering
\caption{Polarization observation log of S4 0954+65.}
\label{tab:pol_log} 
\begin{tabular}{ccccccc} 
\hline\hline 
Date of observation  & Number of  & Duration      & Min PD (\%) & Max PD (\%) &  Min PA ($^{\circ}$) & Max PA ($^{\circ}$) \\
dd-mm-yyyy  &  data points & (in $\sim$hr)  &              &            &               &          \\
\hline
15-03-2022 & 3  & 1.94   &  8.40$\pm$1.20  &  12.60$\pm$2.70  &   117.50$\pm$4.50  &  132.80$\pm$6.90  \\ 
11-04-2022 & 15  & 7.18   &  4.20$\pm$0.80  &  14.20$\pm$0.80  &   58.60$\pm$3.70  &  81.80$\pm$6.30  \\ 
12-04-2022 & 9  & 4.49   &  4.00$\pm$1.20  &  7.50$\pm$1.00  &   71.00$\pm$3.50  &  124.30$\pm$8.00  \\ 
13-04-2022 & 12  & 6.19   &  3.00$\pm$0.80  &  9.00$\pm$1.70  &   10.80$\pm$6.20  &  169.40$\pm$7.10  \\ 
15-04-2022 & 9  & 3.01   &  4.40$\pm$0.90  &  7.90$\pm$0.90  &   132.00$\pm$6.20  &  153.70$\pm$4.70  \\ 
02-06-2022 & 7  & 4.65   &  16.30$\pm$2.20  &  22.90$\pm$3.20  &   97.80$\pm$5.00  &  113.30$\pm$7.50  \\ 
03-06-2022 & 6  & 3.98   &  10.80$\pm$2.20  &  21.90$\pm$4.40  &   84.10$\pm$5.70  &  100.20$\pm$3.10  \\ 
04-06-2022 & 7  & 3.97   &  19.00$\pm$2.40  &  27.90$\pm$3.30  &   91.40$\pm$2.30  &  101.10$\pm$3.70  \\ 
05-06-2022 & 2  & 0.54   &  36.10$\pm$2.30  &  39.20$\pm$1.10  &   75.10$\pm$2.30  &  76.00$\pm$1.50  \\ 
18-06-2022 & 2  & 0.54   &  12.70$\pm$3.80  &  26.00$\pm$3.30  &   115.10$\pm$8.80  &  118.10$\pm$3.40  \\ 
20-06-2022 & 2  & 0.22   &  26.00$\pm$2.10  &  27.00$\pm$2.40  &   78.70$\pm$2.40  &  79.00$\pm$2.30  \\ 
14-02-2023 & 1  & -      &  17.00$\pm$1.90  & 17.00$\pm$1.90   &   141.4$\pm$3.60  & 141.4$\pm$3.60 \\
15-02-2023 & 4  & 1.97   &  16.80$\pm$2.60  &  22.40$\pm$4.00  &   141.70$\pm$6.20  &  147.00$\pm$5.10  \\ 
16-02-2023 & 4  & 1.44   &  8.70$\pm$1.80  &  14.80$\pm$1.10  &   117.20$\pm$8.00  &  132.10$\pm$3.00  \\ 
17-02-2023 & 1  &  -	    &  19.70$\pm$0.50 & 19.70$\pm$0.50   &   138.50$\pm$1.20  &   138.50$\pm$1.20 \\
18-02-2023 & 1  &  -     &  22.30$\pm$1.50 & 22.30$\pm$1.50   &   108.50$\pm$2.20   &   108.50$\pm$2.20 \\	
19-02-2023 & 4  & 1.61   &  17.40$\pm$3.90  &  19.40$\pm$2.90  &   85.40$\pm$1.00  &  94.30$\pm$3.10  \\ 
22-03-2023 & 9  & 3.48   &  30.30$\pm$0.50  &  33.30$\pm$1.10  &   102.90$\pm$1.40  &  107.00$\pm$1.10  \\ 
23-03-2023 & 9  & 3.33   &  22.60$\pm$0.80  &  25.10$\pm$0.50  &   114.50$\pm$0.70  &  117.50$\pm$0.70  \\ 
24-03-2023 & 13  & 4.45   &  14.90$\pm$0.20  &  24.00$\pm$0.20  &   123.80$\pm$0.30  &  134.20$\pm$0.50  \\ 
27-04-2023 & 12  & 3.91   &  18.70$\pm$0.90  &  22.60$\pm$0.90  &   85.80$\pm$1.20  &  90.10$\pm$1.20  \\ 
28-04-2023 & 11  & 2.81   &  27.50$\pm$0.70  &  31.50$\pm$1.00  &   95.60$\pm$0.70  &  100.00$\pm$0.70  \\ 
\hline
\end{tabular}
\end{table*}

\begin{figure}
\centering
\includegraphics[height=8cm,width=8cm]{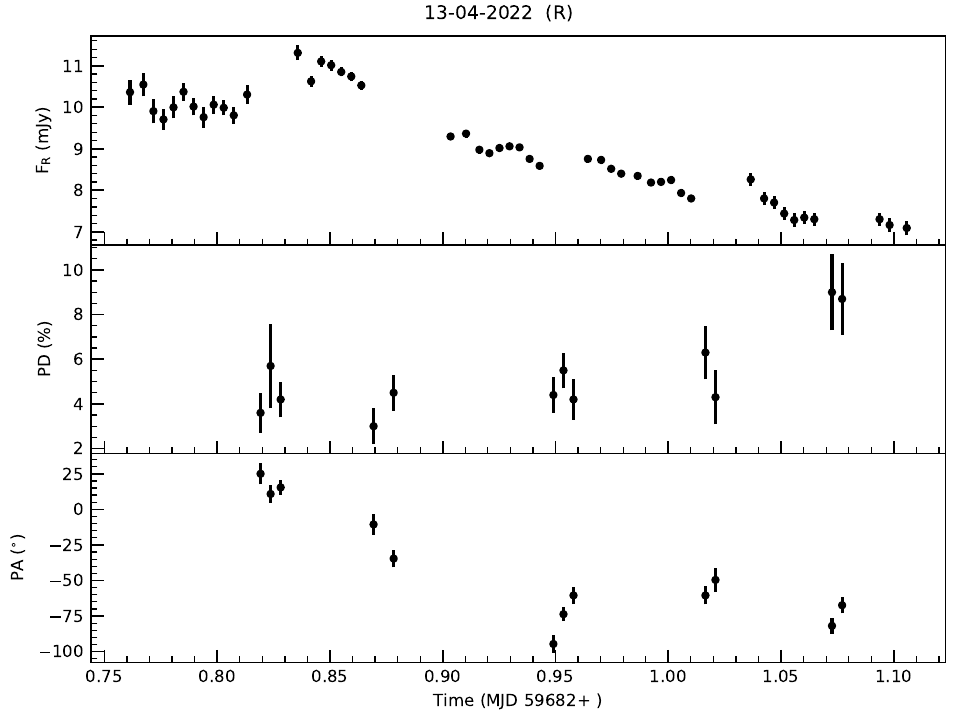}
\caption{\label{fig:sample_flux_pol}A sample plot showing variation of PD and PA with optical flux on intraday timescale. All such IDV plots are shown in Figure \ref{fig:idv_op_pol}. }
\end{figure}

\begin{figure}
\centering
\includegraphics[height=8cm,width=8cm]{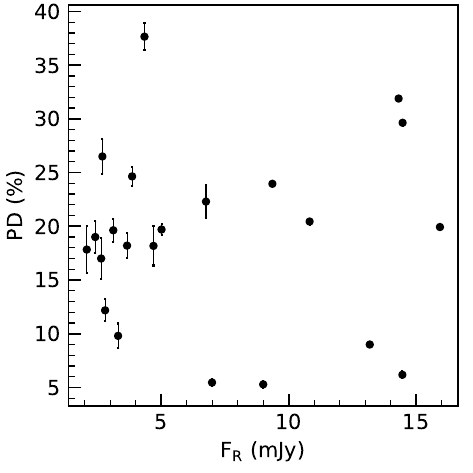}
\caption{\label{fig:flux_pol}The degree of polarization versus optical R-band flux, which indicates that there is no correlation between them.}
\end{figure}

\subsection{Optical flares}\label{subsec:opt_flare}
The source exhibited several high-flux stages during the course of our monitoring period, as seen in Figure \ref{fig:ltv_lc}. We identified five such flaring epochs (F1, F2, F3, F4, and F5) each having at least three data points in each filter. The duration of these epochs, observed changes in R-mag, and the minimum variability timescales are given in Table \ref{tab:flare_analysis}. The maximum change in R-mag, $\Delta$mag = 2.69, was recorded during flare F5, which also has the shortest variability timescale of $\sim$ 21 hr. 

For each of these flaring epochs, we generated the relative SED, shown in Figure \ref{fig:flare_sed}, to investigate the spectral variability. The large uncertainty in the data point corresponding to the B-band in each SED is due to the smaller number of observations in the B-band.  The derived optical spectral indices for these periods are listed in Table \ref{tab:flare_analysis}. The spectral indices are comparable within the uncertainties.

\subsection{Polarization variability}
The magnetic field almost certainly plays an important role in the flux variability of blazars. To obtain information on the magnetic field we also performed optical R-band polarimetric observations of S4 0954$+$65 from 2022 March 15 to 2023 April 28 using the 60 cm telescope at the Belogradchik Observatory (telescope A in Table \ref{tab:telescopes}). The log of our polarimetry observations is given in Table \ref{tab:pol_log}. We can see fluctuations in both PD and PA despite the noisy and sparse nature of the data on IDV timescales, see Figure \ref{fig:sample_flux_pol}. The minimum and maximum values of PD and PA for each night are given in Table \ref{tab:pol_log}. 

We found significant changes in the PD over the course of a night ($\Delta$ PD > 3  $\sigma_{\Delta PD}$) on 5 of the 22 nights.
The maximum significant change in PD was 10$\pm$1\% observed on 2022 April 11. We also noticed a change in PA by $\sim$120 degrees in $\sim$ 3 hr on 2022 April 13  (see Figure \ref{fig:sample_flux_pol}). For the full monitoring period, the values of PD range from $\sim$3\% to 39\%,  while the PA varied between $\sim$11$^{\circ}$ and $\sim$169$^{\circ}$. 

\subsection{Correlation between optical flux and polarization} \label{subsec:flux_pol}
 For the nights when we had more than 5 polarimetry readings, we display both PD and PA with R-band flux (in mJy) in Figure \ref{fig:idv_op_pol}. We see no obvious correlations or trends between the optical flux and polarization on IDV timescales. However, as our IDV polarization data is sparse and has large error bars, this is unsurprising.

For the entire span of our observations, we plotted PD against optical flux in Figure \ref{fig:flux_pol} to investigate their potential correlation. We found no correlation between PD and optical flux, as is evident from the plot, and this is quantified through the values of the correlation coefficient, $r = -0.01$, and the null hypothesis, $p = 0.97$. \cite{2021MNRAS.504.5629R} also observed no correlation between optical brightness and PD for S4 0954$+$65 using data limited to  2019 and 2020. The absence of correlation between optical flux and PD has also been reported in several other blazars \citep[e.g.][]{2009arXiv0912.3664I,2016MNRAS.462.4267J}.  

\begin{figure}
\centering
\includegraphics[height=8cm,width=8cm]{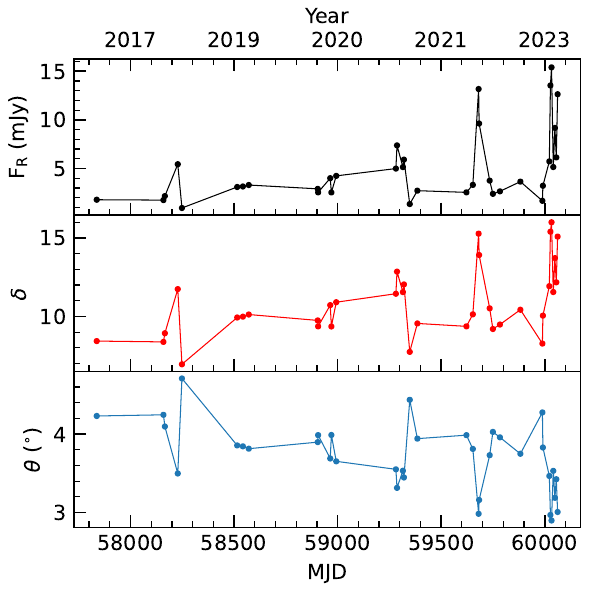}
\caption{\label{fig:delta_var}The temporal variation of the weekly binned R-band flux (top panel) together with the corresponding variation in the Doppler factor (middle panel), and the viewing angle (bottom panel) of the emission region based on the model of \citet{2021MNRAS.504.5629R}. }
\end{figure}
\section{Discussion and Conclusions}\label{sec:diss_con}
The blazar S4 0954$+$65 is well recognized for its extremely variable optical properties over a range of periods. In this study, we used optical photometry and polarization measurements spanning $\sim$6 years to further explore its optical flares. During our monitoring campaign, the blazar S4 0954$+$65 exhibited fluctuations in optical flux and PD over a range of timescales. We found statistically significant variations in 59 out of 138 IDV light curves. The variability amplitudes varied from $\sim$ 6\% to $\sim$ 57\% on IDV timescales. On 2017 March 25, we observed violent optical variability with a magnitude change of 0.58 within 2.64 hr and the corresponding minimum variability timescale of 18.21$\pm$4.87 min which is similar to the variability timescale of 17.10$\pm$6.18 min reported by \cite{2023MNRAS.520.2633B}. Accepting the black hole mass estimate for S4 0954$+$65 of $3.3~\times~10^8~M_\odot$ \citep{2004ApJ...602..103F}, the corresponding event horizon crossing timescale is $\sim$ 27 minutes. Hence the detected variability timescale is less than the event horizon crossing timescale, supporting the expectation that substructures within the jet are responsible for the fastest changes. Using the minimum variability timescale, we can constrain the size of the emission region as
\begin{equation}
    R \leq c \tau \frac{\delta}{1+z} \leq 2.36 \times 10^{14} \left(\frac{\delta }{10}\right)\text{ cm.}
\end{equation}
Here, $\delta$ is the Doppler factor which depends on the viewing angle, $\theta$, and the velocity, $\beta c$, of the jet as   $\delta = 1/(\Gamma[1 - \beta \cos \theta])$, where the bulk Lorentz factor $\Gamma
 = 1/\sqrt{(1 - \beta^2)}$.  
 Assuming a typical value of $\delta$ = 10 \citep{2022ApJS..260...12W}, the size of the emitting region $R \leq 2.36 \times 10^{14}$ cm. Such a compact (< 10$^{-3}$ pc) optical emission region has also been reported for the blazar S5 0716$+$714 by \cite{2021MNRAS.501.1100R}. 
 
This indicates that the IDV fluctuations originate from very compact ($R < 10^{-3}$ pc) regions. We also noticed variations in PD and PA on IDV timescales. The maximum variation in PD was $\sim$ 10\% and the PA changed over different ranges. We detected a rapid PA rotation of $\sim$ 120 degrees within $\sim$ 3 hr on 2022 April 13. In literature, the fastest PA rotation to date was reported by \cite{2018A&A...619A..45M} in blazar S5 0716$+$714, showing a change of 300 degrees in PA  in just 3.6 hr.

The stochastic acceleration of particles in a turbulent region within the jet can account for such rapid fluctuations originating from relatively compact regions \citep{2014ApJ...780...87M,2016ApJ...820...12P,2021ApJ...912..109K}. In such scenarios,  the randomness of the magnetic field direction within turbulent cells would produce random PA rotations.  Variant models such as the ``striped blazar jet'' scenario of \cite{2021MNRAS.502.1145Z} can also produce rapid particle acceleration in small volumes that yield flux changes similar to the observed IDV.

Throughout the course of our observations, the maximum change in the source brightness was $\sim$ 4 magnitudes. These flux variations on longer timescales in blazars depend on a number of parameters, including propagating shocks and changes in the injected spectral index, magnetic field, Doppler factor, and/or the density of particles within the emission region. The linear relationship we observed between different waveband light curves showed that the SED of the variable component remains constant during the monitoring period, which is further confirmed by the lack of a significant correlation between V$-$I colour and I-mag. To investigate the role of the magnetic field in these longer-term optical variations, we examined the correlation between PD and optical flux for the entire duration of these observations. The lack of a substantial relationship between these quantities indicates that the flux changes in S4 0954$+$65 are not primarily caused by the magnetic field.

The long-term flux variations in blazar S4 0954$+$65 can very possibly be explained by changes in the viewing angle, and hence in the Doppler factor, using a helical jet model \citep[e.g.,][]{1999A&A...347...30V,2021MNRAS.504.5629R}. In this scenario, the relativistic plasma flows continuously in an inhomogeneous helical jet such that each part of the jet, located at a given distance from the jet apex, produces a constant flux controlled by its local physical parameters such as magnetic field, optical depth, and particle density. The twisting of the helical jet causes flux variations over time, while all other jet parameters are considered to remain constant. The observed flux depends on the Doppler factor as F$^{obs}_\nu$ = $\delta^{2+\alpha}$ F$^{em}_\nu$, where $\alpha$ is the power-law spectral index and F$^{em}_\nu$ is the emitted flux. So, whenever the emitting region aligns closely to the observer's line of sight there is an increase in the $\delta$ and hence a flare (increased flux) is observed. 

If the long-term variations in the optical light curve of S4 0954$+$65 are only due to geometrical reasons such as a change in the orientation of the emitting region and consequent change in the Doppler factor, $\delta$, then $\delta = \delta_{\rm max}(F_{\nu}/F_{\nu, \rm max})^{1/2+\alpha}$. Taking $\alpha = 1.37$ as obtained in Section \ref{subsec:op_spec_var} and a tentative value of  $\delta_{\rm max} = 16$ \citep{2021MNRAS.504.5629R}, the Doppler factor should vary from $\delta_{\rm min} \sim 7$ to $\delta_{\rm max} = 16$ to explain the variations in the weekly binned R-band light curve (see Figure \ref{fig:delta_var}). Using the definition of $\delta$, the corresponding change in the viewing angle can also be estimated as follows:
\begin{equation}
    \theta = \text{arccos} \left (\frac{\Gamma \delta - 1 }{\delta \sqrt{\Gamma^2 - 1}}    \right ) .
\end{equation}
Adopting $\Gamma = 38.8$ from \cite{2022ApJS..260...12W}, we find $\theta_{\rm min}~\sim ~2.9^{\circ}$ for $\delta_{\rm max}~=~16$ and $\theta_{\rm max} ~\sim~4.7^{\circ}$ for $\delta_{\rm min}~=~7$ (Figure \ref{fig:delta_var}). This helical jet model has been previously used to explain the long-term optical flux variations in S4 0954$+$65 \citep{2021MNRAS.504.5629R}.

In the present work, we investigated the physical mechanisms causing the optical flux and polarization fluctuations of the TeV blazar S4 0954$+$65 on IDV and LTV timescales. By analysing an extensive IDV data set, we found that the IDV flux and polarization variations could be explained by the acceleration of the particles in the turbulent medium within the relativistic jet. On LTV timescales, the spectral variability and correlation between optical brightness and PD provided hints that the changes in the spectral index and magnetic field are not the primary factors responsible for the long-term flux variations in S4 0954$+$65. We discussed the change in the Doppler factor as a possible cause for the LTV variations and estimated the corresponding variations in the viewing angle of the emitting region. However, in order to fully comprehend the long-term flux variations of S4 0954$+$65, a further study employing a multi-wavelength data set is necessary, which will be the subject of our future work.

\section{Summary} \label{sec:sum}
We studied the flux, colour, and polarization fluctuations of blazar S4 0954$+$65 on diverse timescales from 2017 to 2023 using multiband optical photometry and R-band polarimetry observations. Our key findings are summarized as follows:
   \begin{enumerate}
      \item On IDV timescales, we found significant flux variation in 59 out of 138 light curves. The variability amplitudes ranged from  5.7$\pm$0.4 to 56.5$\pm$11.4\%.  
      \item The observed minimum variability timescale was 18.21$\pm$4.87 minutes indicating a compact (10$^{-4}$ pc) emitting region. 
      \item The acceleration of particles by shock in a turbulent plasma may cause flux variations on IDV timescales. 
      \item On longer timescales the brightness of S4 0954$+$65 varied by $\sim$ 4 mag, while no spectral variability was detected. The power-law spectral index for relative SED was found to be 1.37$\pm$0.04. 
      \item We observed a PA rotation of $\sim$120 degrees in $\sim$ 3 hr. 
      \item No correlation was detected between the optical flux and PD. 
      \item Long-term flux variations may be caused by the change in the Doppler factor of the emission region.
   \end{enumerate}

\begin{acknowledgements}
We thank the anonymous referee for their thoughtful comments that helped  improve the manuscript. Part of this work was supported by the Polish Funding Agency National Science Centre, project 2017/26/A/ST9/00756 (MAESTRO 9). This project has received funding from the European Research Council (ERC) under the European Union’s Horizon 2020 research and innovation program (grant agreement No. [951549]). This research was partially supported by the Bulgarian National Science
Fund of the Ministry of Education and Science under grants KP-06-H38/4
(2019), KP-06-KITAJ/2 (2020) and KP-06-H68/4 (2022).

\end{acknowledgements}

\bibliographystyle{aa} 
\bibliography{master} 


%

\onecolumn
\begin{appendix} 

\section{Observation log}
\begin{longtable}{cccc}
\caption{\label{tab:opt_log}Optical photometric observation log.}\\
\hline
\hline
Obs date   & Telescope  &   Duration (hr) & B,V,R,I \\
\hline
\endfirsthead
\caption{continued}\\
\hline
Obs date   & Telescope  &   Duration (hr) & B,V,R,I \\
\hline
\endhead
\hline
\endfoot
\hline
\endlastfoot
21-03-2017 &  A  & 0.07  &  3,2,2,2 \\ 
22-03-2017 &  A  & 2.96  &  0,27,26,27 \\ 
25-03-2017 &  A  & 3.10  &  0,27,27,27 \\ 
28-03-2017 &  A  & 3.89  &  0,26,26,27 \\ 
09-02-2018 &  A  & 3.30  &  0,30,30,30 \\ 
16-02-2018 &  A  & 1.73  &  0,12,11,12 \\ 
18-04-2018 &  A  & 3.36  &  4,28,30,28 \\ 
19-04-2018 &  A  & 3.82  &  4,34,34,34 \\ 
21-04-2018 &  A  & 3.71  &  5,29,30,30 \\ 
08-05-2018 &  A  & 3.26  &  0,30,30,29 \\ 
11-05-2018 &  A  & 2.58  &  0,22,22,23 \\ 
01-02-2019 &  A  & 1.38  &  2,12,12,12 \\ 
27-02-2019 &  A  & 1.87  &  2,17,17,17 \\ 
28-02-2019 &  A  & 2.73  &  3,25,25,23 \\ 
28-03-2019 &  A  & 2.15  &  0,21,19,20 \\ 
24-02-2020 &  A  & 5.40  &  6,44,46,42 \\ 
25-02-2020 &  A  & 3.62  &  5,33,32,32 \\ 
26-02-2020 &  A  & 2.57  &  3,24,23,24 \\ 
27-02-2020 &  A  & 3.01  &  2,28,27,27 \\ 
22-04-2020 &  A  & 4.62  &  7,40,38,38 \\ 
23-04-2020 &  A  & 6.48  &  6,56,57,56 \\ 
24-04-2020 &  A  & 5.26  &  5,42,42,41 \\ 
25-04-2020 &  A  & 3.59  &  2,10,9,9 \\ 
26-04-2020 &  A  & 0.43  &  1,5,4,3 \\ 
27-04-2020 &  A  & 2.86  &  3,26,25,26 \\ 
30-04-2020 &  A  & 6.97  &  7,54,51,49 \\ 
23-05-2020 &  A  & 2.06  &  2,19,19,19 \\ 
24-05-2020 &  A  & 1.46  &  2,15,14,14 \\ 
07-03-2021 &  A  & 3.55  &  6,26,26,26 \\ 
08-03-2021 &  A  & 4.70  &  6,38,37,37 \\ 
13-03-2021 &  A  & 4.56  &  8,34,34,34 \\ 
10-04-2021 &  A  & 0.07  &  3,2,2,2 \\ 
11-04-2021 &  A  & 0.04  &  0,2,2,2 \\ 
15-04-2021 &  A  & 0.11  &  0,2,2,2 \\ 
16-04-2021 &  A  & 0.11  &  0,2,1,1 \\ 
13-05-2021 &  A  & 0.04  &  0,2,1,2 \\ 
14-05-2021 &  A  & 0.04  &  0,2,2,2 \\ 
17-06-2021 &  A  & 0.07  &  3,2,2,2 \\ 
20-06-2021 &  A  & 0.07  &  3,2,3,1 \\ 
21-06-2021 &  A  & 0.04  &  0,2,2,2 \\ 
12-02-2022 &  A  & 0.33  &  0,2,2,2 \\ 
14-03-2022 &  A  & 4.06  &  5,20,19,19 \\ 
11-04-2022 &  A  & 8.51  &  10,55,53,54 \\ 
12-04-2022 &  A  & 8.39  &  7,54,53,53 \\ 
13-04-2022 &  A  & 8.27  &  9,49,48,48 \\ 
14-04-2022 &  A  & 3.50  &  4,25,25,26 \\ 
15-04-2022 &  A  & 4.97  &  4,30,30,31 \\ 
02-06-2022 &  A  & 4.87  &  0,35,37,36 \\ 
03-06-2022 &  A  & 4.96  &  0,37,34,34 \\ 
04-06-2022 &  A  & 4.29  &  0,31,34,34 \\ 
05-06-2022 &  A  & 1.07  &  0,9,9,9 \\ 
18-06-2022 &  A  & 1.65  &  0,13,12,13 \\ 
20-06-2022 &  A  & 0.22  &  0,2,2,2 \\ 
23-07-2022 &  A  & 0.03  &  1,2,2,1 \\ 
29-10-2022 &  B  & 3.67  &  106,0,203,0 \\ 
31-10-2022 &  B  & 2.36  &  68,0,242,0 \\ 
13-02-2023 &  A  & 0.22  &  0,2,2,2 \\ 
14-02-2023 &  A  & 0.22  &  0,3,3,3 \\ 
15-02-2023 &  A  & 2.41  &  0,12,12,12 \\ 
16-02-2023 &  A  & 1.54  &  0,8,8,8 \\ 
17-02-2023 &  A  & 0.22  &  0,2,2,2 \\ 
18-03-2023 &  A  & 0.36  &  2,2,2,2 \\ 
19-03-2023 &  A  & 1.93  &  3,8,8,8 \\ 
22-03-2023 &  A  & 3.70  &  6,15,16,15 \\ 
23-03-2023 &  A  & 3.33  &  5,16,18,18 \\ 
24-03-2023 &  A  & 4.67  &  6,22,22,22 \\ 
26-03-2023 &  C  & 5.01  &  0,0,7,0 \\ 
28-03-2023 &  C  & 8.24  &  0,5,6,4 \\ 
29-03-2023 &  C  & 8.23  &  0,5,5,5 \\ 
30-03-2023 &  C  & 7.04  &  0,5,5,5 \\ 
04-04-2023 &  C  & 8.24  &  0,5,5,5 \\ 
05-04-2023 &  C  & 8.23  &  0,6,6,6 \\ 
06-04-2023 &  C  & 8.24  &  0,5,5,4 \\ 
08-04-2023 &  C  & 7.03  &  0,6,6,6 \\ 
09-04-2023 &  C  & 4.64  &  0,4,4,3 \\ 
11-04-2023 &  C  & 7.03  &  0,5,4,4 \\ 
13-04-2023 &  C  & 0.00  &  0,1,1,1 \\ 
15-04-2023 &  C  & 6.99  &  0,3,4,3 \\ 
16-04-2023 &  C  & 7.04  &  0,6,5,6 \\ 
17-04-2023 &  C  & 7.04  &  0,7,7,6 \\ 
18-04-2023 &  C  & 3.44  &  0,3,3,3 \\ 
19-04-2023 &  C  & 5.87  &  0,4,3,2 \\ 
20-04-2023 &  C  & 7.17  &  0,6,6,6 \\ 
21-04-2023 &  C  & 6.00  &  0,3,3,3 \\ 
22-04-2023 &  C  & 1.23  &  0,2,2,2 \\ 
23-04-2023 &  C  & 7.19  &  0,6,6,5 \\ 
24-04-2023 &  C  & 5.99  &  0,6,6,6 \\ 
27-04-2023 &  A  & 4.24  &  0,16,15,15 \\ 
28-04-2023 &  A  & 2.81  &  0,12,12,12 \\ 

\end{longtable}

\section{Results of variability analyses.}
    
\begin{longtable}{cccccccccc}
\caption{\label{tab:var_res}Results of IDV analyses of the TeV blazar S4 0954$+$65. The values of variability amplitude, $\Delta$mag, and $\tau_{var}$ are only mentioned for the variable light curves. For non-variable light curves, we put a '-'. Also, when $\tau_{var} < \sigma$ ($\tau_{var}$) we use a '-' for $\tau_{var}$. A 
$^*$ denotes that the detected variability timescale is greater than the duration of the light curve. }\\
\hline
\hline
Observation date & Start MJD  & Band & DoF & $F_{enh}$ & $F_c$  & Status & Amplitude & $\Delta$ mag & $\tau_{\rm var}$\\
dd-mm-yyyy &                  &      &    $\nu_1$,$\nu_2$ &           &        &        &   (\%)    & &  (in min)   \\ 
\hline
\endfirsthead
\caption{continued} \\
\hline
Observation date & Start MJD  & Band & DoF & $F_{enh}$ & $F_c$  & Status & Amplitude & $\Delta$ mag & $\tau_{\rm var}$\\
dd-mm-yyyy &                  &      &    $\nu_1$,$\nu_2$ &           &        &        &    (\%)  &&   (in min)   \\ 
\hline
\endhead
\hline
\endfoot
\hline
\endlastfoot
22-03-2017  &  57834.83372   &   V &  26, 52 & 0.71 & 2.14 & NV & - &  - & - \\ 
            &  57834.83524   &   R &  25, 50 & 0.98 & 2.17 & NV & - &  - & - \\ 
	    &  57834.83671   &   I &  26, 52 & 0.55 & 2.14 & NV & - &  - & - \\ 
	    
25-03-2017  &  57837.81406   &   V &  26, 52 & 5.24 & 2.14 & V  &  56.5$\pm$11.4  &  0.58$\pm$0.11  &  18.21$\pm$4.87 \\ 
	    &  57837.81553   &   R &  26, 52 & 3.06 & 2.14 & V  &  32.3$\pm$8.3  &  0.33$\pm$0.08  &  -\\ 
	    &  57837.81700   &   I &  26, 52 & 3.04 & 2.14 & V  &  35.4$\pm$8.4  &  0.36$\pm$0.08  &  - \\ 
	    
28-03-2017  &  57840.76318   &   V &  25, 50 & 0.12 & 2.17 & NV & - &  - & - \\ 
	    &  57840.76471   &   R &  25, 50 & 0.07 & 2.17 & NV & - &  - & - \\ 
	    &  57840.76618   &   I &  25, 50 & 0.14 & 2.17 & NV & - &  - & - \\ 
	    
09-02-2018  &  58158.92130   &   V &  29, 58 & 0.43 & 2.05 & NV & - &  - & - \\ 
	    &  58158.92277   &   R &  29, 58 & 0.20 & 2.05 & NV & - &  - & - \\ 
	    &  58158.92422   &   I &  29, 58 & 0.38 & 2.05 & NV & - &  - & - \\ 
	    
16-02-2018  &  58165.87456   &   V &  11, 22 & 0.41 & 3.18 & NV & - &  - & - \\ 
	    &  58165.87608   &   R &  10, 20 & 0.57 & 3.37 & NV & - &  - & - \\ 
	    &  58165.87753   &   I &  10, 20 & 0.71 & 3.37 & NV & - &  - & - \\ 
	    
18-04-2018  &  58226.76861   &   V &  27, 54 & 4.36 & 2.11 & V  &  29.1$\pm$3.5 &  0.29$\pm$0.03  &  - \\ 
	    &  58226.76421   &   R &  29, 58 & 2.33 & 2.05 & V  &  28.5$\pm$2.5 &  0.29$\pm$0.03  &  - \\ 
	    &  58226.76567   &   I &  27, 54 & 2.85 & 2.11 & V  &  28.2$\pm$2.4 &  0.28$\pm$0.02  &  114.81$\pm$32.51 \\ 
	    
19-04-2018  &  58227.77932   &   V &  33, 66 & 2.48 & 1.96 & V  &  16.9$\pm$3.3 &  0.17$\pm$0.03  &  64.21$\pm$19.44 \\ 
	    &  58227.78079   &   R &  33, 66 & 4.28 & 1.96 & V  &  19.1$\pm$2.5 &  0.19$\pm$0.02  &  104.77$\pm$33.64 \\ 
	    &  58227.78225   &   I &  33, 66 & 2.35 & 1.96 & V  &  14.8$\pm$2.3 &  0.15$\pm$0.02  &  - \\ 
	    
21-04-2018  &  58229.77088   &   V &  28, 56 & 2.11 & 2.08 & V  &  28.8$\pm$3.7 &  0.29$\pm$0.04  &  - \\ 
	    &  58229.77234   &   R &  29, 58 & 2.14 & 2.05 & V  &  24.3$\pm$2.4 &  0.24$\pm$0.02  &  137.57$\pm$39.77 \\ 
	    &  58229.77380   &   I &  29, 58 & 1.84 & 2.05 & NV & - &  - & - \\ 
	    
08-05-2018  &  58246.78161   &   V &  29, 58 & 0.27 & 2.05 & NV & - &  - & - \\ 
	    &  58246.78308   &   R &  29, 58 & 0.08 & 2.05 & NV & - &  - & - \\ 
	    &  58246.79042   &   I &  28, 56 & 0.08 & 2.08 & NV & - &  - & - \\ 
	    
11-05-2018  &  58249.79575   &   V &  19, 38 & 1.24 & 2.42 & NV & - &  - & - \\ 
	    &  58249.78250   &   R &  21, 42 & 0.60 & 2.32 & NV & - &  - & - \\ 
	    &  58249.78396   &   I &  22, 44 & 0.66 & 2.28 & NV & - &  - & - \\ 
	    
01-02-2019  &  58515.11557   &   V &  11, 22 & 0.69 & 3.18 & NV & - &  - & - \\ 
	    &  58515.11704   &   R &  11, 22 & 0.92 & 3.18 & NV & - &  - & - \\ 
	    &  58515.11850   &   I &  11, 22 & 1.99 & 3.18 & NV & - &  - & - \\ 
	    
27-02-2019  &  58541.99293   &   V &  16, 32 & 0.55 & 2.62 & NV & - &  - & - \\ 
	    &  58541.99440   &   R &  16, 32 & 0.39 & 2.62 & NV & - &  - & - \\ 
	    &  58541.99586   &   I &  16, 32 & 1.47 & 2.62 & NV & - &  - & - \\ 
	    
28-02-2019  &  58542.98566   &   V &  24, 48 & 0.55 & 2.20 & NV & - &  - & - \\ 
	    &  58542.98712   &   R &  24, 48 & 0.75 & 2.20 & NV & - &  - & - \\ 
	    &  58542.98858   &   I &  22, 44 & 0.37 & 2.28 & NV & - &  - & - \\ 
	    
28-03-2019  &  58570.96037   &   V &  19, 38 & 1.76 & 2.42 & NV & - &  - & - \\ 
	    &  58570.96189   &   R &  18, 36 & 2.38 & 2.48 & NV & - &  - & - \\ 
	    &  58570.96334   &   I &  19, 38 & 2.84 & 2.42 & V  &  9.7$\pm$3.2 &  0.10$\pm$0.03  &  - \\ 
	    
24-02-2020  &  58903.84796   &   V &  43, 86 & 0.71 & 1.81 & NV & - &  - & - \\ 
	    &  58903.84943   &   R &  45, 90 & 0.73 & 1.79 & NV & - &  - & - \\ 
	    &  58903.85681   &   I &  41, 82 & 0.70 & 1.84 & NV & - &  - & - \\ 
	    
25-02-2020  &  58904.92771   &   V &  31, 62 & 0.97 & 2.01 & NV & - &  - & - \\ 
	    &  58904.92918   &   R &  31, 62 & 0.82 & 2.01 & NV & - &  - & - \\ 
	    &  58904.93065   &   I &  31, 62 & 0.38 & 2.01 & NV & - &  - & - \\ 
	    
26-02-2020  &  58905.84214   &   V &  22, 44 & 1.53 & 2.28 & NV & - &  - & - \\ 
	    &  58905.84361   &   R &  22, 44 & 1.04 & 2.28 & NV & - &  - & - \\ 
	    &  58905.84508   &   I &  23, 46 & 3.04 & 2.24 & V  &  21.8$\pm$3.4 &  0.22$\pm$0.03  &  - \\ 
	    
27-02-2020  &  58906.87792   &   V &  26, 52 & 0.54 & 2.14 & NV & - &  - & - \\ 
	    &  58906.87939   &   R &  26, 52 & 0.46 & 2.14 & NV & - &  - & - \\ 
	    &  58906.88117   &   I &  26, 52 & 0.49 & 2.14 & NV & - &  - & - \\ 
	    
22-04-2020  &  58961.80124   &   V &  39, 78 & 2.56 & 1.86 & V  &  10.2$\pm$3.9 &  0.11$\pm$0.04  &  - \\ 
	    &  58961.80271   &   R &  37, 74 & 3.52 & 1.89 & V  &  11.4$\pm$2.3 &  0.12$\pm$0.02  &  - \\ 
	    &  58961.80418   &   I &  37, 74 & 3.27 & 1.89 & V  &  11.2$\pm$3.7 &  0.12$\pm$0.04  &  -  \\ 
	    
23-04-2020  &  58962.78927   &   V &  55,110 & 0.52 & 1.69 & NV &- &  - & - \\ 
	    &  58962.79074   &   R &  56,112 & 0.90 & 1.68 & NV & - &  - & - \\ 
	    &  58962.79221   &   I &  55,110 & 0.64 & 1.69 & NV & - &  - & - \\ 
	    
24-04-2020  &  58963.79951   &   V &  41, 82 & 1.02 & 1.84 & NV & - &  - & - \\ 
	    &  58963.80098   &   R &  41, 82 & 0.93 & 1.84 & NV & - &  - & - \\ 
	    &  58963.80247   &   I &  40, 80 & 0.69 & 1.85 & NV & - &  - & - \\ 
	    
25-04-2020  &  58964.81051   &   V &   9, 18 & 1.10 & 3.60 & NV & - &  - & - \\ 
27-04-2020  &  58966.79419   &   V &  25, 50 & 0.27 & 2.17 & NV & - &  - & - \\ 
	    &  58966.79566   &   R &  24, 48 & 0.23 & 2.20 & NV & - &  - & - \\ 
	    &  58966.79713   &   I &  25, 50 & 0.28 & 2.17 & NV & - &  - & - \\ 
	    
30-04-2020  &  58969.77789   &   V &  52,104 & 0.32 & 1.72 & NV & - &  - & - \\ 
	    &  58969.77936   &   R &  50,100 & 0.38 & 1.74 & NV & - &  - & - \\ 
	    &  58969.78083   &   I &  48, 96 & 0.30 & 1.75 & NV & - &  - & - \\ 
	    
23-05-2020  &  58992.83843   &   V &  18, 36 & 0.96 & 2.48 & NV & - &  - & - \\ 
	    &  58992.83990   &   R &  18, 36 & 0.67 & 2.48 & NV & - &  - & - \\ 
	    &  58992.84137   &   I &  18, 36 & 0.66 & 2.48 & NV & - &  - & - \\ 
	    
24-05-2020  &  58993.88975   &   V &  13, 26 & 1.18 & 2.90 & NV & - &  - & - \\ 
	    &  58993.89123   &   R &  13, 26 & 1.70 & 2.90 & NV & - &  - & - \\ 
	    &  58993.89270   &   I &  13, 26 & 1.16 & 2.90 & NV & - &  - & - \\ 
	    
07-03-2021  &  59280.87602   &   V &  25, 50 & 2.43 & 2.17 & V  &  13.4$\pm$4.8 &  0.14$\pm$0.05  &  - \\ 
	    &  59280.87749   &   R &  25, 50 & 4.13 & 2.17 & V  &  15.3$\pm$3.5 &  0.16$\pm$0.03  &  -  \\ 
	    &  59280.87896   &   I &  25, 50 & 16.67 & 2.17 & V  &  20.8$\pm$4.0 &  0.21$\pm$0.04  &  - \\ 
	    
08-03-2021  &  59281.86245   &   V &  36, 72 & 5.61 & 1.91 & V  &  29.4$\pm$3.8 &  0.30$\pm$0.04  &  - \\ 
	    &  59281.86392   &   R &  36, 72 & 6.28 & 1.91 & V  &  26.5$\pm$2.3 &  0.27$\pm$0.02  &  -  \\ 
	    &  59281.86539   &   I &  36, 72 & 18.07 & 1.91 & V  &  26.8$\pm$2.8 &  0.27$\pm$0.03  &  -  \\ 
	    
13-03-2021  &  59286.77670   &   V &  33, 66 & 4.31 & 1.96 & V  &  13.5$\pm$4.1 &  0.14$\pm$0.04  &  - \\ 
	    &  59286.77817   &   R &  33, 66 & 3.34 & 1.96 & V  &  10.2$\pm$3.3 &  0.11$\pm$0.03  &  - \\ 
	    &  59286.77964   &   I &  33, 66 & 1.99 & 1.96 & V  &  9.0$\pm$2.7 &  0.09$\pm$0.03  &  - \\ 
	    
14-03-2022  &  59652.94300   &   V &  18, 36 & 0.45 & 2.48 & NV & - &  - & - \\ 
	    &  59652.94500   &   R &  18, 36 & 0.47 & 2.48 & NV & - &  - & - \\ 
	    &  59652.94600   &   I &  18, 36 & 0.51 & 2.48 & NV & - &  - & - \\ 
	    
11-04-2022  &  59680.76247   &   B &   9, 18 & 26.71 & 3.60 & V  &  55.4$\pm$16.1 &  0.57$\pm$0.16  &  - \\ 
	    &  59680.76539   &   V &  53,106 & 31.53 & 1.71 & V  &  47.9$\pm$4.5 &  0.48$\pm$0.04  &  - \\ 
	    &  59680.76831   &   R &  52,104 & 28.16 & 1.72 & V  &  40.9$\pm$1.5 &  0.41$\pm$0.01  &  -  \\ 
	    &  59680.77124   &   I &  53,106 & 26.83 & 1.71 & V  &  39.5$\pm$1.9 &  0.40$\pm$0.02  &  106.98$\pm$30.28 \\ 
	    
12-04-2022  &  59681.75519   &   V &  52,104 & 6.46 & 1.72 & V  &  31.9$\pm$4.9 &  0.32$\pm$0.05  &  - \\ 
	    &  59681.75666   &   R &  52,104 & 5.96 & 1.72 & V  &  29.7$\pm$3.1 &  0.30$\pm$0.03  &  - \\ 
	    &  59681.75814   &   I &  52,104 & 3.56 & 1.72 & V  &  29.2$\pm$3.3 &  0.29$\pm$0.03  &  - \\ 
	    
13-04-2022  &  59682.75987   &   V &  47, 94 & 18.17 & 1.76 & V  & 54.3$\pm$4.3 &  0.54$\pm$0.04  &  - \\ 
	    &  59682.76134   &   R &  47, 94 & 16.99 & 1.76 & V  & 50.6$\pm$3.2 &  0.51$\pm$0.03  &  - \\ 
	    &  59682.76281   &   I &  47, 94 & 11.38 & 1.76 & V  & 45.6$\pm$3.1 &  0.46$\pm$0.03  &  - \\ 
	    
14-04-2022  &  59683.75852   &   V &  23, 46 & 3.23 & 2.24 & V  &  20.1$\pm$15.8 &  0.22$\pm$0.14  &  -\\ 
	    &  59683.75999   &   R &  23, 46 & 3.41 & 2.24 & V  &  27.3$\pm$8.9 &  0.28$\pm$0.09  &  - \\ 
	    &  59683.76147   &   I &  24, 48 & 3.63 & 2.20 & V  &  23.2$\pm$4.5 &  0.24$\pm$0.04  &  - \\ 
	    
15-04-2022  &  59684.80767   &   V &  28, 56 & 5.41 & 2.08 & V  &  23.2$\pm$4.1 &  0.24$\pm$0.04  &  -  \\ 
	    &  59684.80469   &   R &  29, 58 & 8.74 & 2.05 & V  &  22.9$\pm$2.6 &  0.23$\pm$0.03  &  96.92$\pm$28.58 \\ 
	    &  59684.78242   &   I &  30, 60 & 4.10 & 2.03 & V  &  25.7$\pm$4.8 &  0.26$\pm$0.05  &  - \\ 
	    
02-06-2022  &  59732.82031   &   V &  34, 68 & 1.57 & 1.95 & NV & - &  - & - \\ 
	    &  59732.82190   &   R &  35, 70 & 2.06 & 1.93 & V  &  39.9$\pm$4.8 &  0.40$\pm$0.05  &  -\\ 
	    &  59732.82337   &   I &  35, 70 & 1.64 & 1.93 & NV & - &  - & - \\ 
	    
03-06-2022  &  59733.80929   &   V &  36, 72 & 0.59 & 1.91 & NV & - &  - & - \\ 
	    &  59733.81088   &   R &  33, 66 & 2.24 & 1.96 & V  &  31.5$\pm$12.3 &  0.32$\pm$0.12  &  -  \\ 
	    &  59733.81236   &   I &  33, 66 & 1.19 & 1.96 & NV & - &  - & - \\ 
	    
04-06-2022  &  59734.83396   &   V &  30, 60 & 1.97 & 2.03 & NV & - &  - & - \\ 
	    &  59734.83554   &   R &  32, 64 & 1.99 & 1.98 & V  &  39.7$\pm$5.5  &  0.40$\pm$0.06  &  - \\ 
	    &  59734.83701   &   I &  32, 64 & 1.22 & 1.98 & NV & - &  - & - \\ 
	    
18-06-2022  &  59748.85271   &   V &  12, 24 & 0.85 & 3.03 & NV & - &  - & - \\ 
	    &  59748.85429   &   R &  11, 22 & 0.24 & 3.18 & NV & - &  - & - \\ 
	    &  59748.85578   &   I &  12, 24 & 0.87 & 3.03 & NV & - &  - & - \\ 
	    
29-10-2022  &  59881.99244   &   B & 105,210 & 16.94 & 1.47 & V  &  16.4$\pm$1.5 &  0.17$\pm$0.01  &  - \\ 
	    &  59881.99213   &   R & 202,404 & 23.79 & 1.32 & V  &  10.2$\pm$0.5  &  0.10$\pm$0.01  &  80.06$\pm$23.16\\ 
	    
31-10-2022  &  59883.00814   &   B &  67,134 & 7.58 & 1.61 & V  &  7.6$\pm$0.5 &  0.08$\pm$0.01  &  - \\ 
	    &  59883.00758   &   R & 240,480 & 2.20 & 1.29 & V  &  5.7$\pm$0.4 &  0.06$\pm$0.01  &  35.64$\pm$8.40\\ 
	    
15-02-2023  &  59990.88028   &   V &  11, 22 & 0.32 & 3.18 & NV & - &  - & - \\ 
	    &  59990.88186   &   R &  11, 22 & 1.05 & 3.18 & NV & - &  - & - \\ 
	    &  59990.88333   &   I &  11, 22 & 0.39 & 3.18 & NV & - &  - & - \\ 
	    
22-03-2023  &  60025.92133   &   V &  13, 26 & 0.39 & 2.90 & NV & - &  - & - \\ 
	    &  60025.92280   &   R &  15, 30 & 1.10 & 2.70 & NV & - &  - & - \\ 
	    &  60025.92428   &   I &  14, 28 & 0.52 & 2.79 & NV & - &  - & - \\ 
	    
23-03-2023  &  60026.80009   &   V &  15, 30 & 12.67 & 2.70 & V  &  37.2$\pm$2.4 &  0.37$\pm$0.02  &  -\\ 
	    &  60026.80156   &   R &  17, 34 & 15.50 & 2.54 & V  &  37.6$\pm$1.4 &  0.38$\pm$0.01  &  -\\ 
	    &  60026.80303   &   I &  17, 34 & 9.32 & 2.54 & V  &  35.7$\pm$1.5 &  0.36$\pm$0.02  &  -\\ 
	    
24-03-2023  &  60027.81407   &   V &  21, 42 & 10.64 & 2.32 & V  &  13.8$\pm$2.0 &  0.14$\pm$0.02  &  - \\ 
	    &  60027.81554   &   R &  21, 42 & 10.90 & 2.32 & V  &  13.4$\pm$1.0 &  0.13$\pm$0.01  &  584.21$\pm$184.13$^*$  \\ 
	    &  60027.81701   &   I &  21, 42 & 6.93 & 2.32 & V  &  12.5$\pm$1.3  &  0.13$\pm$0.01  &  - \\ 
	    
27-04-2023  &  60061.79841   &   V &  15, 30 & 2.97 & 2.70 & V  &  8.6$\pm$3.1 &  0.09$\pm$0.03  &  - \\ 
	    &  60061.80000   &   R &  14, 28 & 3.08 & 2.79 & V  &  9.6$\pm$1.5 &  0.10$\pm$0.01  &  502.01$\pm$128.20$^*$ \\ 
	    &  60061.80147   &   I &  14, 28 & 1.49 & 2.79 & NV & - &  - & - \\ 
	    
28-04-2023  &  60062.80072   &   V &  11, 22 & 26.71 & 3.18 & V  &  29.1$\pm$2.8 &  0.29$\pm$0.03  &  171.61$\pm$50.30$^*$\\ 
	    &  60062.80230   &   R &  11, 22 & 21.66 & 3.18 & V  &  28.9$\pm$1.5  &  0.29$\pm$0.01  &  263.09$\pm$37.43$^*$ \\ 
	    &  60062.80377   &   I &  11, 22 & 11.64 & 3.18 & V  &  28.9$\pm$1.9  &   0.29$\pm$0.02  &  -\\              
\end{longtable}


\section{Intraday differential light curves of S4 0954$+$65.}
\begin{figure*}[!h]
\vbox{
\hbox{
\includegraphics[height=6cm,width=6cm]{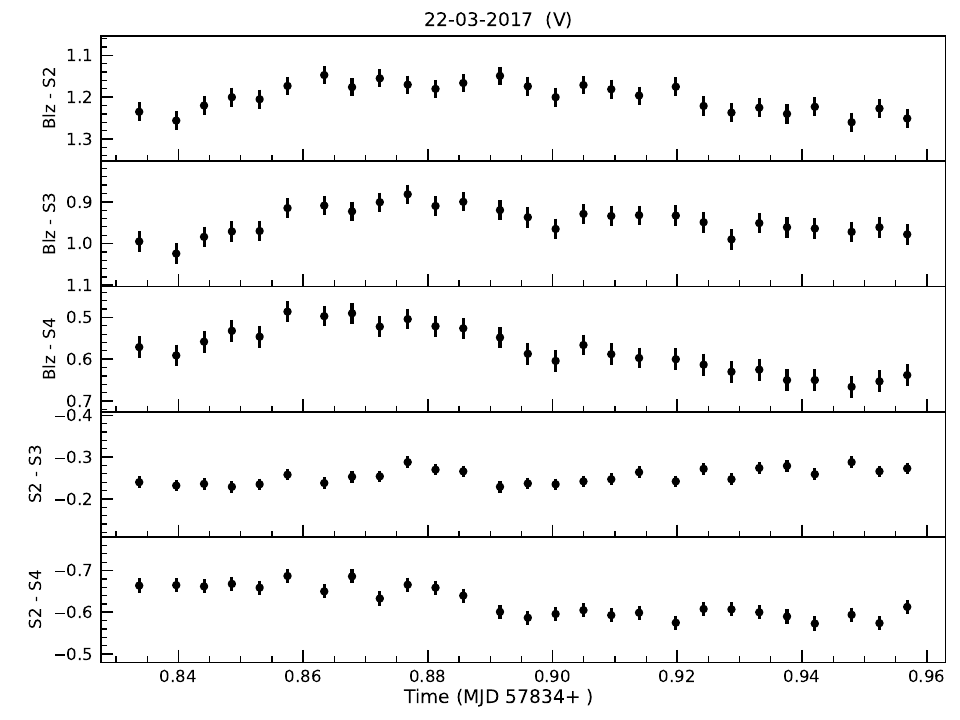}
\includegraphics[height=6cm,width=6cm]{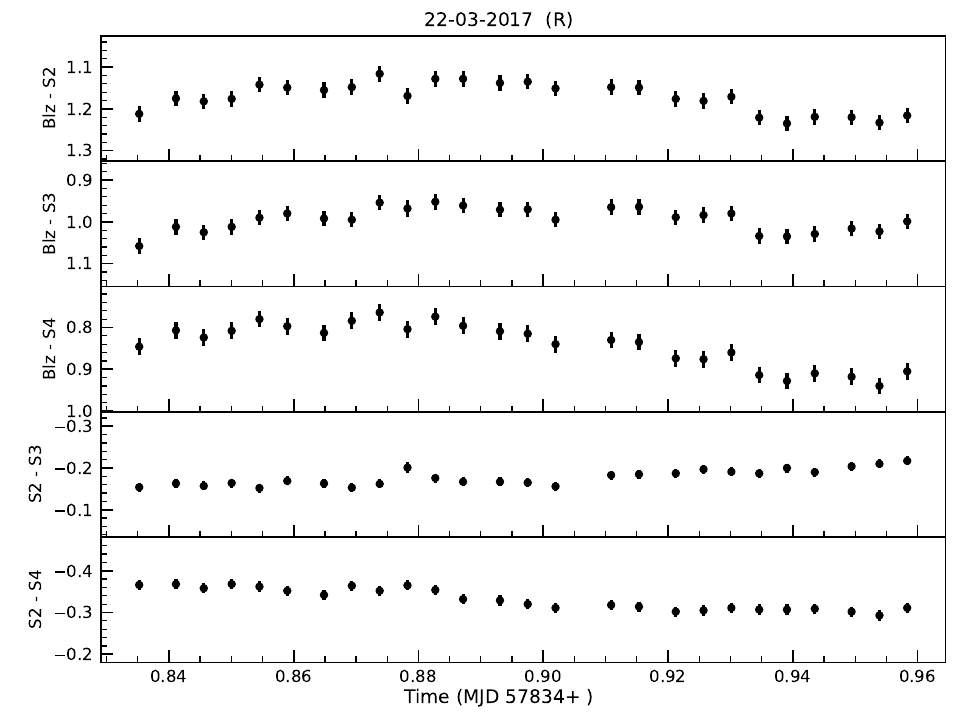}
\includegraphics[height=6cm,width=6cm]{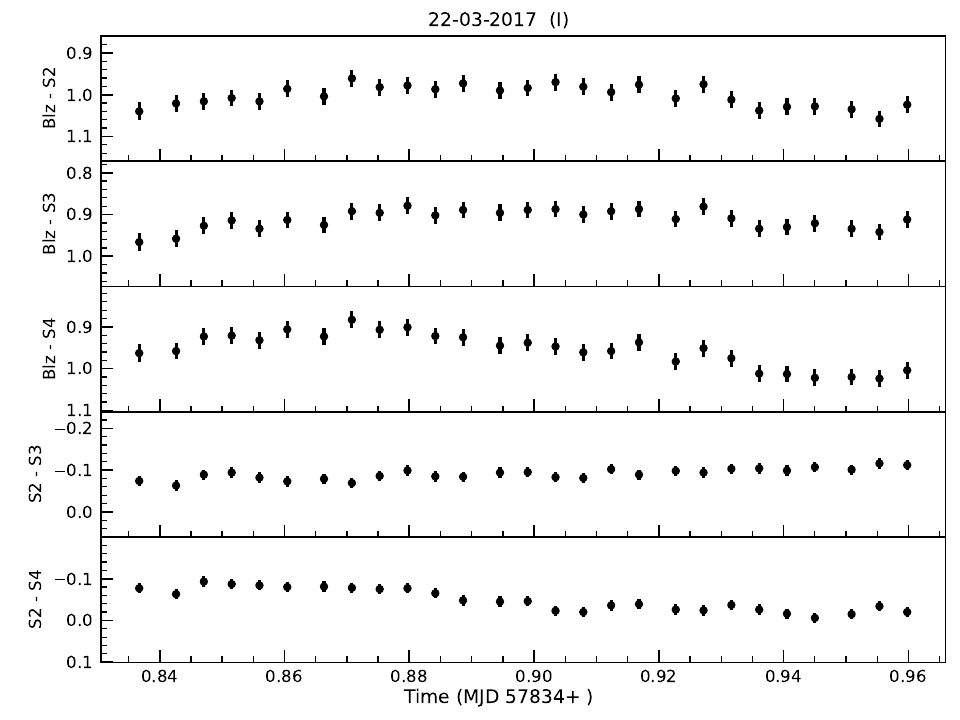}
}
\hbox{
\includegraphics[height=6cm,width=6cm]{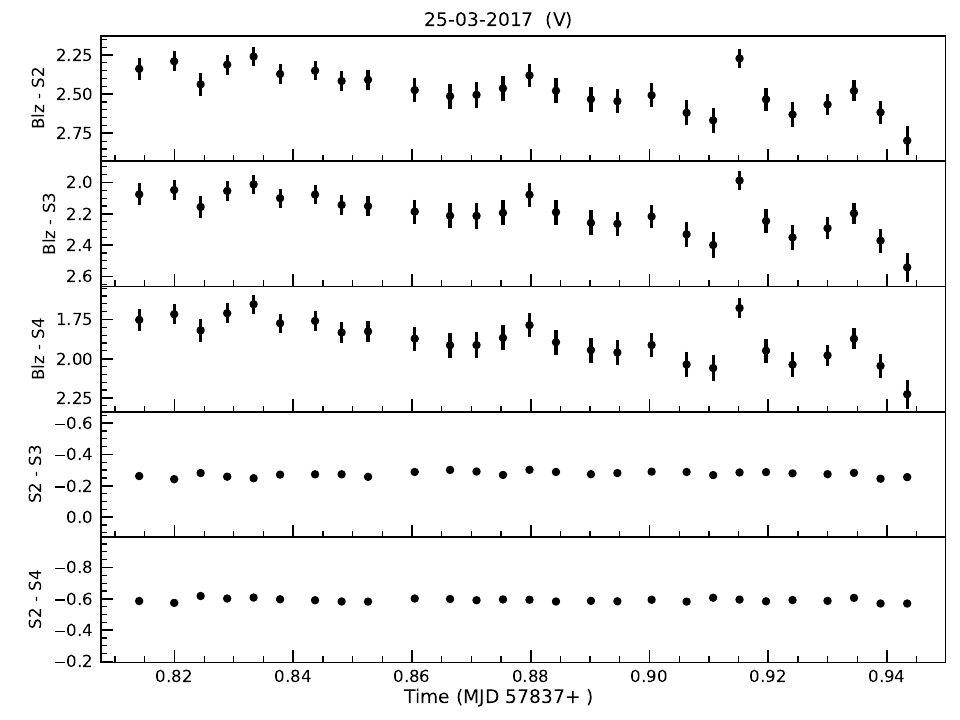}
\includegraphics[height=6cm,width=6cm]{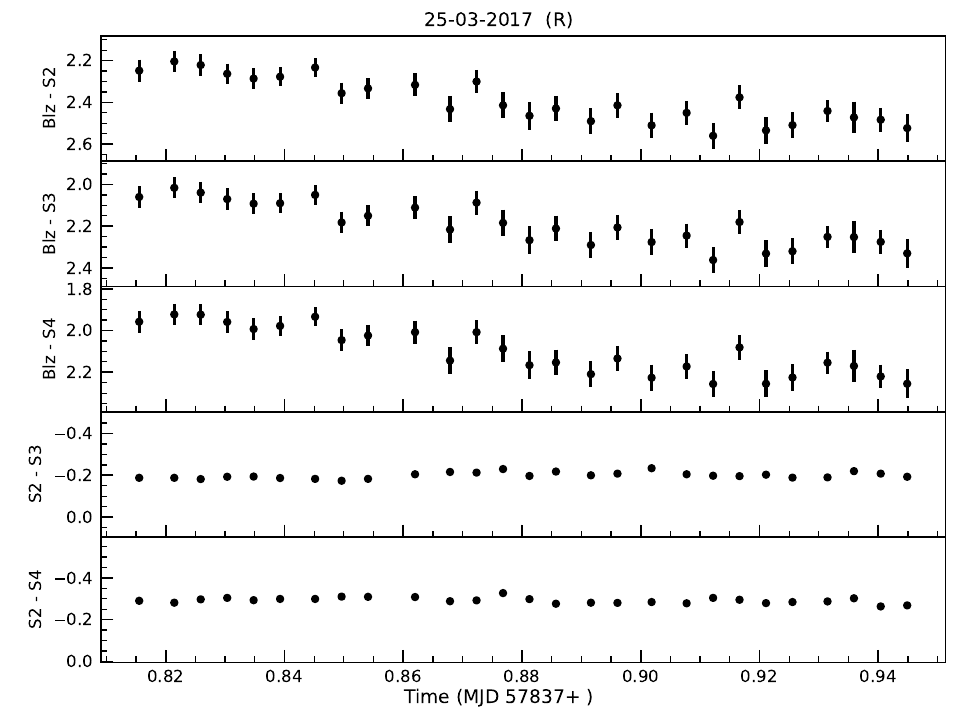}
\includegraphics[height=6cm,width=6cm]{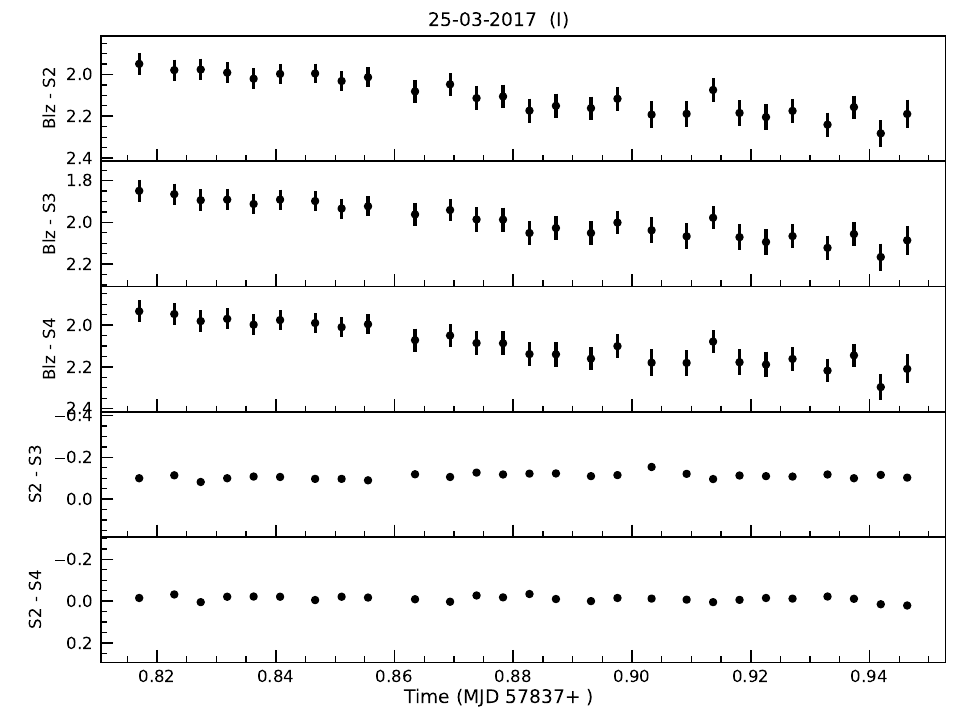}
}
\hbox{
\includegraphics[height=6cm,width=6cm]{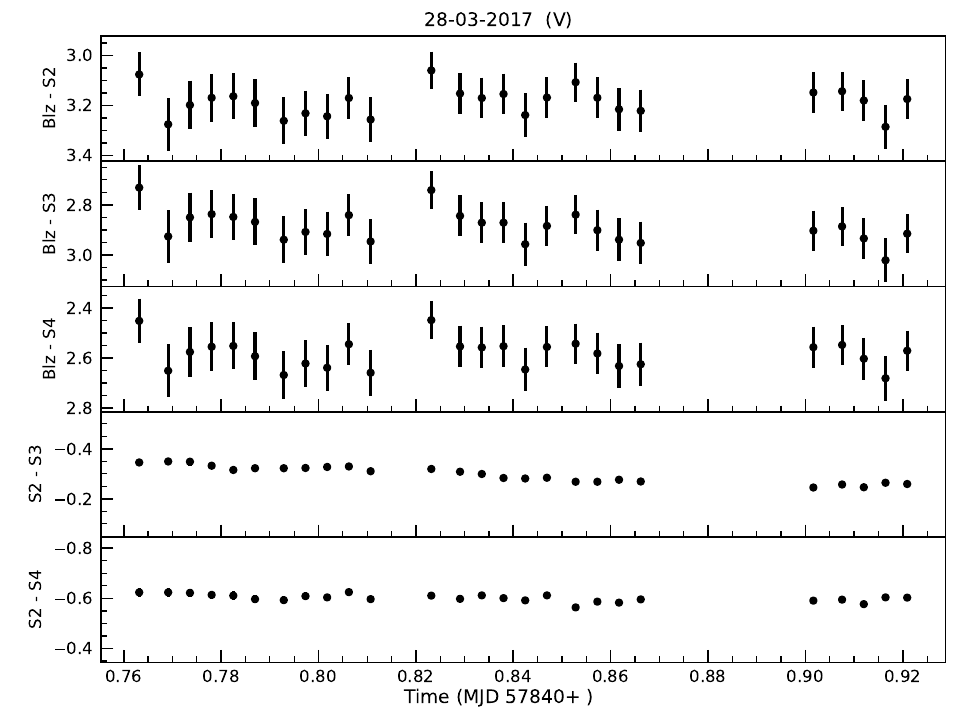}
\includegraphics[height=6cm,width=6cm]{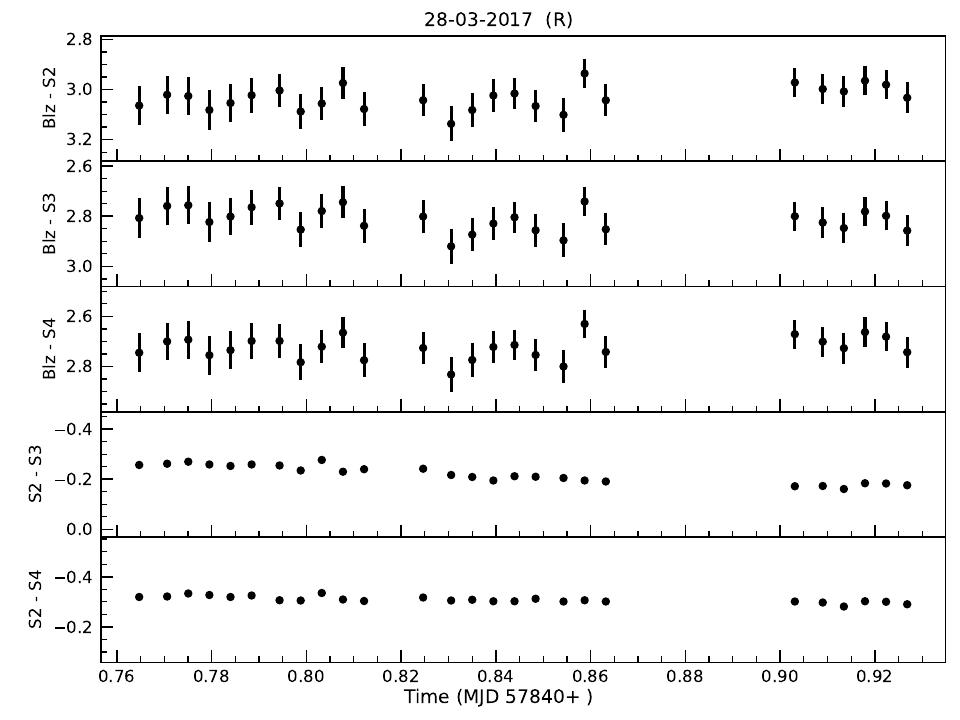}
\includegraphics[height=6cm,width=6cm]{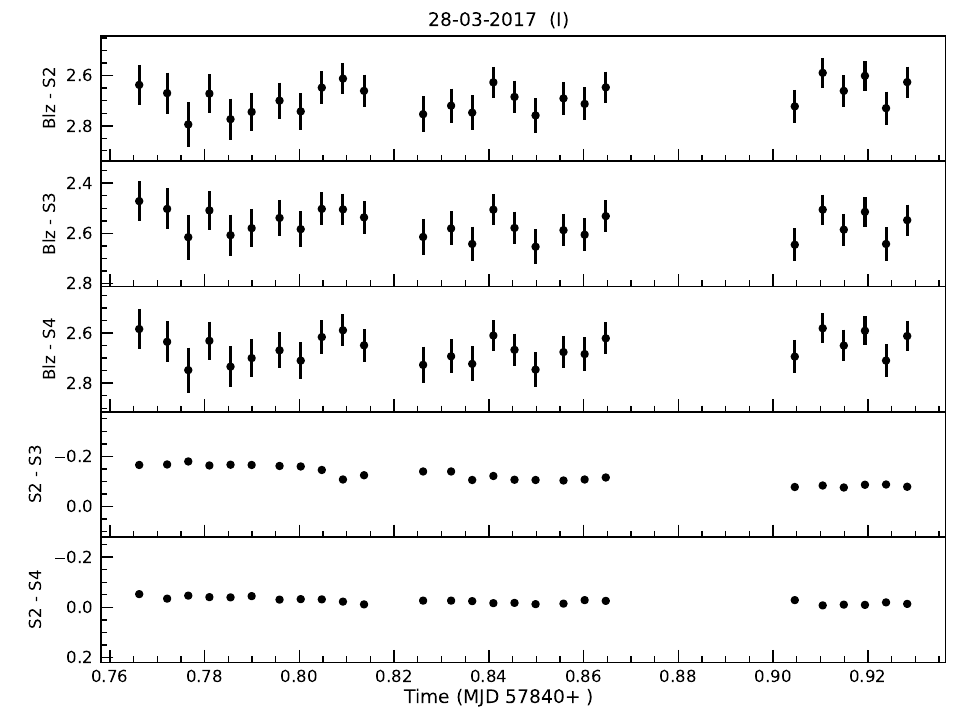}
}
\hbox{
\includegraphics[height=6cm,width=6cm]{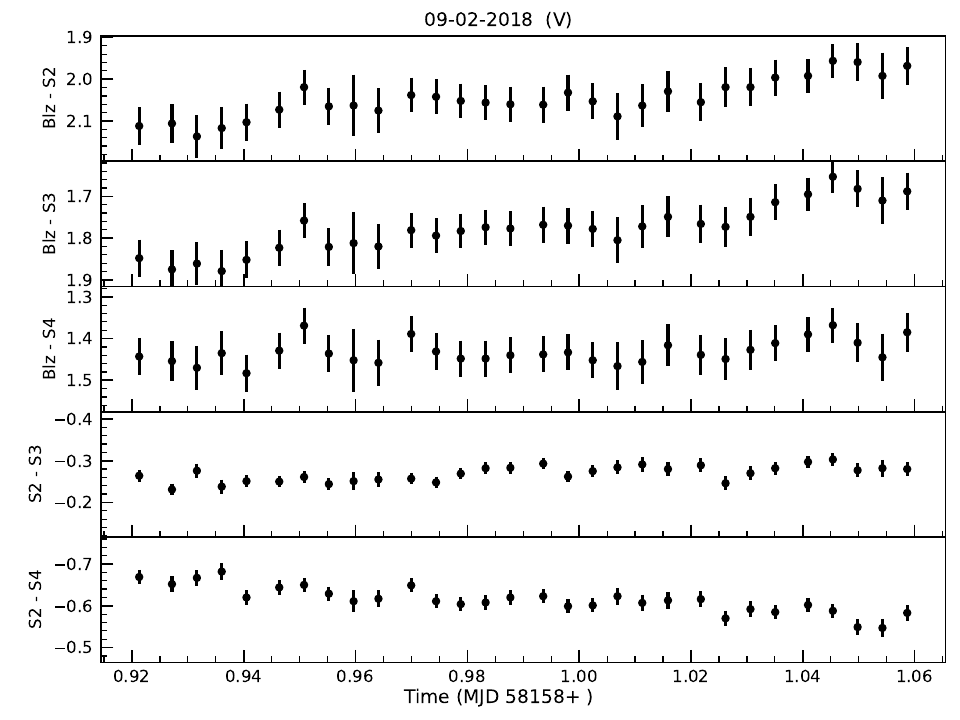}
\includegraphics[height=6cm,width=6cm]{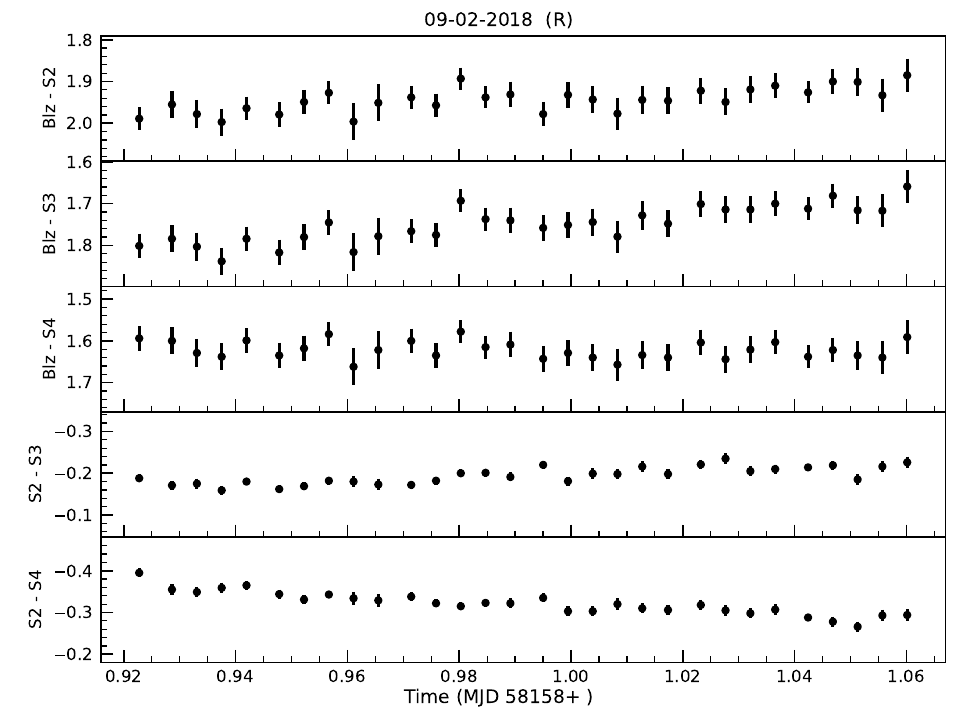}
\includegraphics[height=6cm,width=6cm]{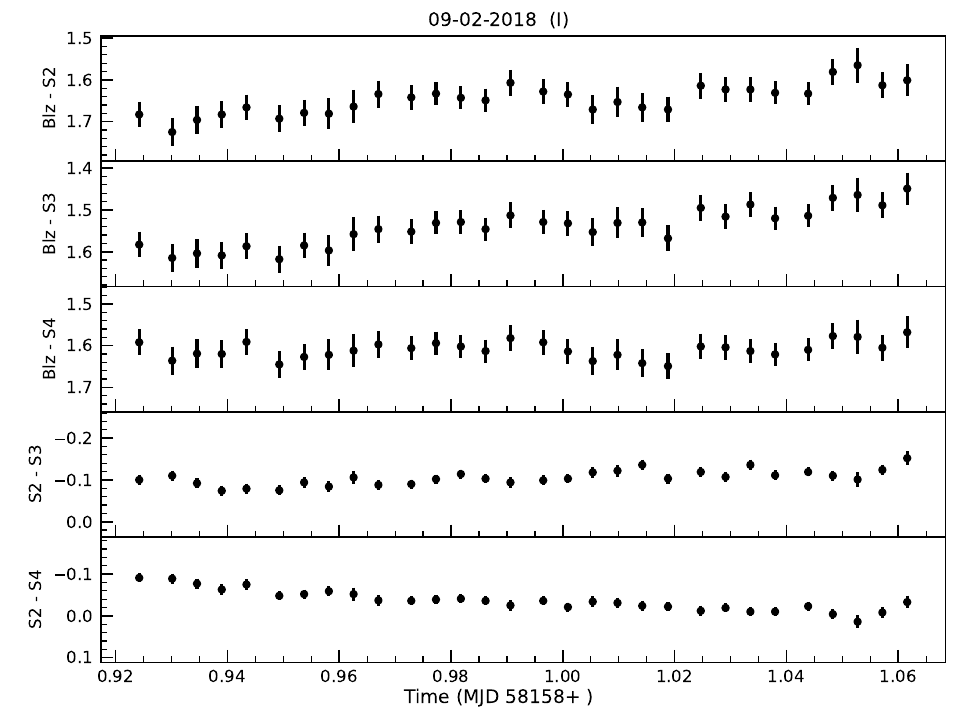}
}
}
\caption{\label{fig:idv_lc1}DLCs of the TeV blazar S4 0954$+$65. Here, S2, S3 and S4 refer to the comparison stars and Blz refers to the blazar. The observation date and the filter name are mentioned at the top of each plot.}
\end{figure*}

\begin{figure*}[!h]
 \ContinuedFloat
\vbox{
\hbox{
\includegraphics[height=6cm,width=6cm]{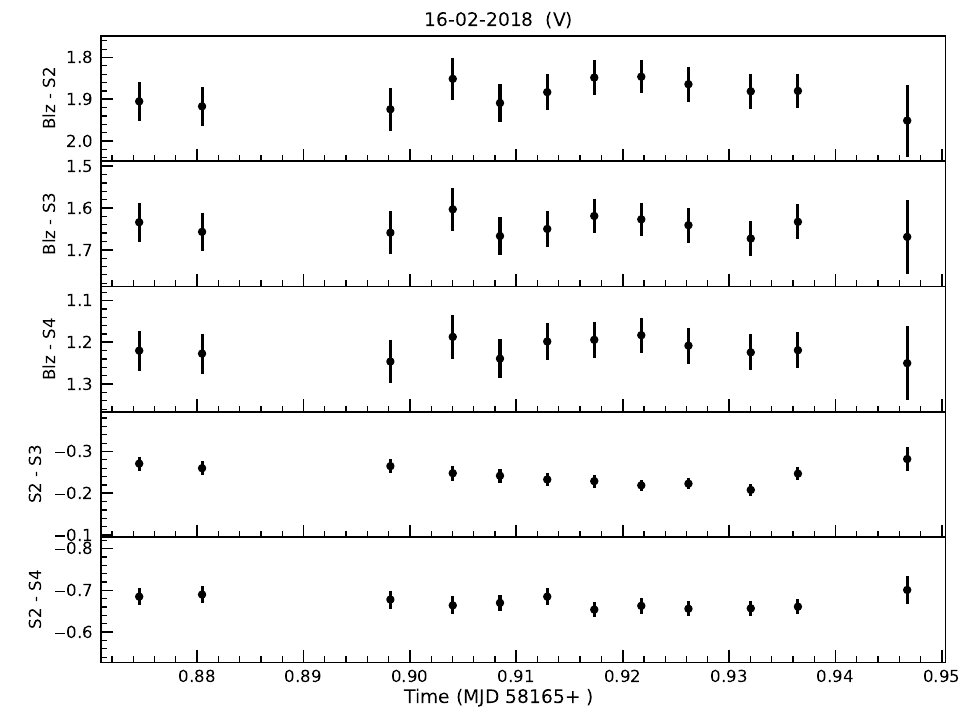}
\includegraphics[height=6cm,width=6cm]{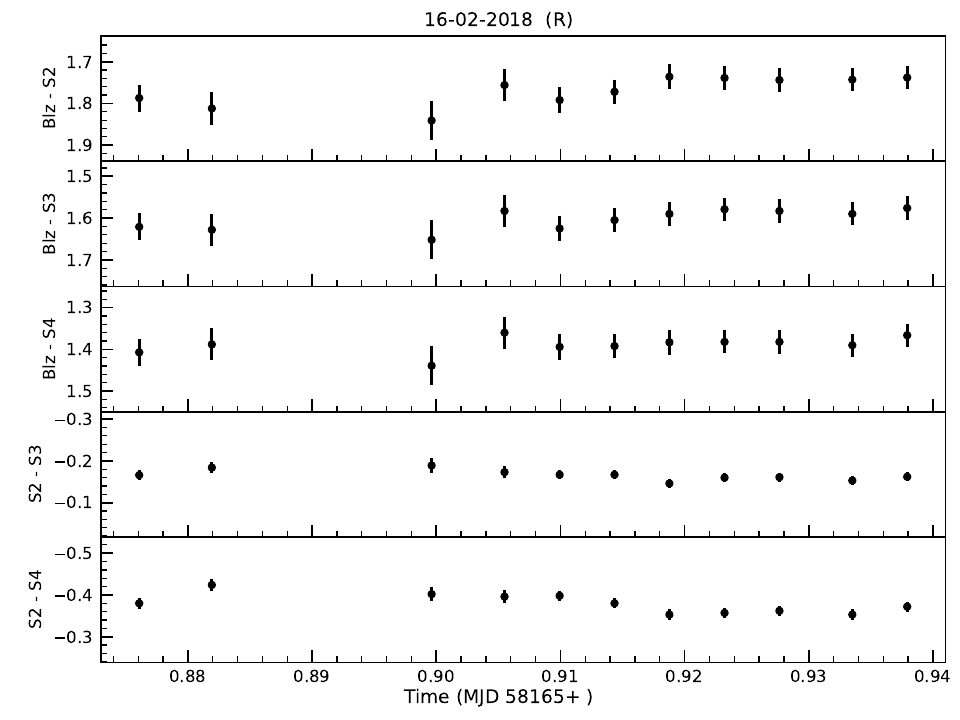}
\includegraphics[height=6cm,width=6cm]{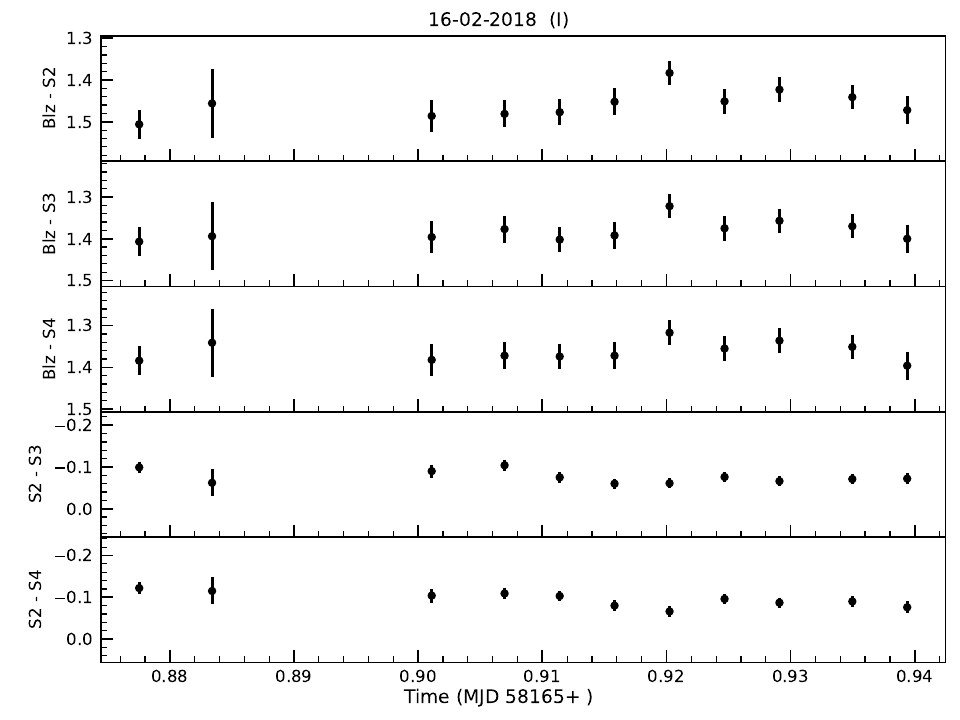}
}

\hbox{
\includegraphics[height=6cm,width=6cm]{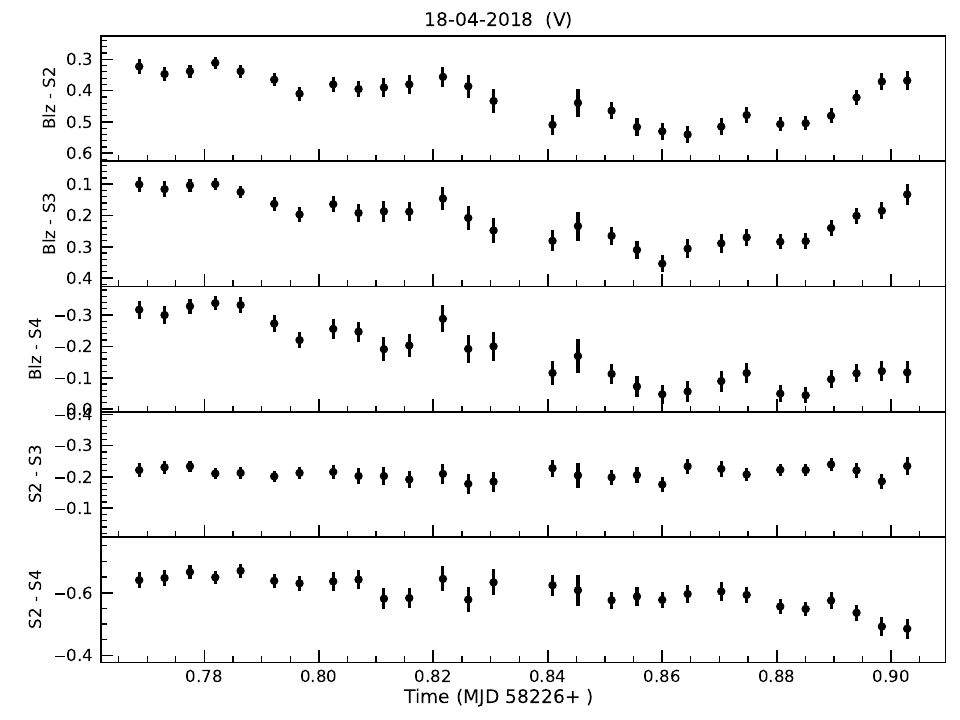}
\includegraphics[height=6cm,width=6cm]{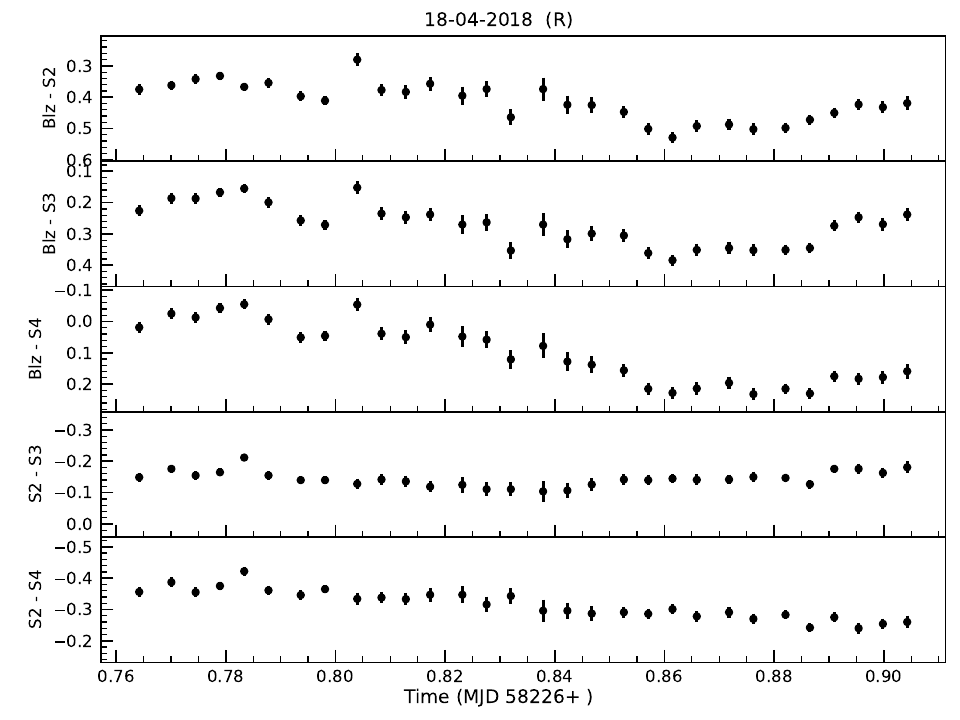}
\includegraphics[height=6cm,width=6cm]{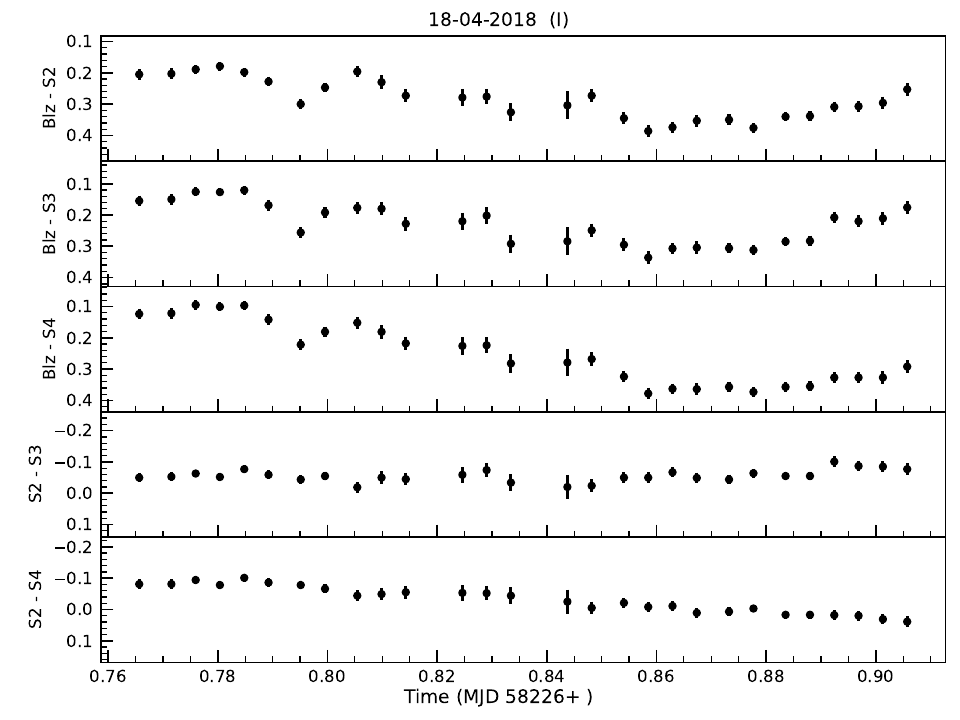}
}

\hbox{
\includegraphics[height=6cm,width=6cm]{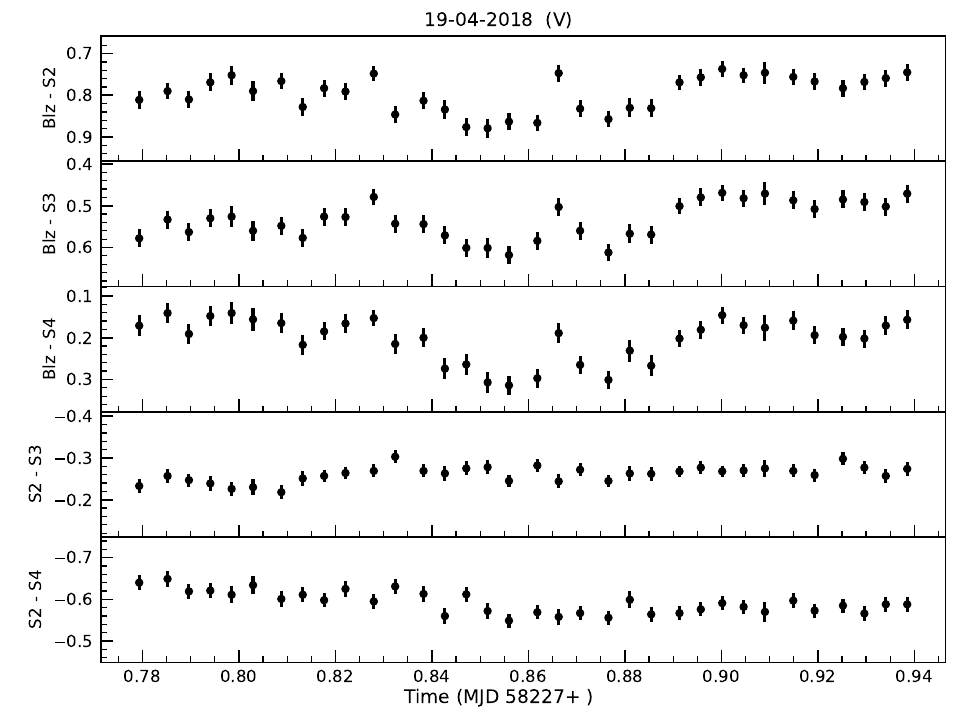}
\includegraphics[height=6cm,width=6cm]{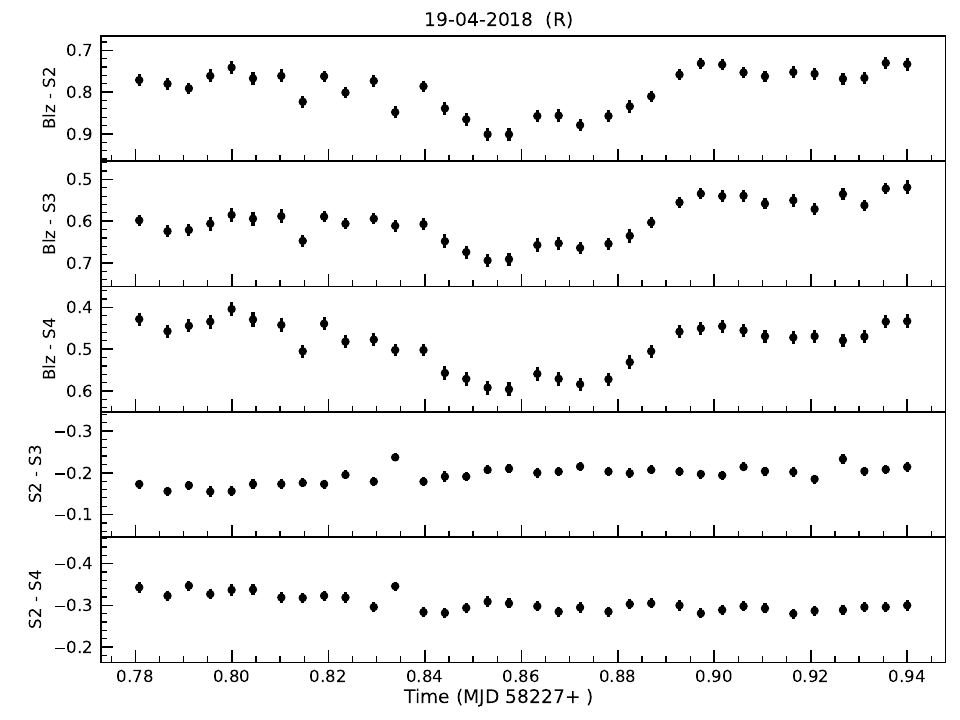}
\includegraphics[height=6cm,width=6cm]{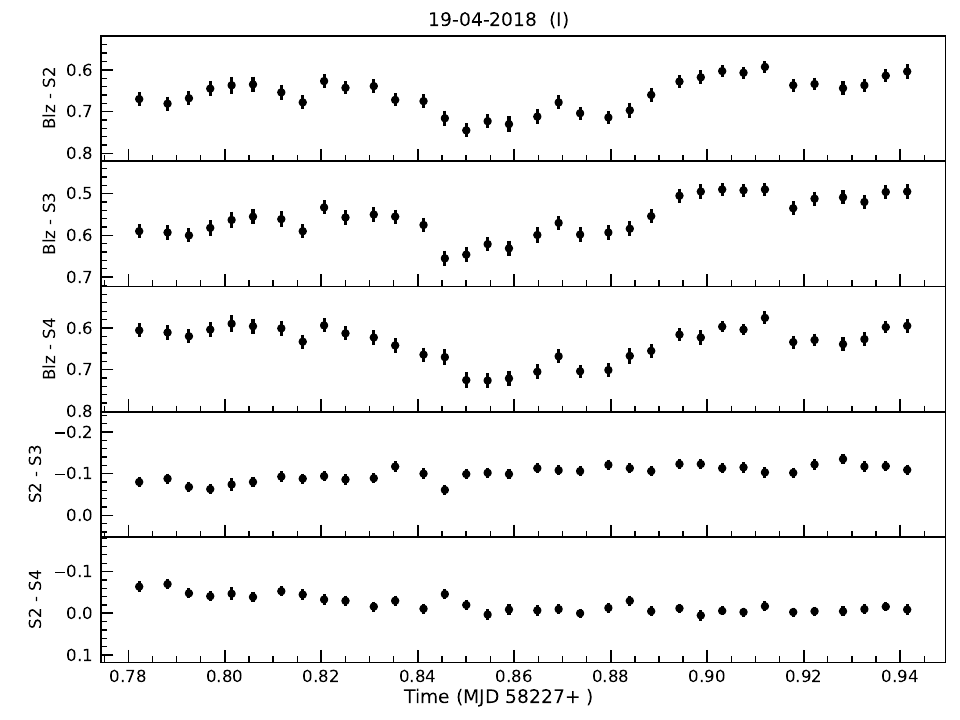}
}
\hbox{
\includegraphics[height=6cm,width=6cm]{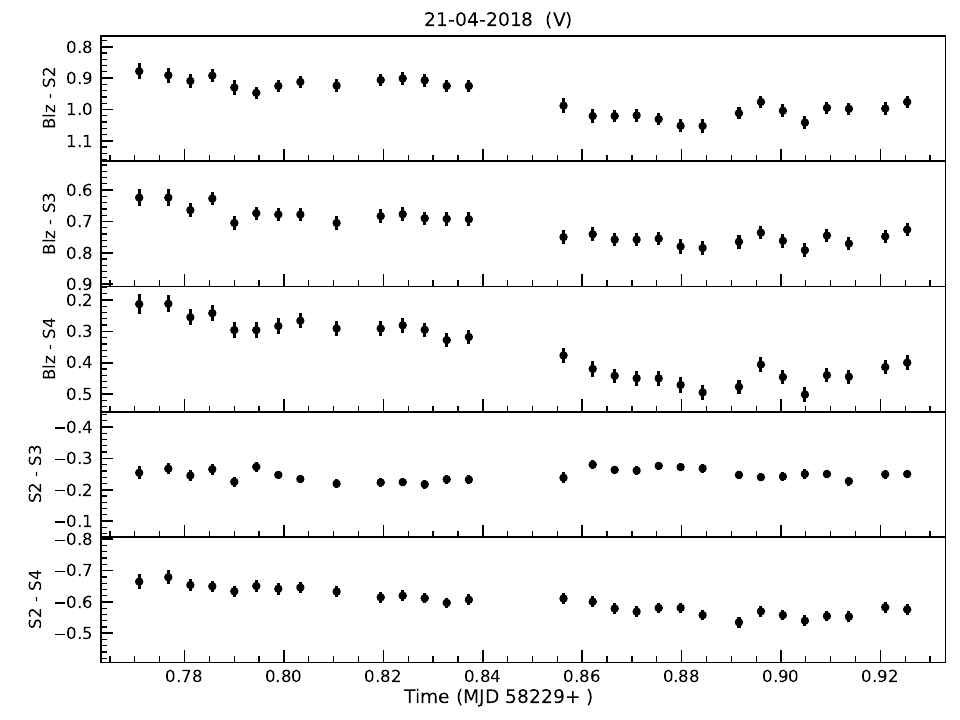}
\includegraphics[height=6cm,width=6cm]{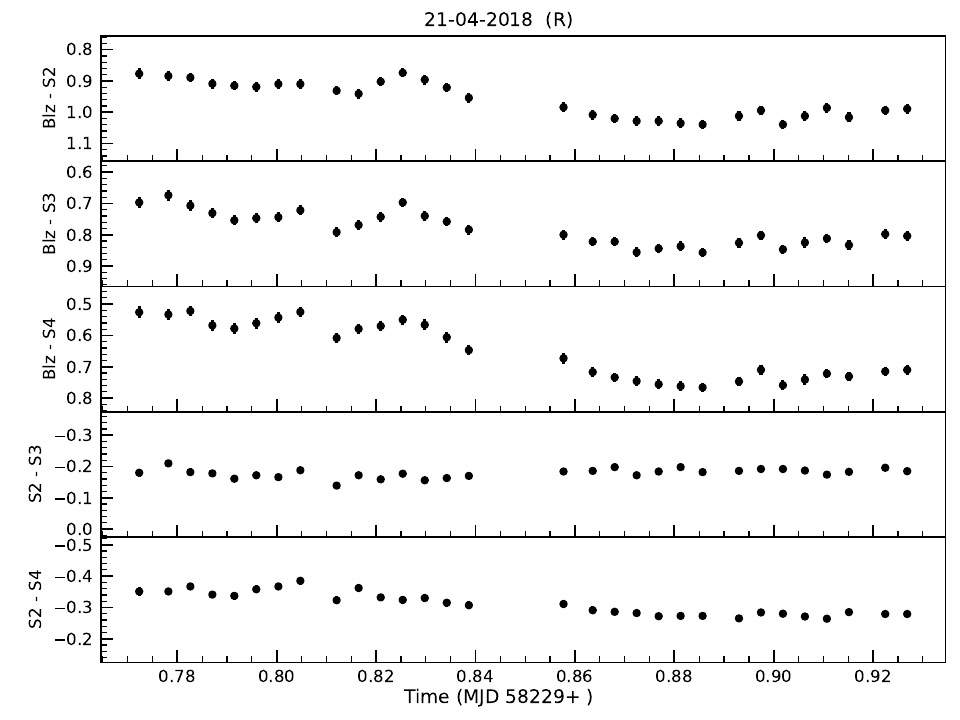}
\includegraphics[height=6cm,width=6cm]{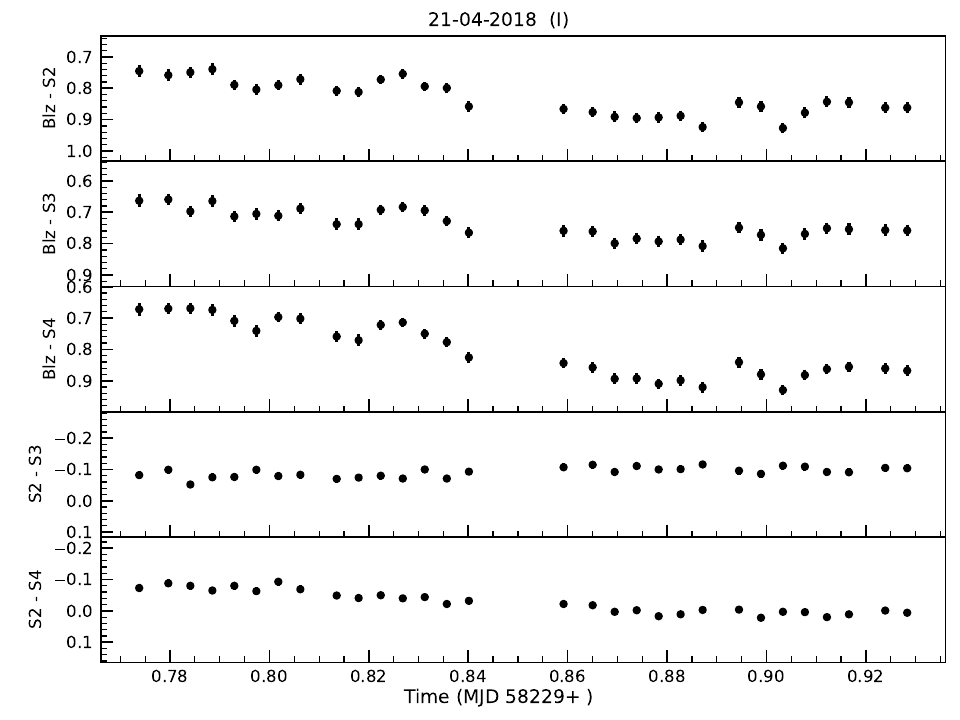}
}
}
\caption{Continued.}
\end{figure*}

\begin{figure*}[h]
 \ContinuedFloat
\vbox{
\hbox{
\includegraphics[height=6cm,width=6cm]{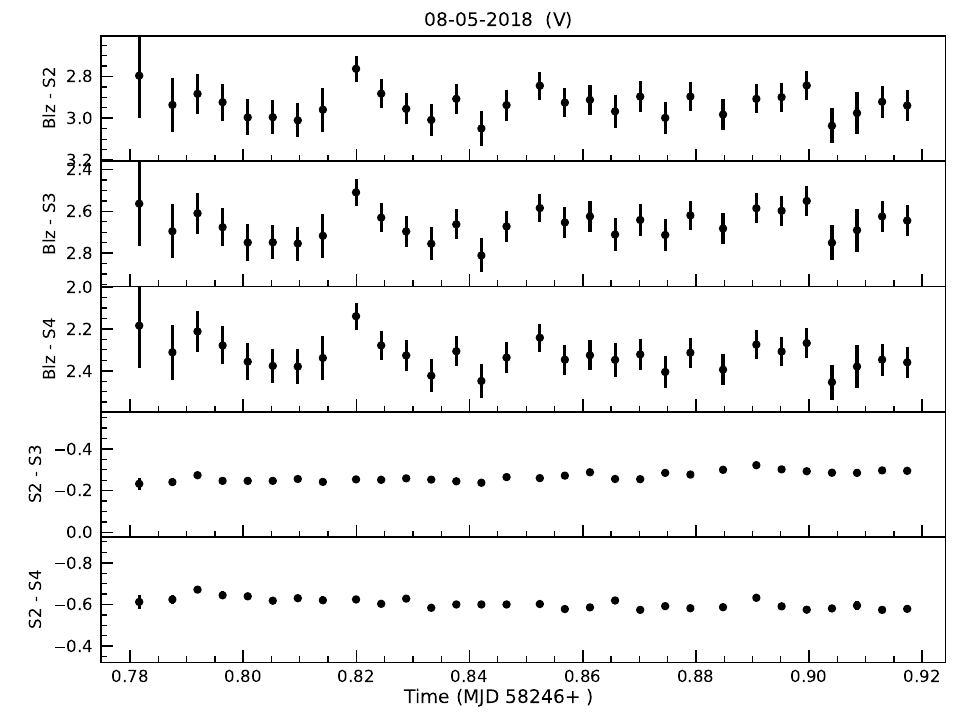}
\includegraphics[height=6cm,width=6cm]{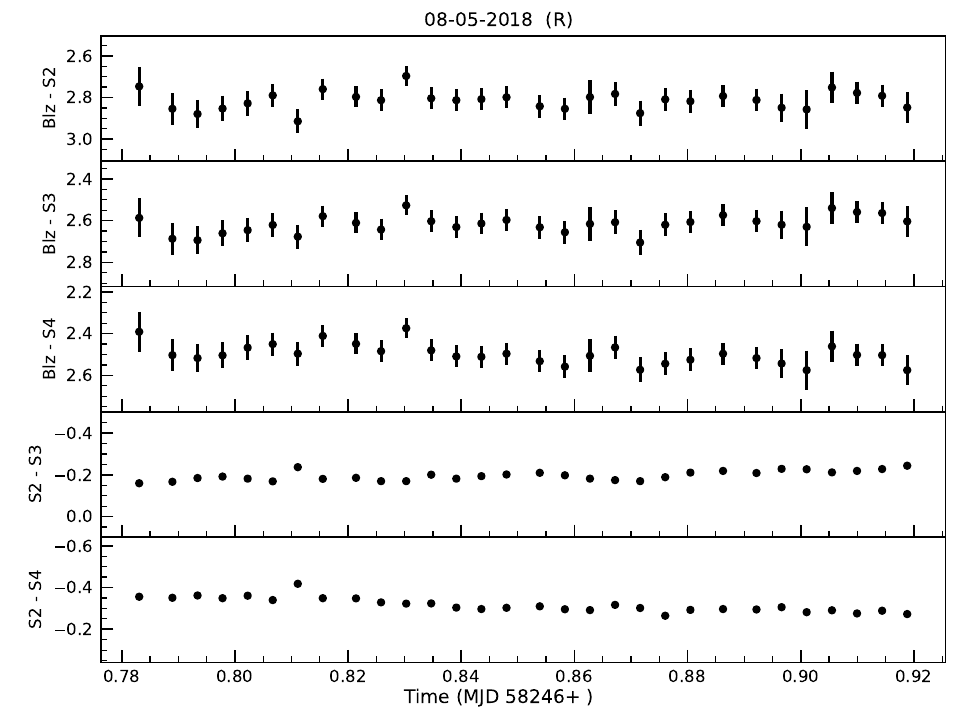}
\includegraphics[height=6cm,width=6cm]{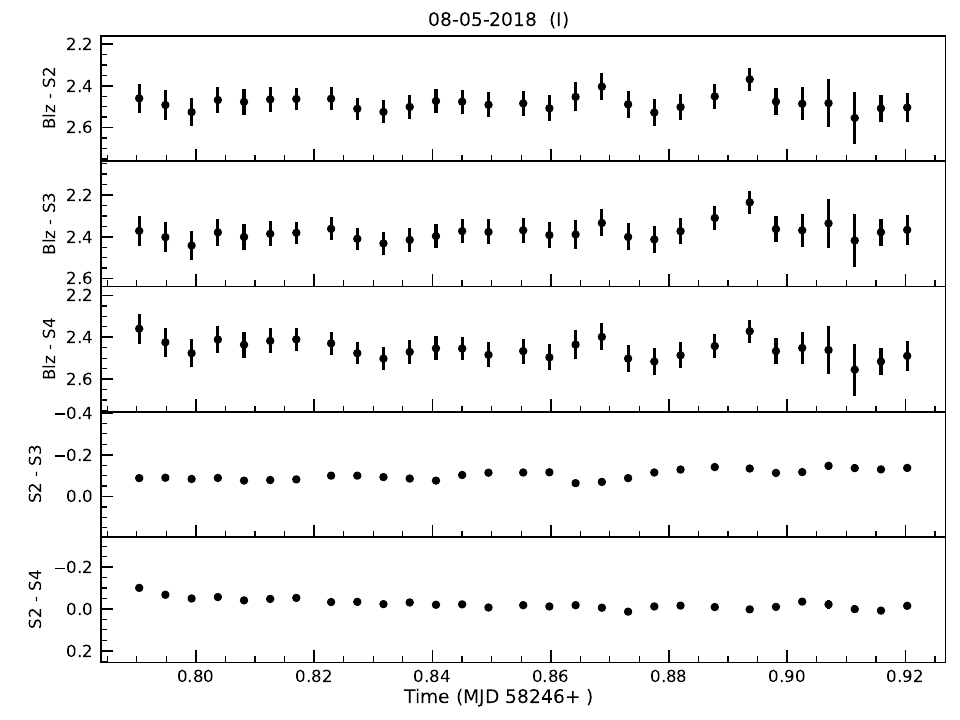}
}

\hbox{
\includegraphics[height=6cm,width=6cm]{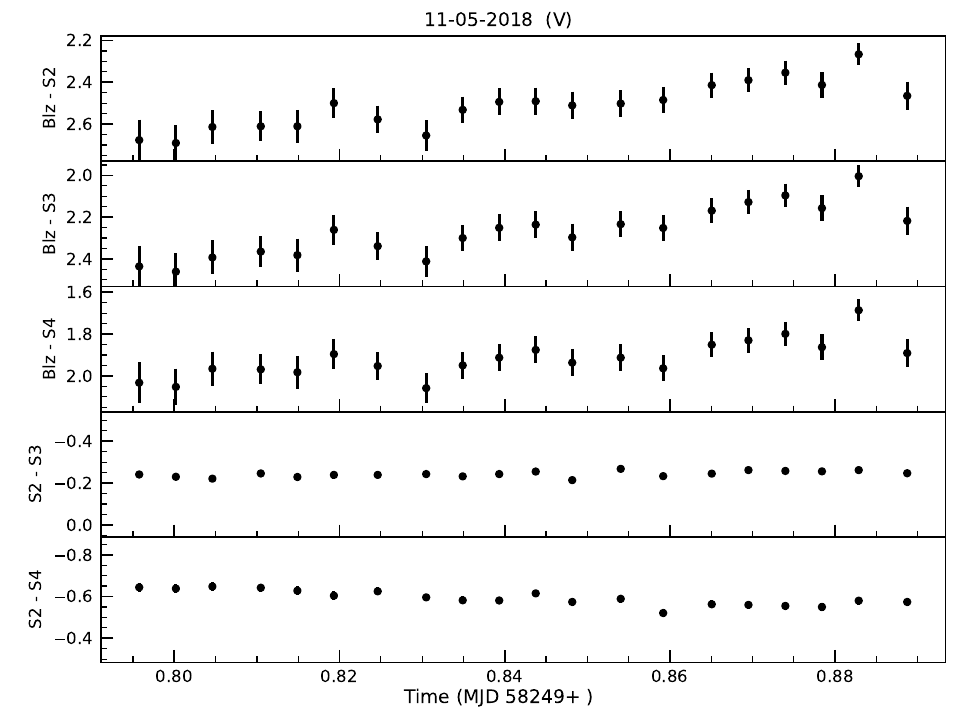}
\includegraphics[height=6cm,width=6cm]{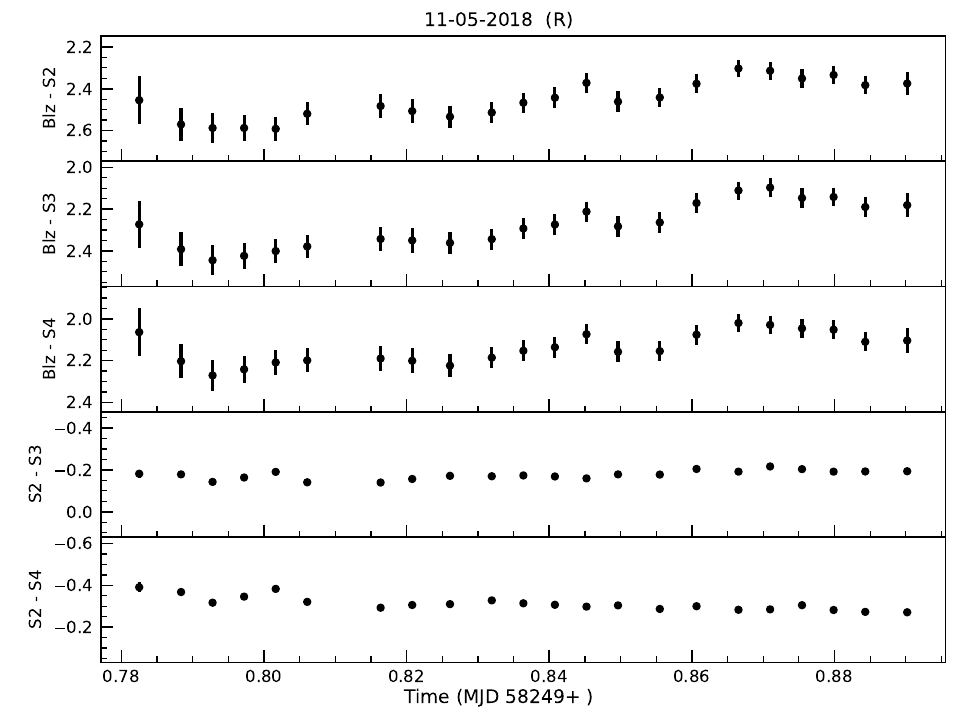}
\includegraphics[height=6cm,width=6cm]{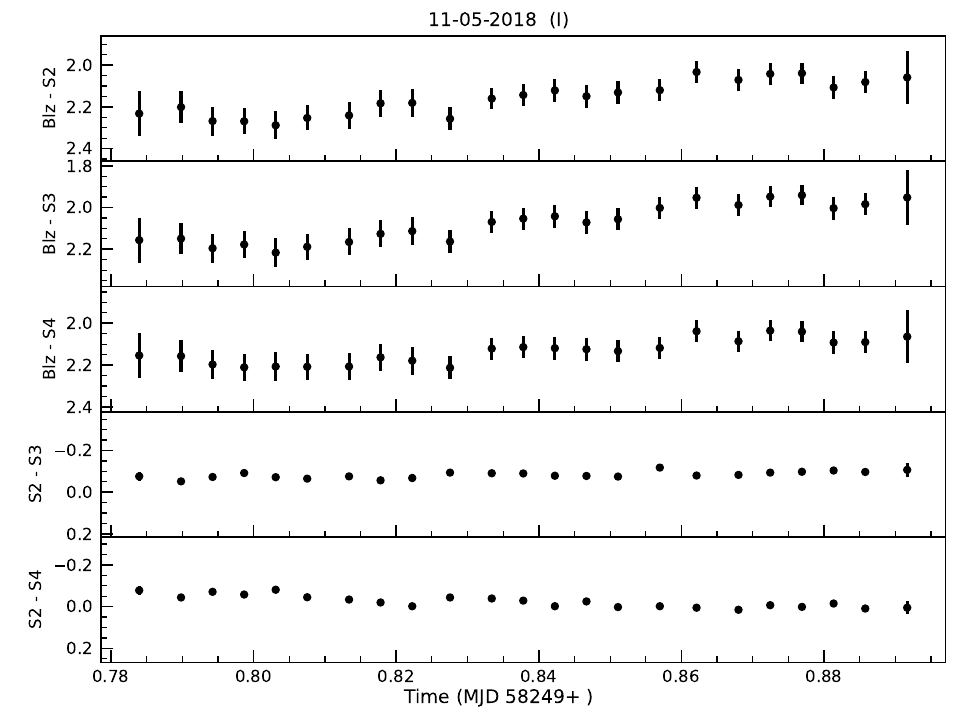}
}

\hbox{
\includegraphics[height=6cm,width=6cm]{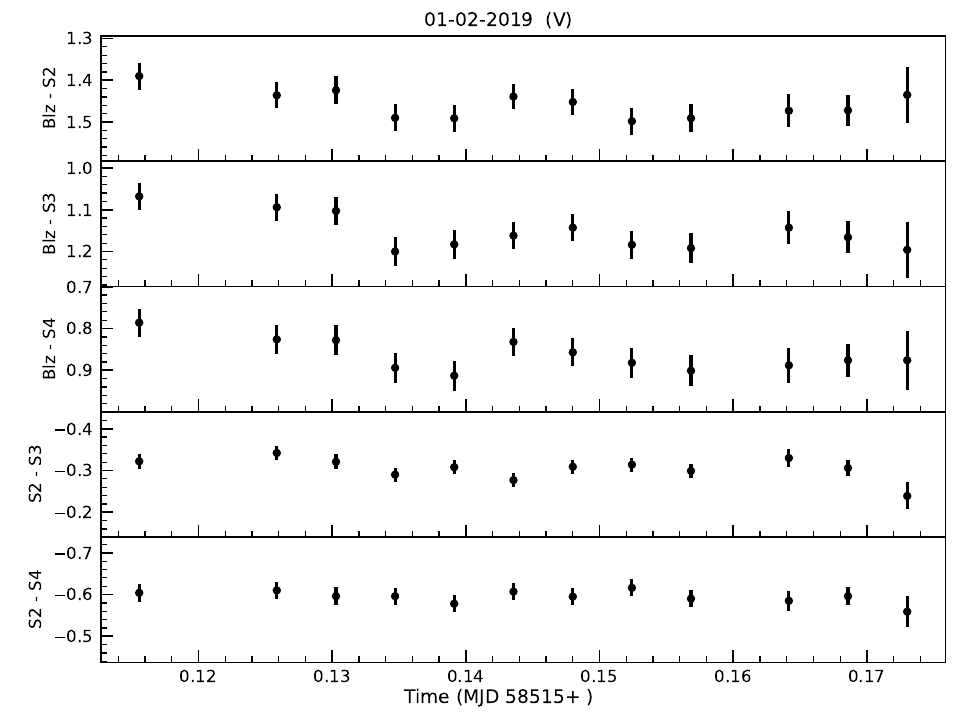}
\includegraphics[height=6cm,width=6cm]{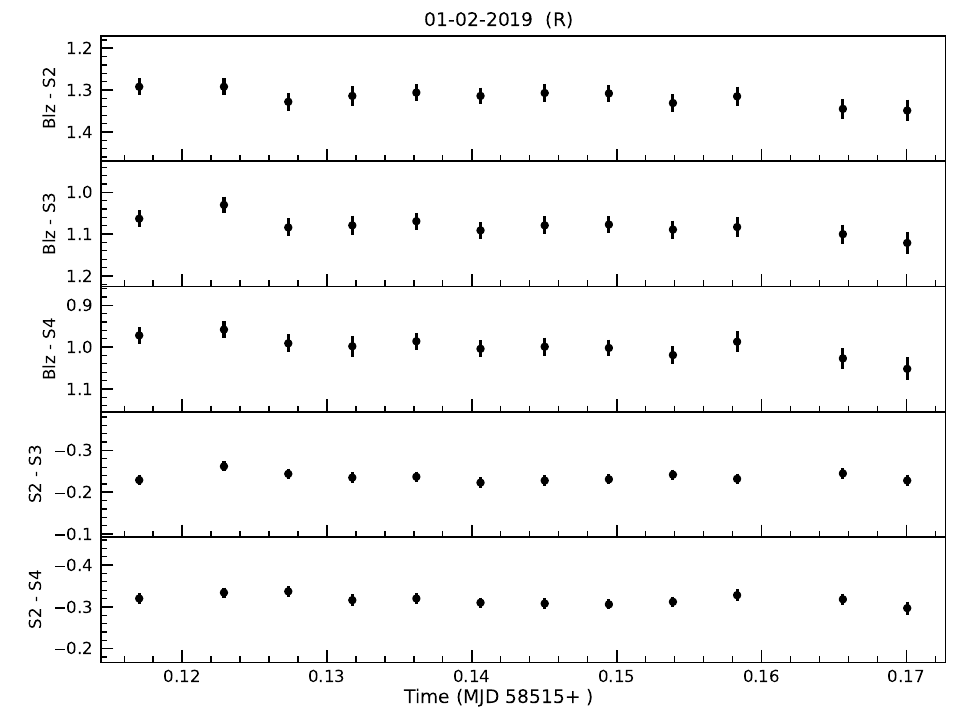}
\includegraphics[height=6cm,width=6cm]{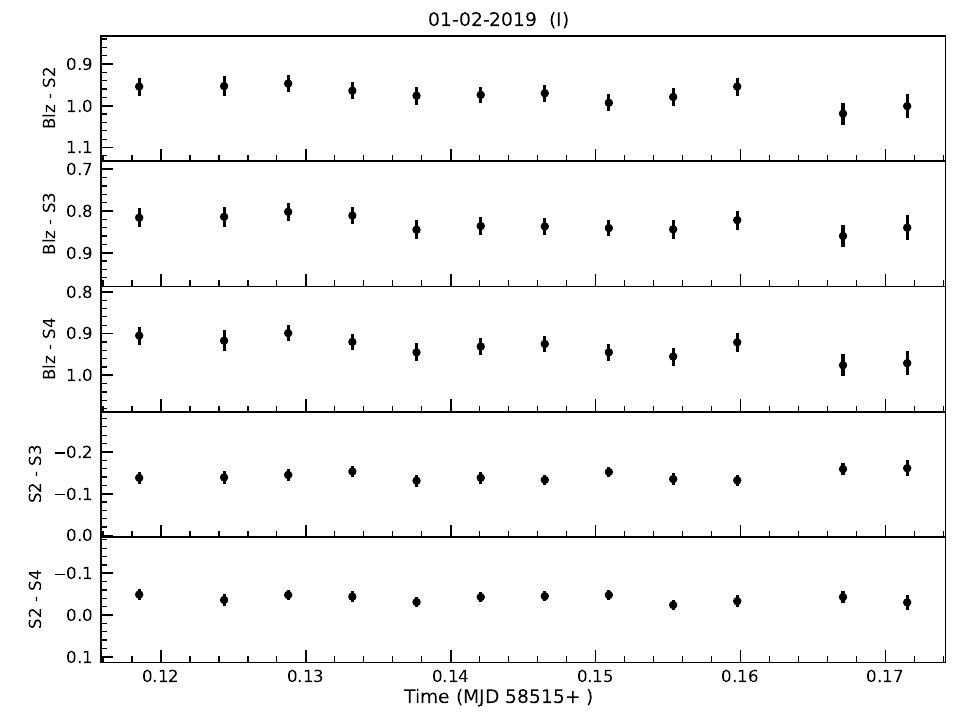}
}
\hbox{
\includegraphics[height=6cm,width=6cm]{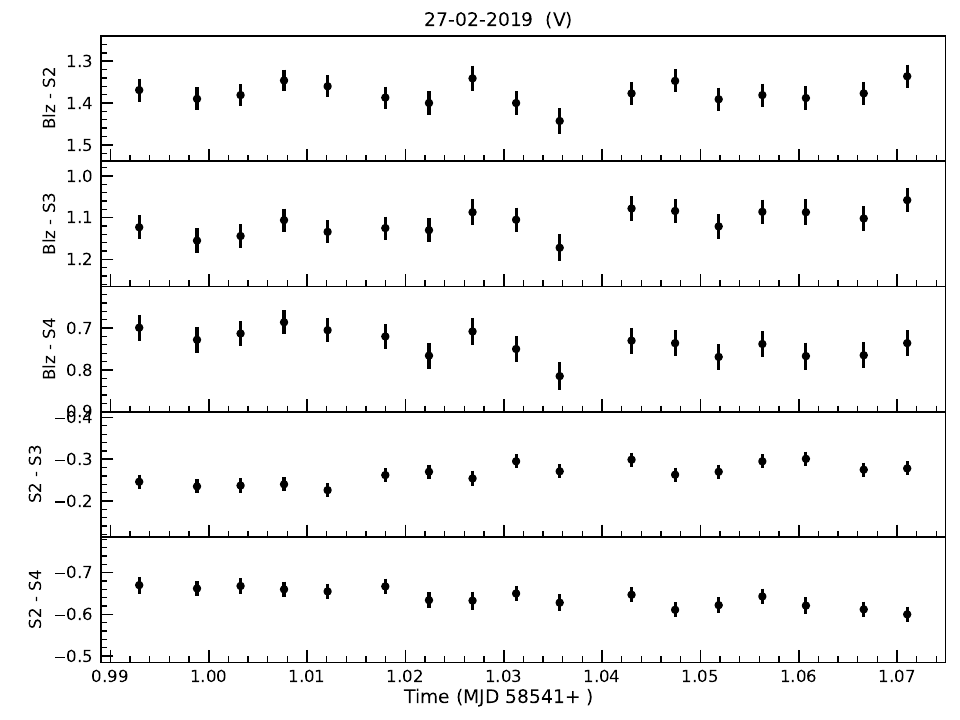}
\includegraphics[height=6cm,width=6cm]{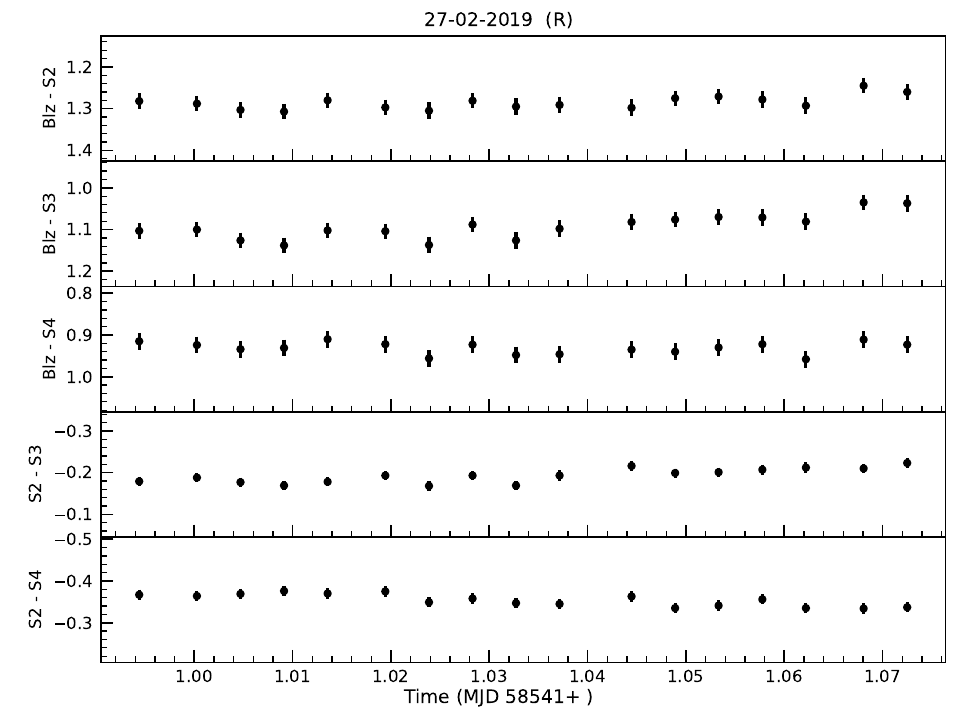}
\includegraphics[height=6cm,width=6cm]{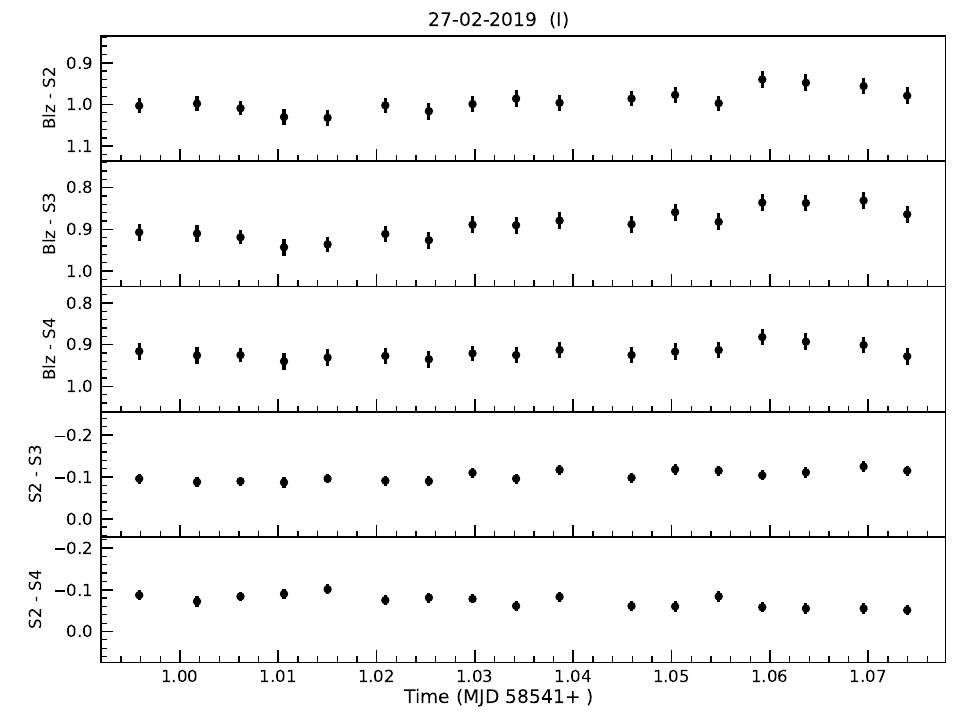}
}
}
\caption{Continued.}
\end{figure*}

\begin{figure*}[h]
 \ContinuedFloat
\vbox{
\hbox{
\includegraphics[height=6cm,width=6cm]{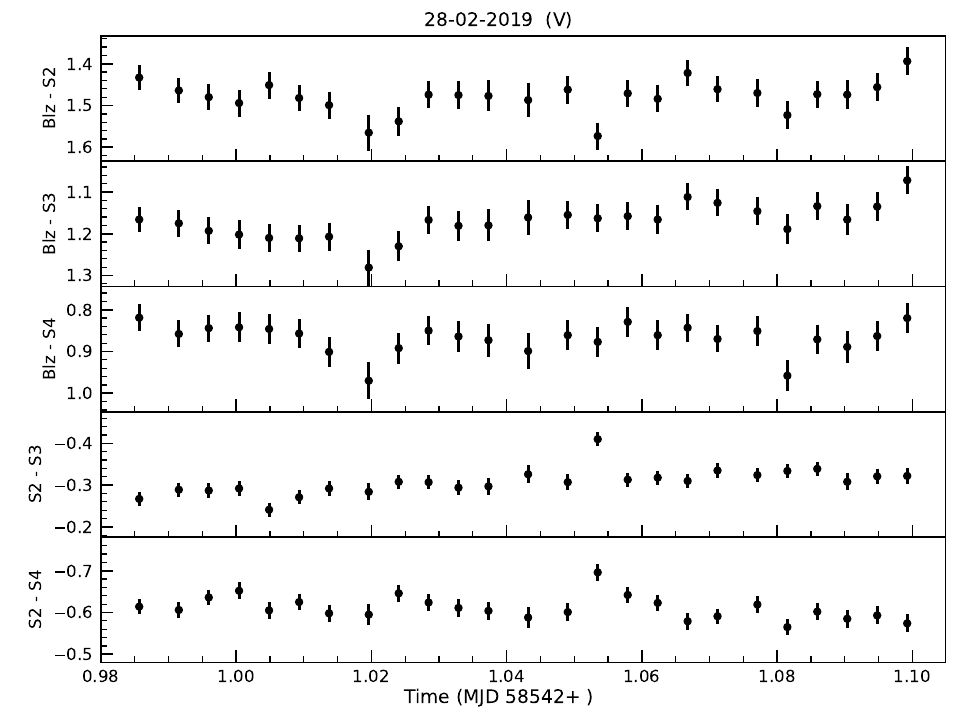}
\includegraphics[height=6cm,width=6cm]{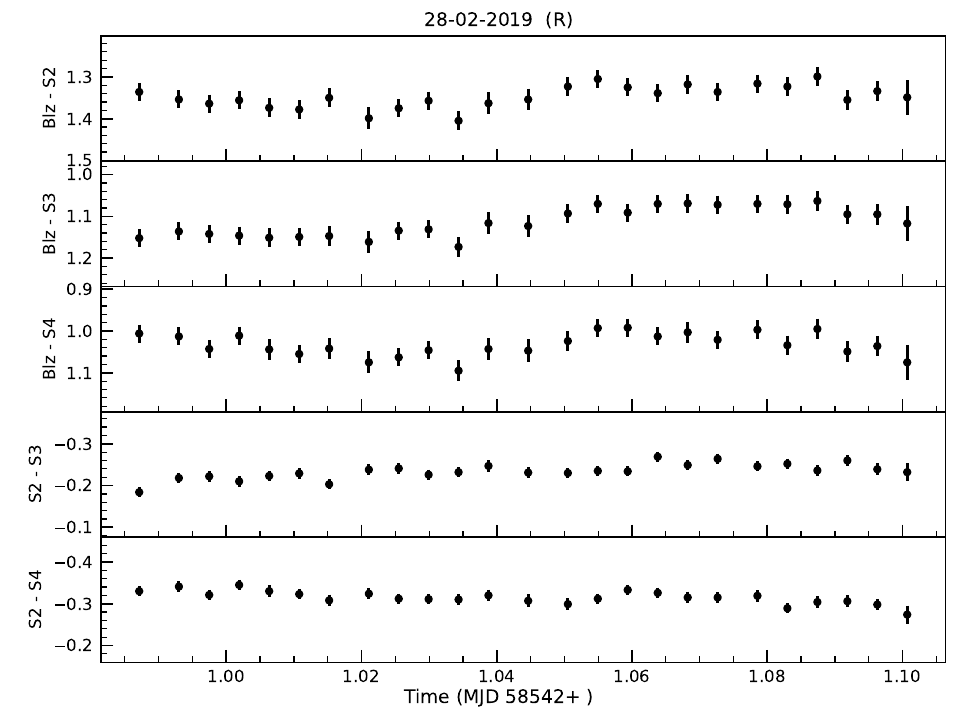}
\includegraphics[height=6cm,width=6cm]{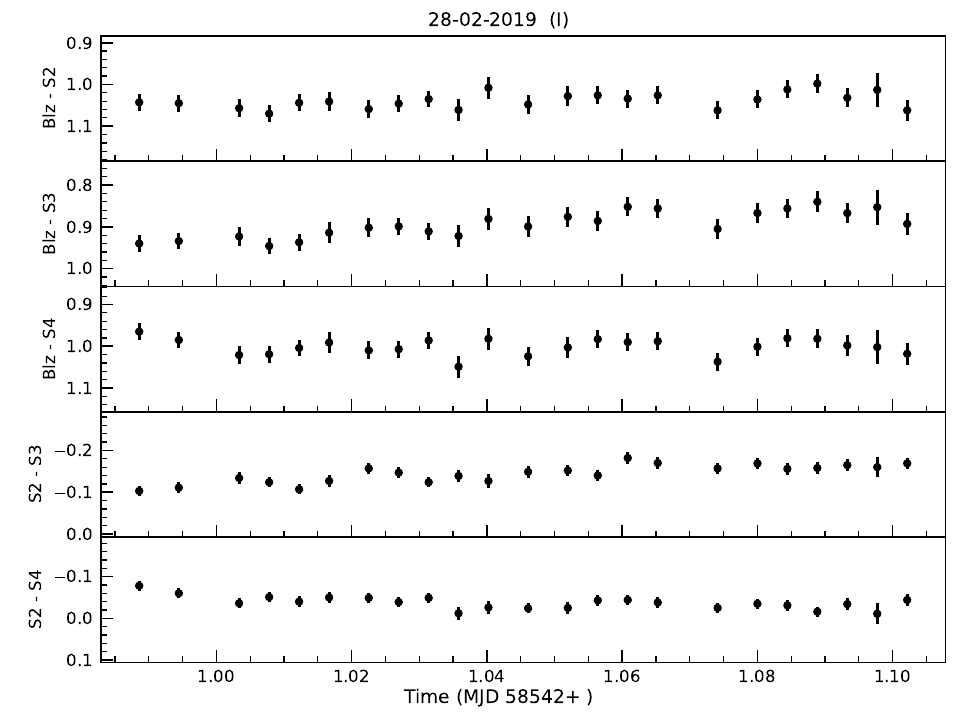}
}

\hbox{
\includegraphics[height=6cm,width=6cm]{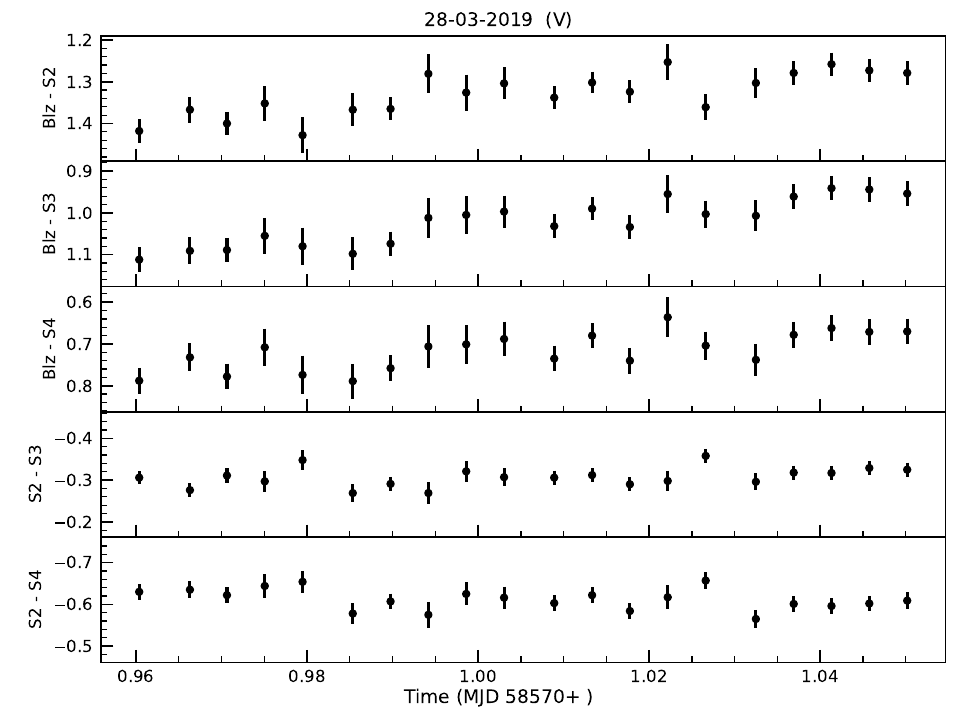}
\includegraphics[height=6cm,width=6cm]{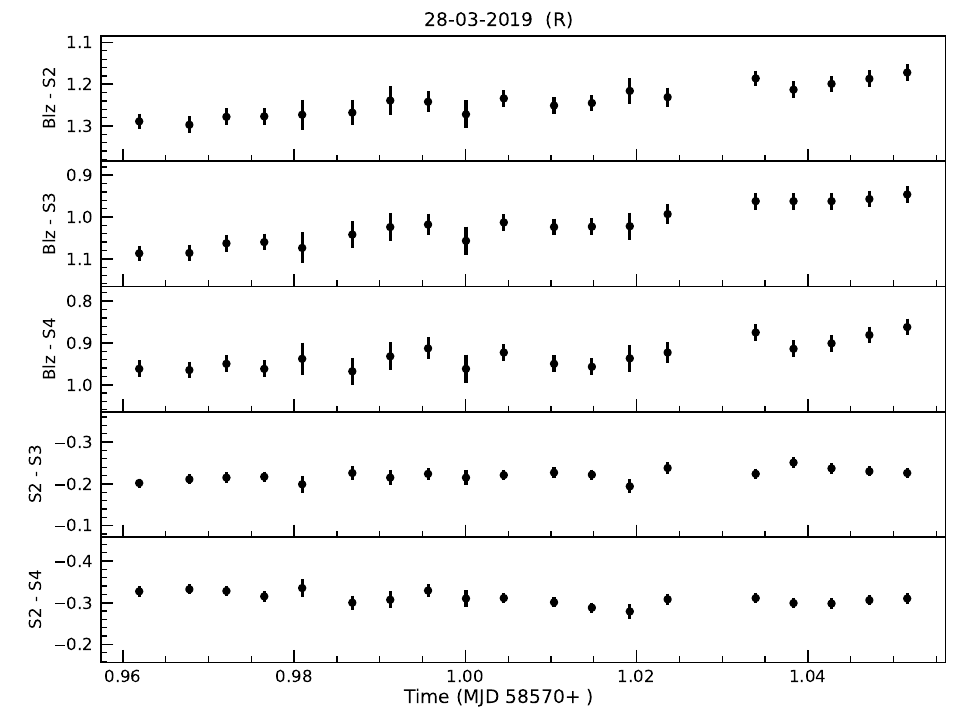}
\includegraphics[height=6cm,width=6cm]{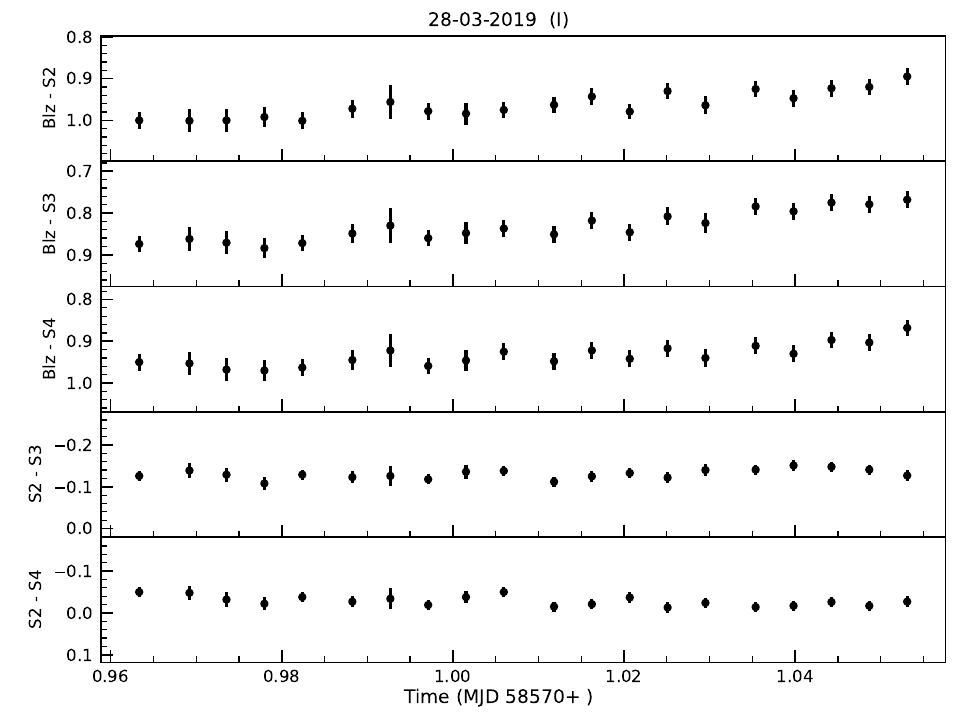}
}

\hbox{
\includegraphics[height=6cm,width=6cm]{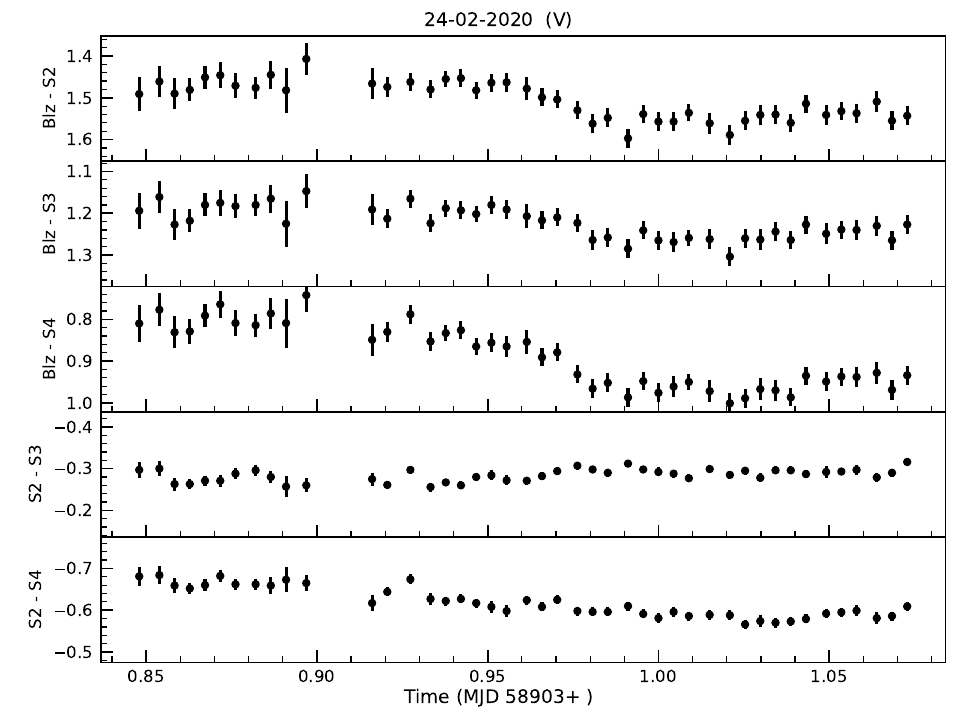}
\includegraphics[height=6cm,width=6cm]{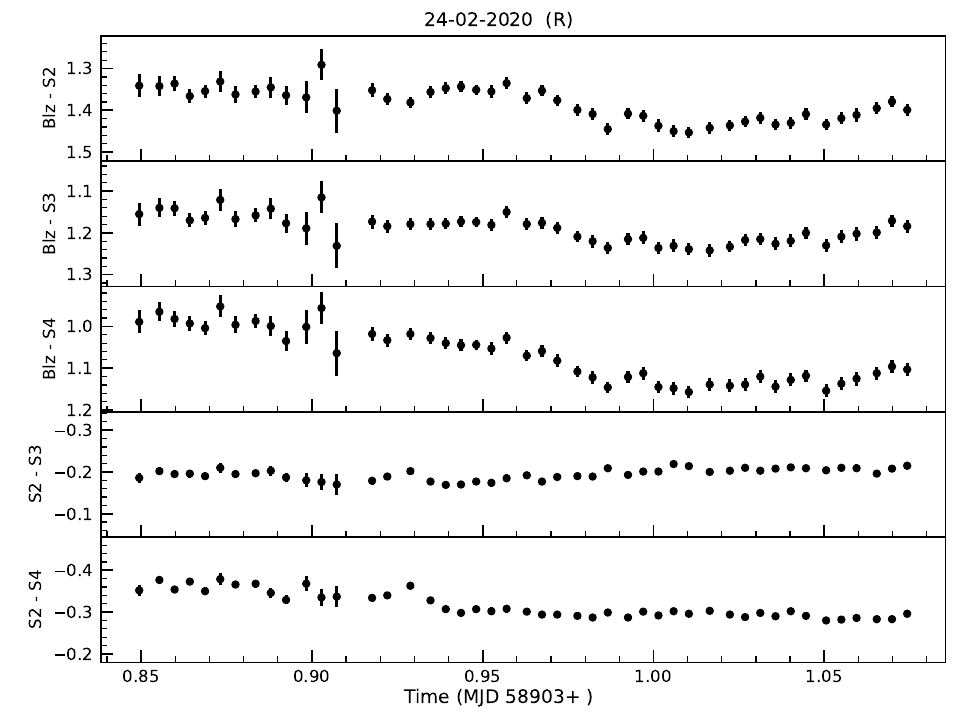}
\includegraphics[height=6cm,width=6cm]{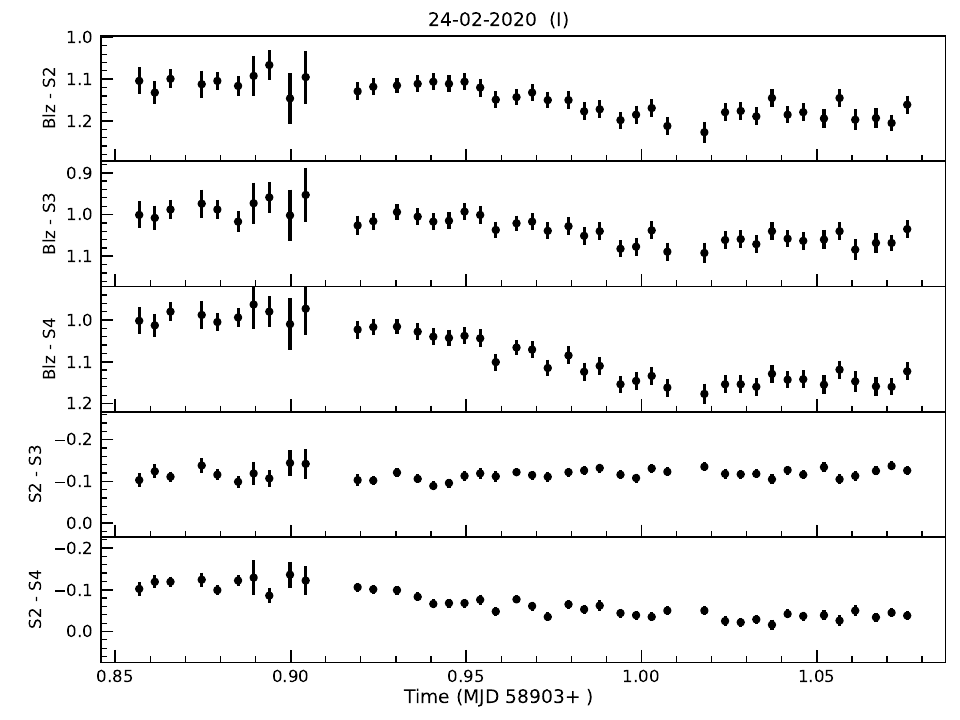}
}
\hbox{
\includegraphics[height=6cm,width=6cm]{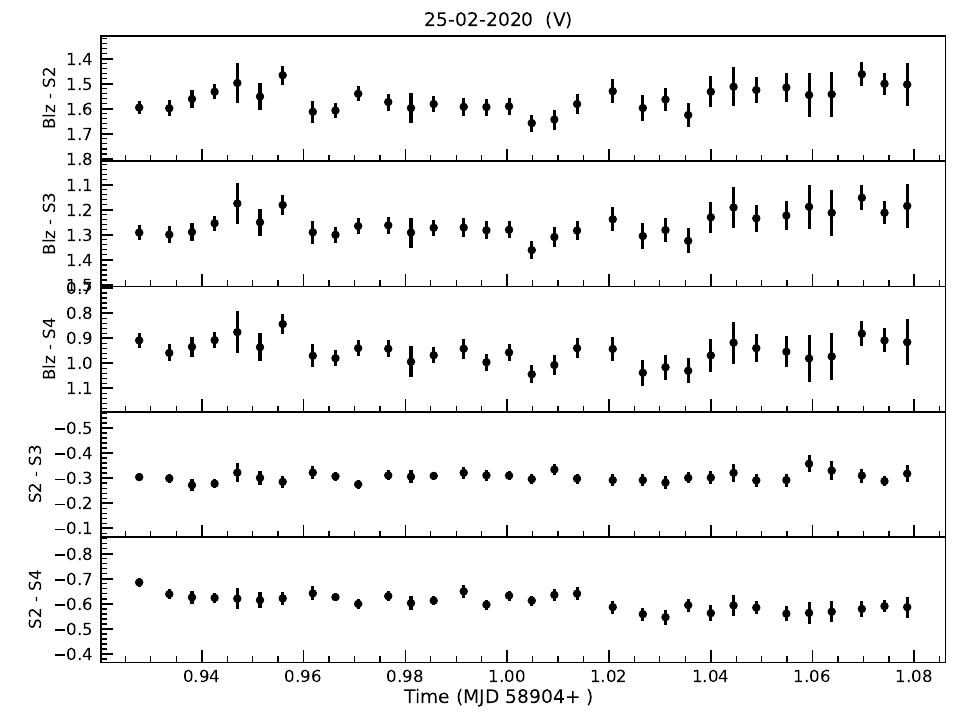}
\includegraphics[height=6cm,width=6cm]{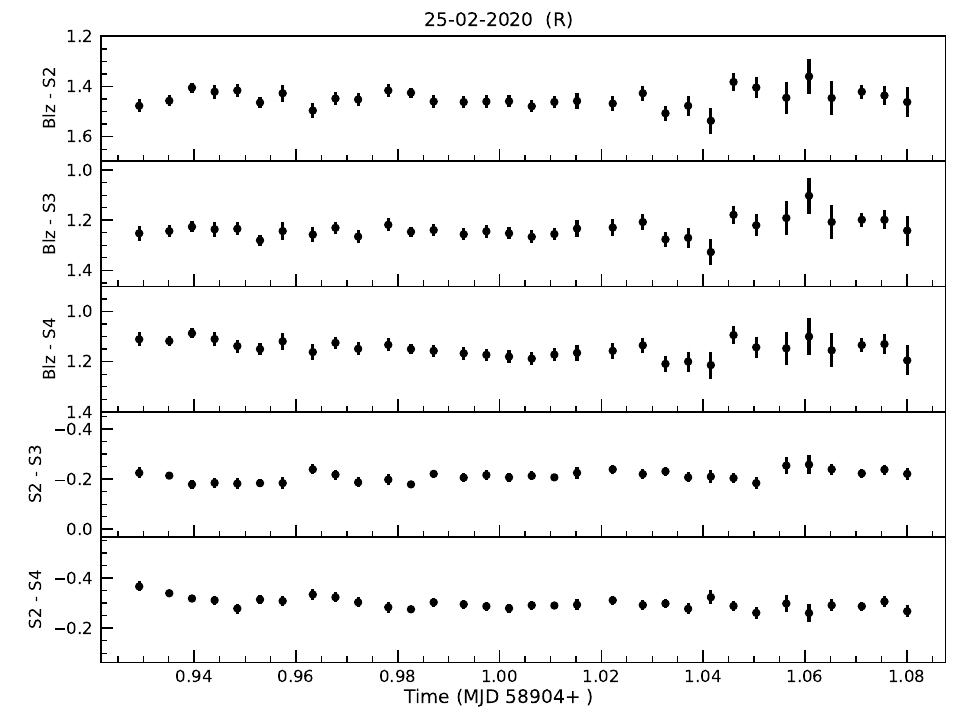}
\includegraphics[height=6cm,width=6cm]{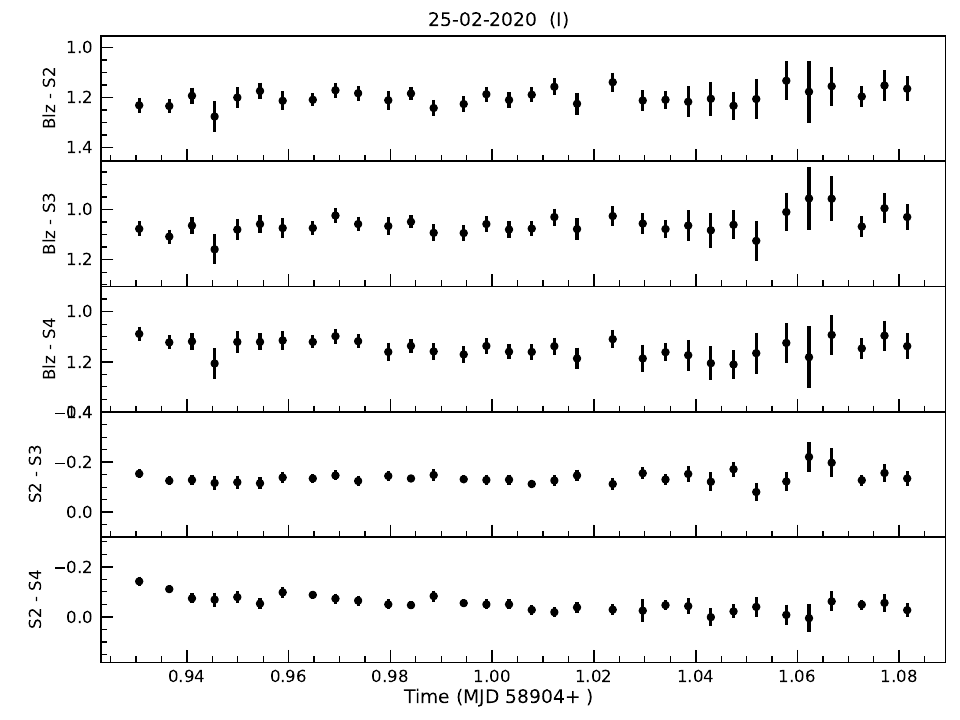}
}
}
\caption{Continued.}
\end{figure*}

\begin{figure*}[h]
 \ContinuedFloat
\vbox{
\hbox{
\includegraphics[height=6cm,width=6cm]{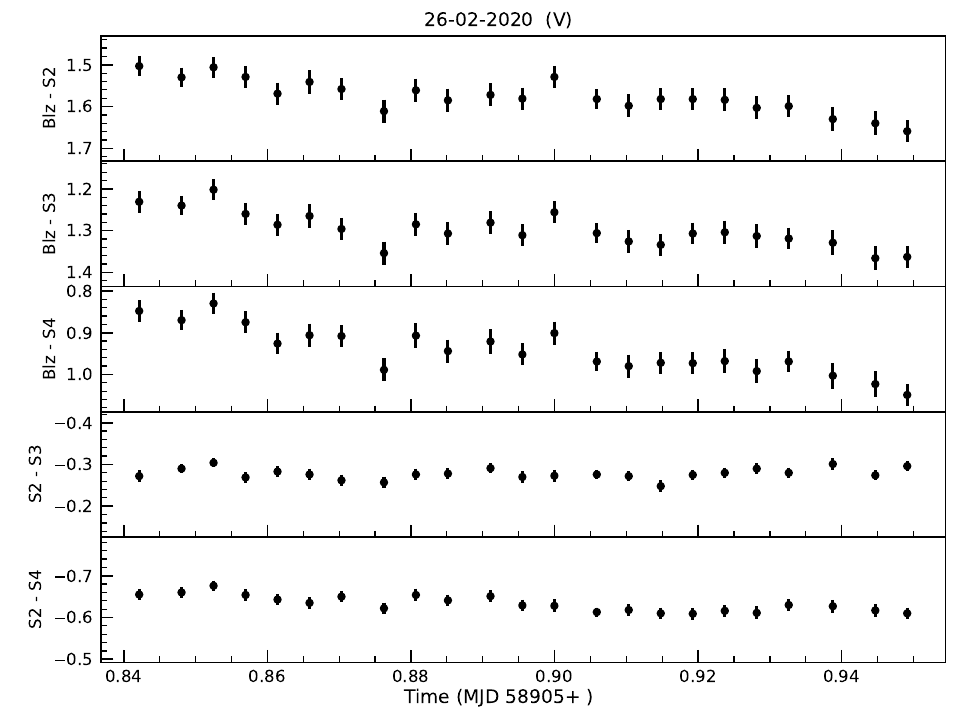}
\includegraphics[height=6cm,width=6cm]{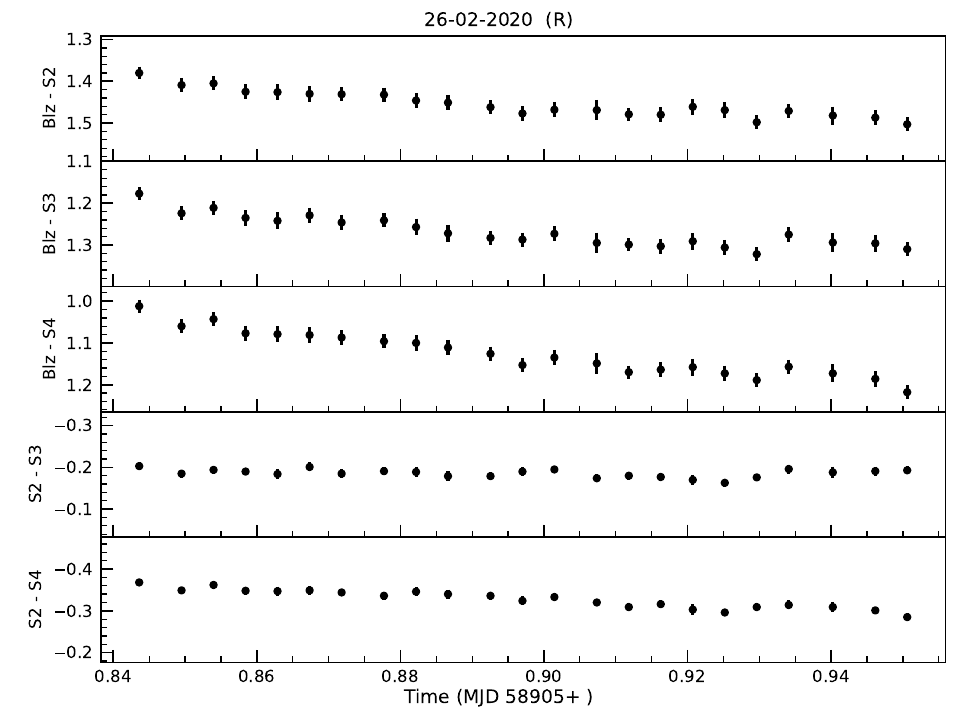}
\includegraphics[height=6cm,width=6cm]{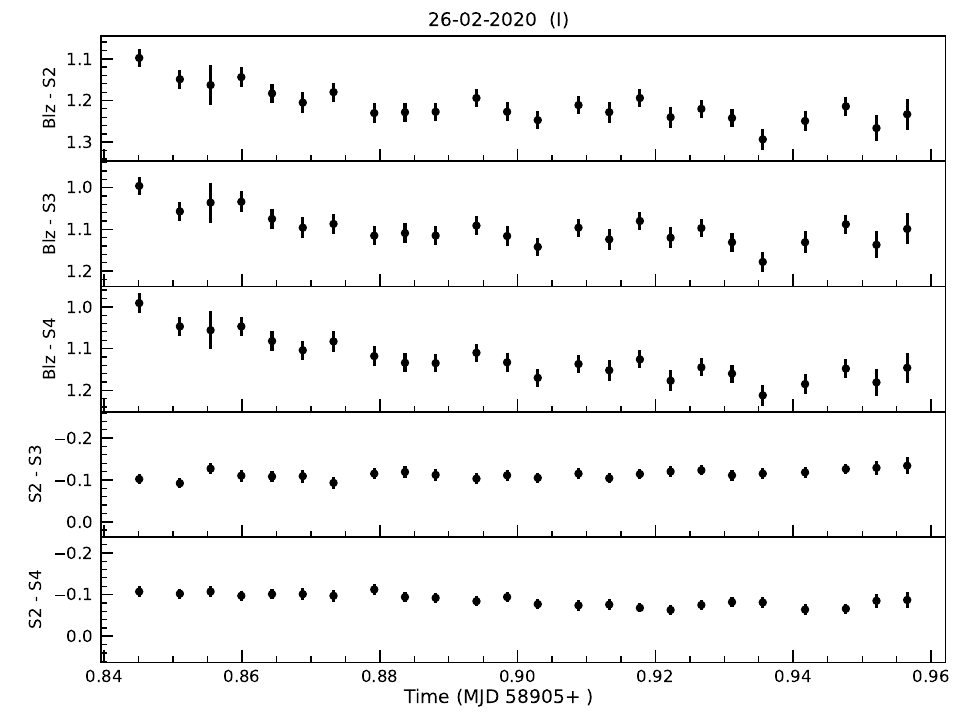}
}

\hbox{
\includegraphics[height=6cm,width=6cm]{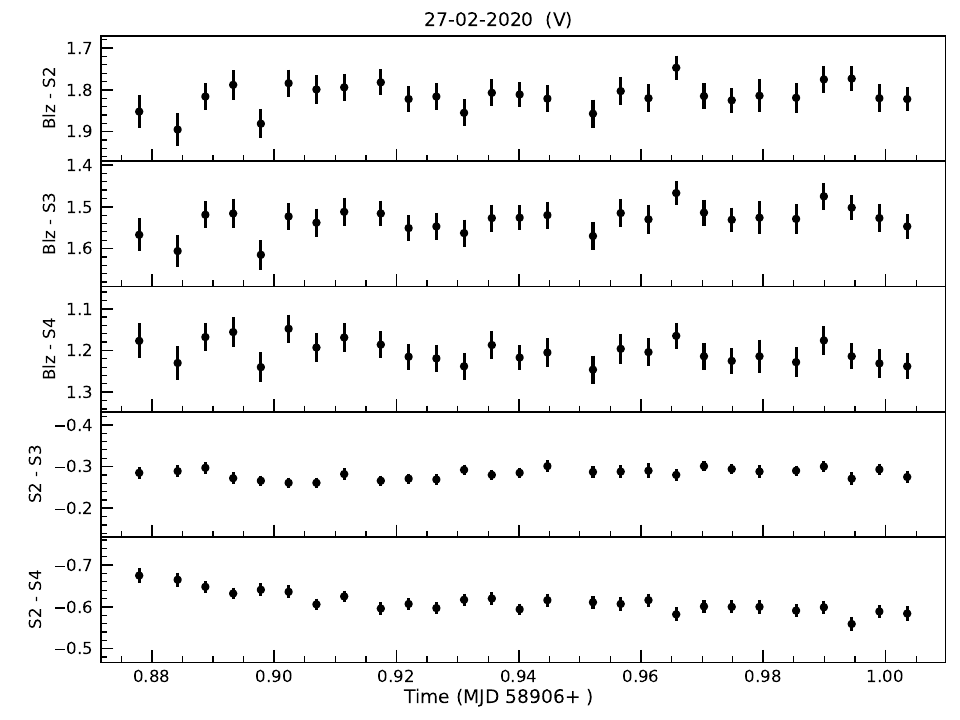}
\includegraphics[height=6cm,width=6cm]{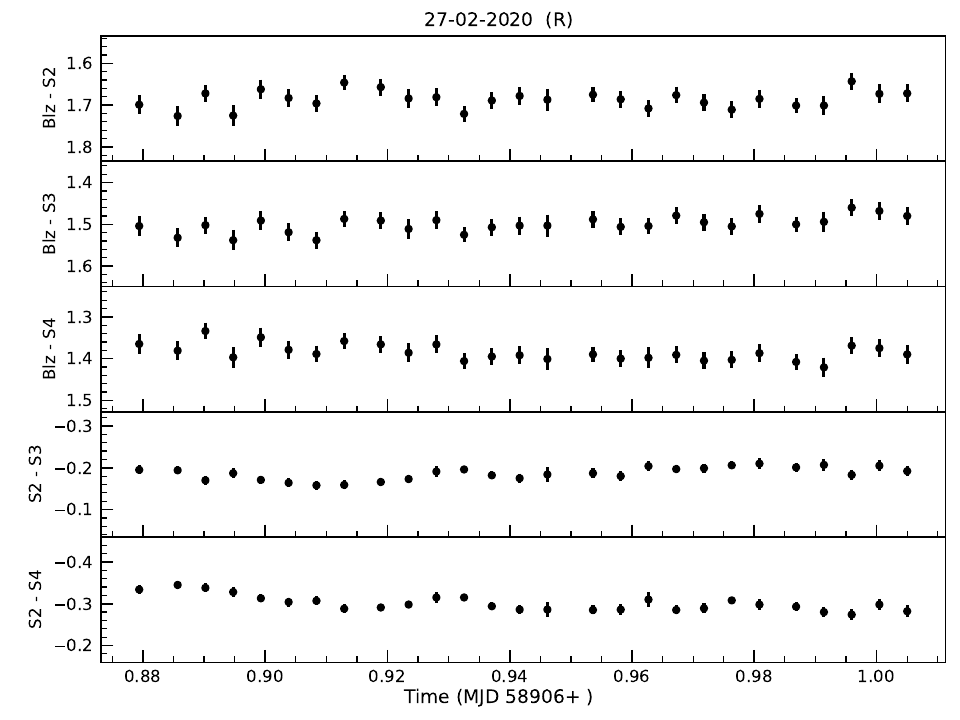}
\includegraphics[height=6cm,width=6cm]{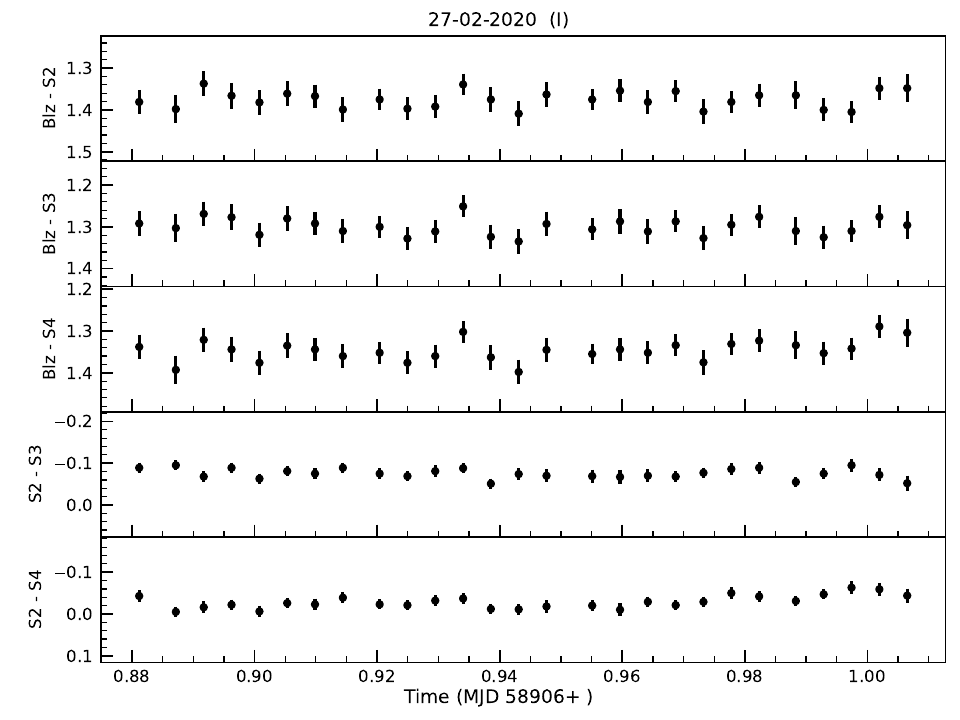}
}

\hbox{
\includegraphics[height=6cm,width=6cm]{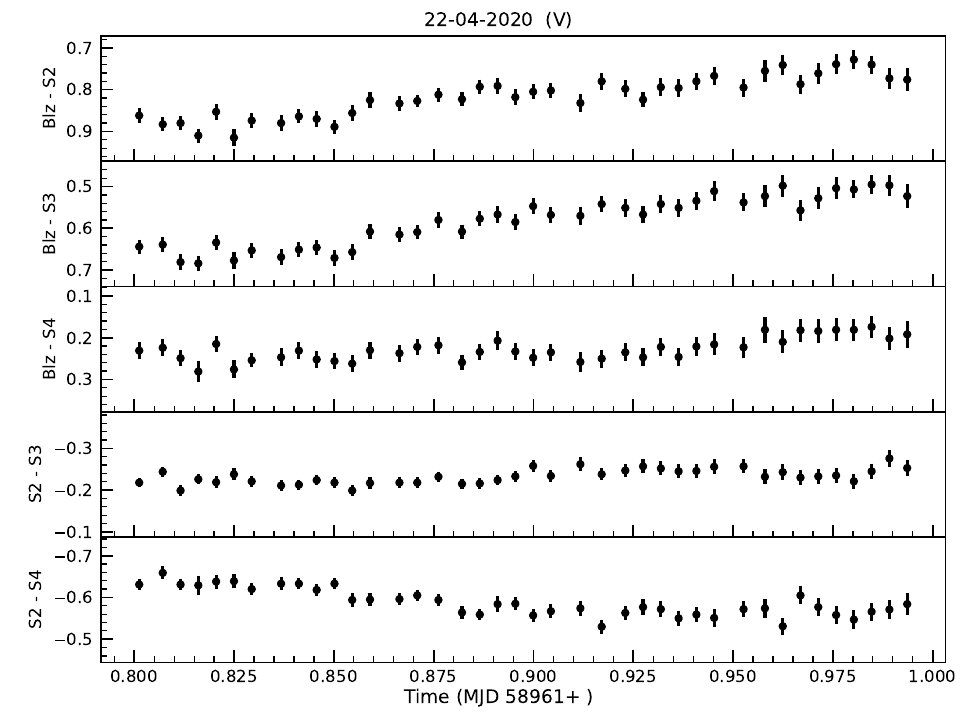}
\includegraphics[height=6cm,width=6cm]{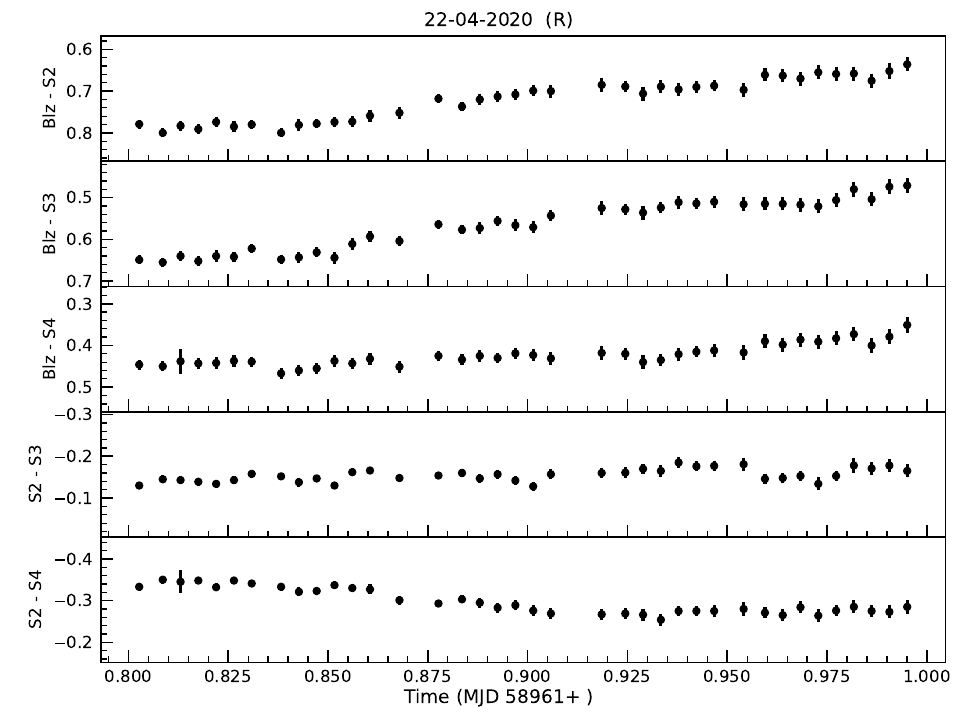}
\includegraphics[height=6cm,width=6cm]{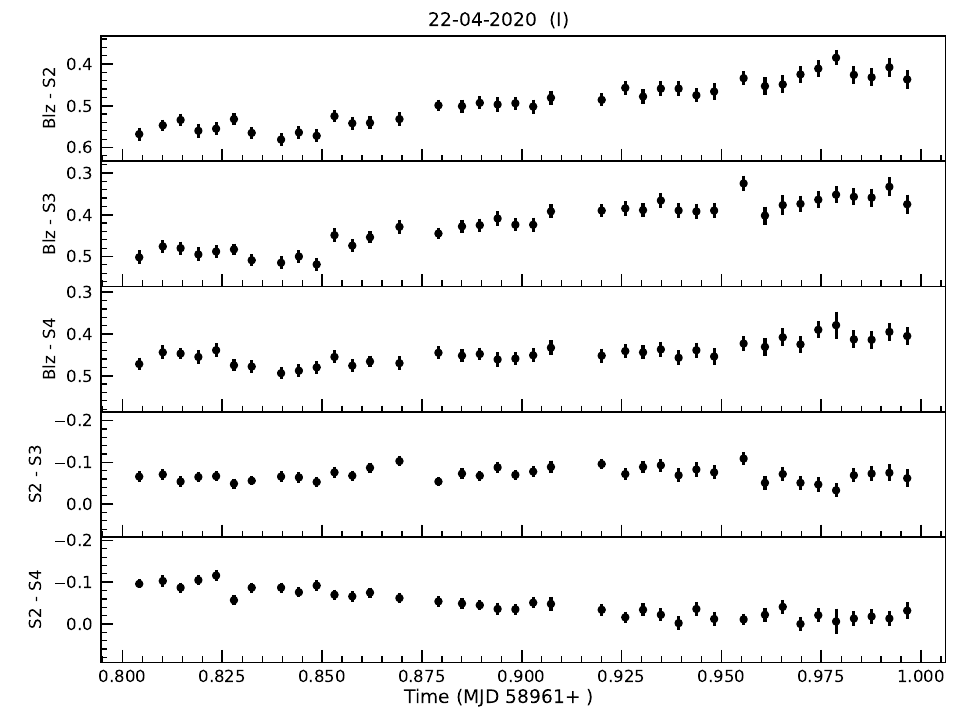}
}
\hbox{
\includegraphics[height=6cm,width=6cm]{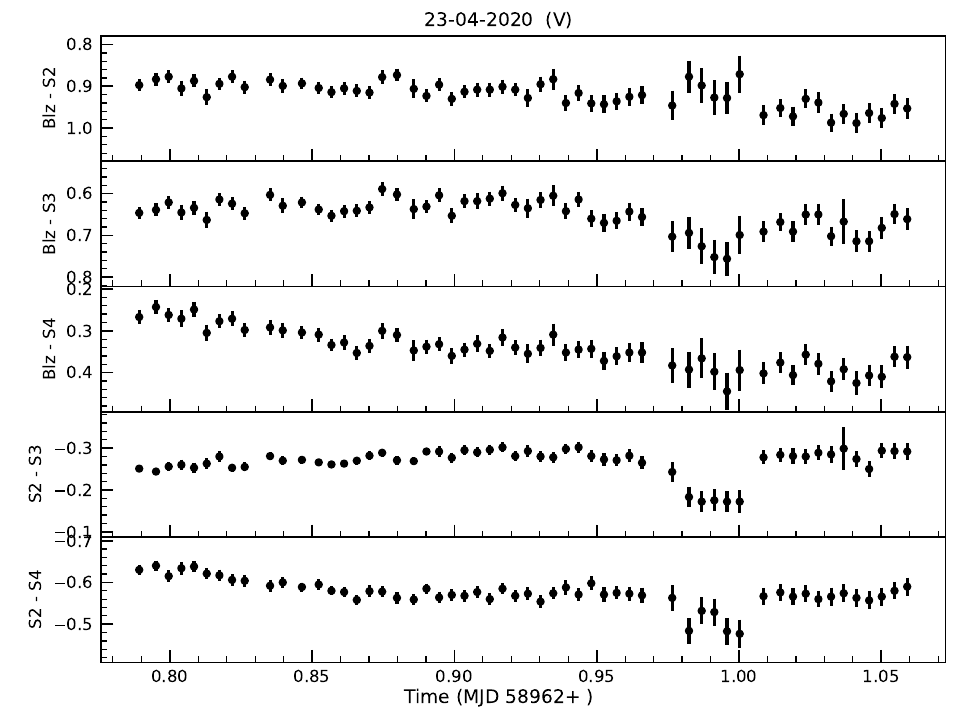}
\includegraphics[height=6cm,width=6cm]{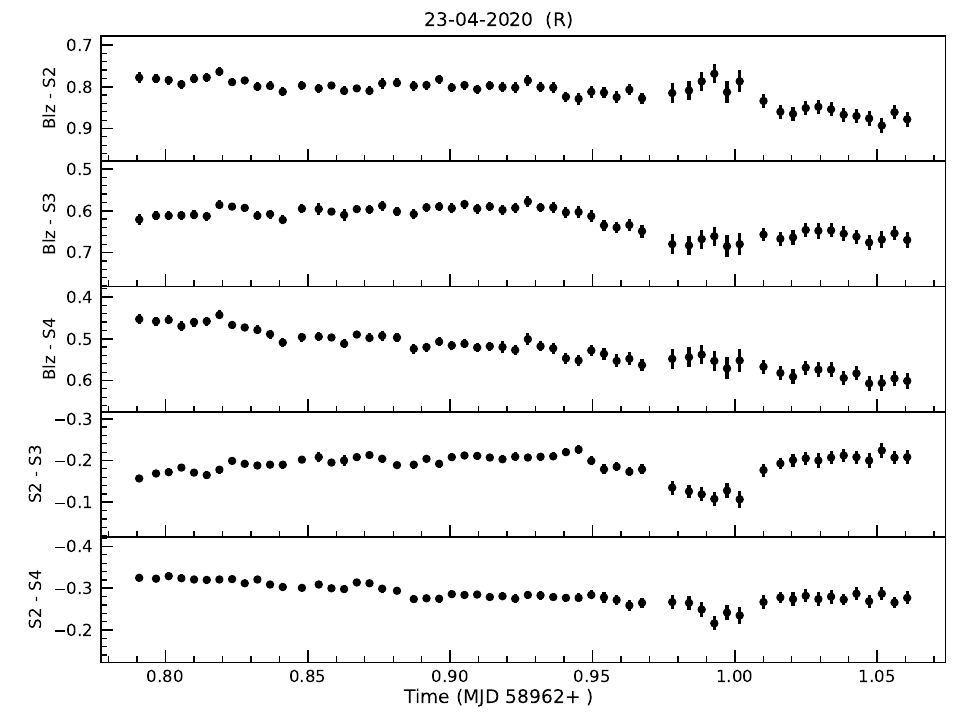}
\includegraphics[height=6cm,width=6cm]{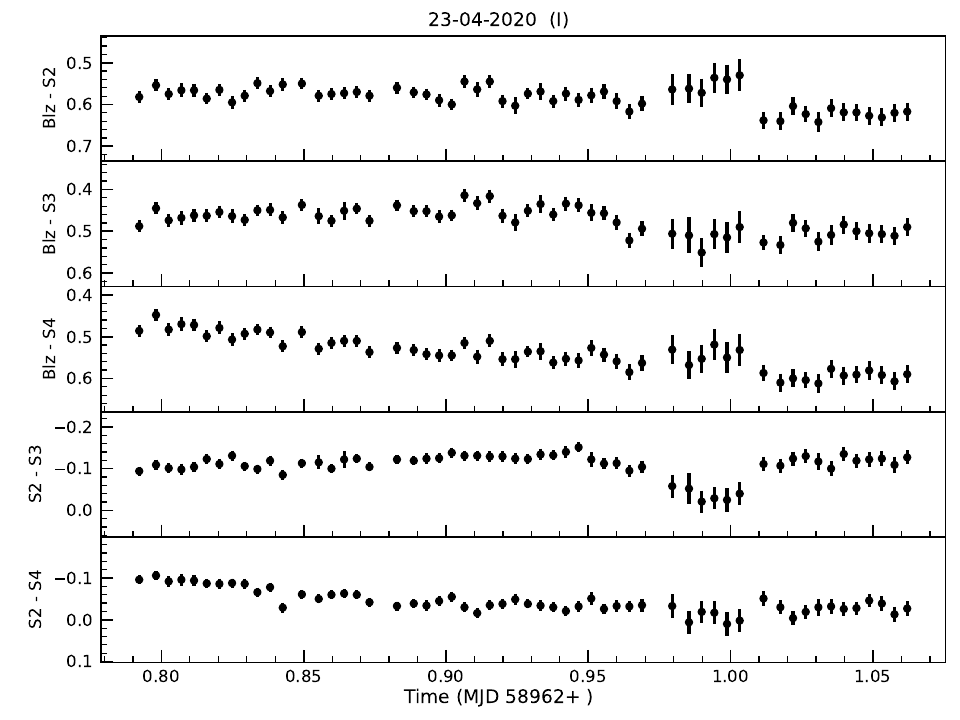}
}
}
\caption{Continued.}
\end{figure*}

\begin{figure*}[h]
 \ContinuedFloat
\vbox{
\hbox{
\includegraphics[height=6cm,width=6cm]{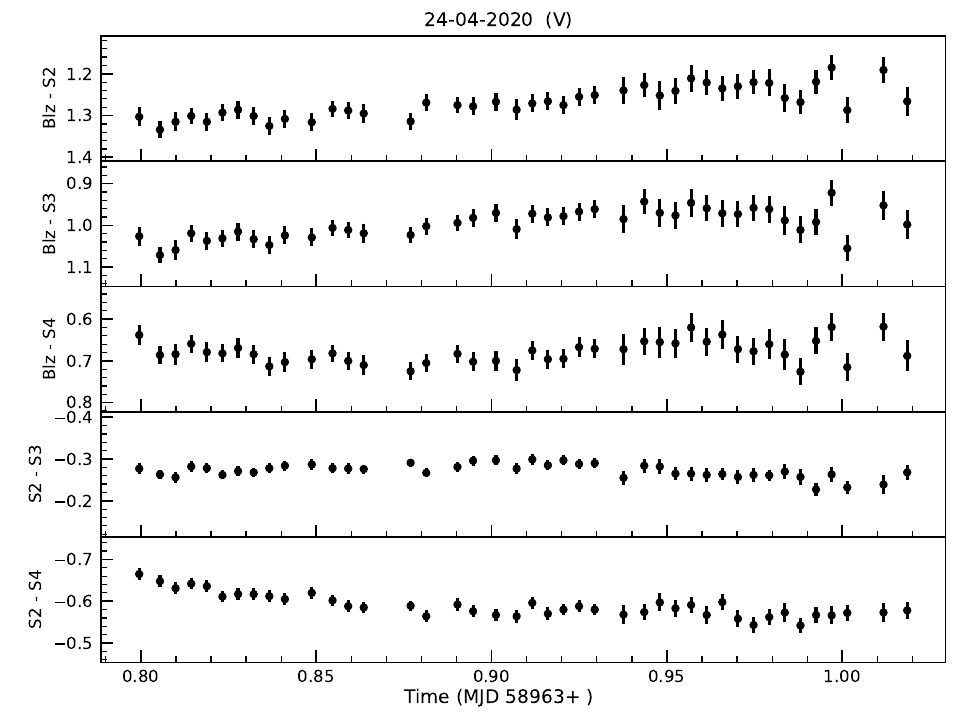}
\includegraphics[height=6cm,width=6cm]{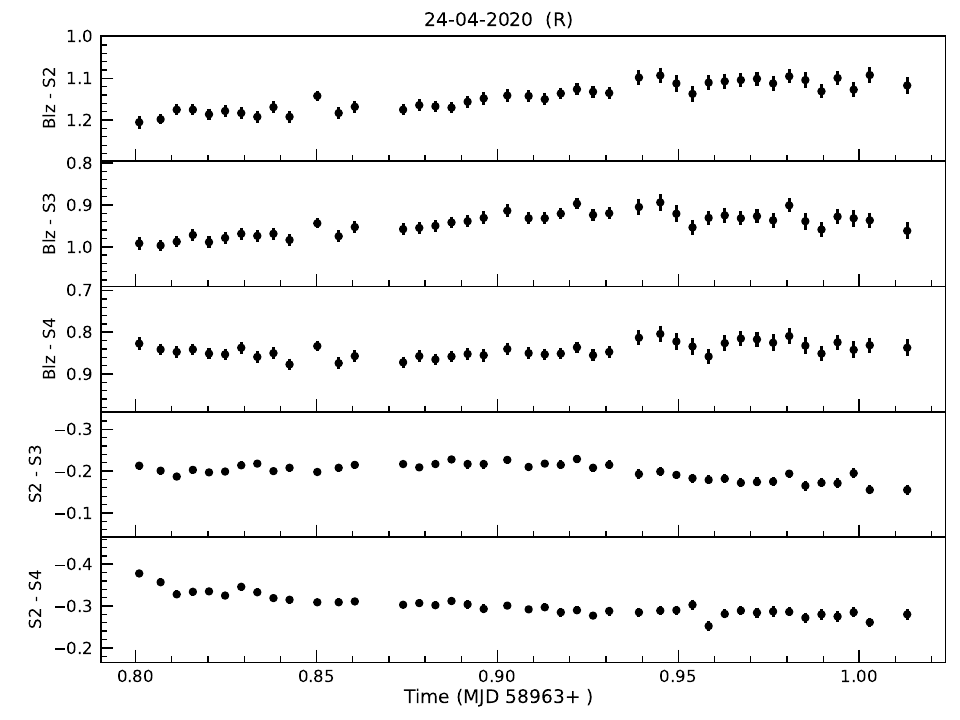}
\includegraphics[height=6cm,width=6cm]{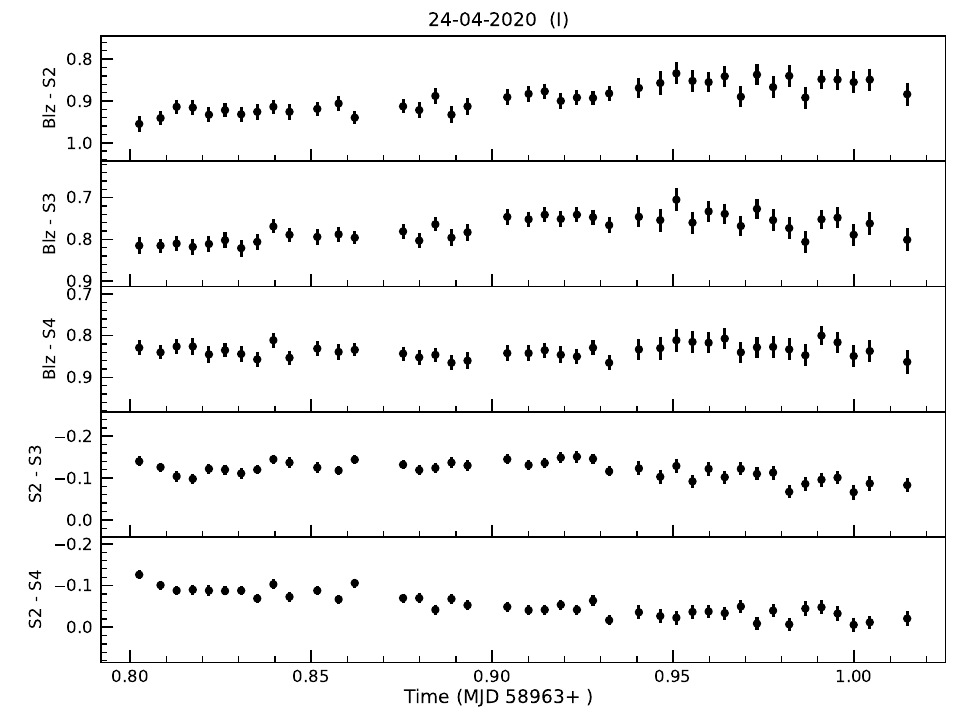}
}

\hbox{
\includegraphics[height=6cm,width=6cm]{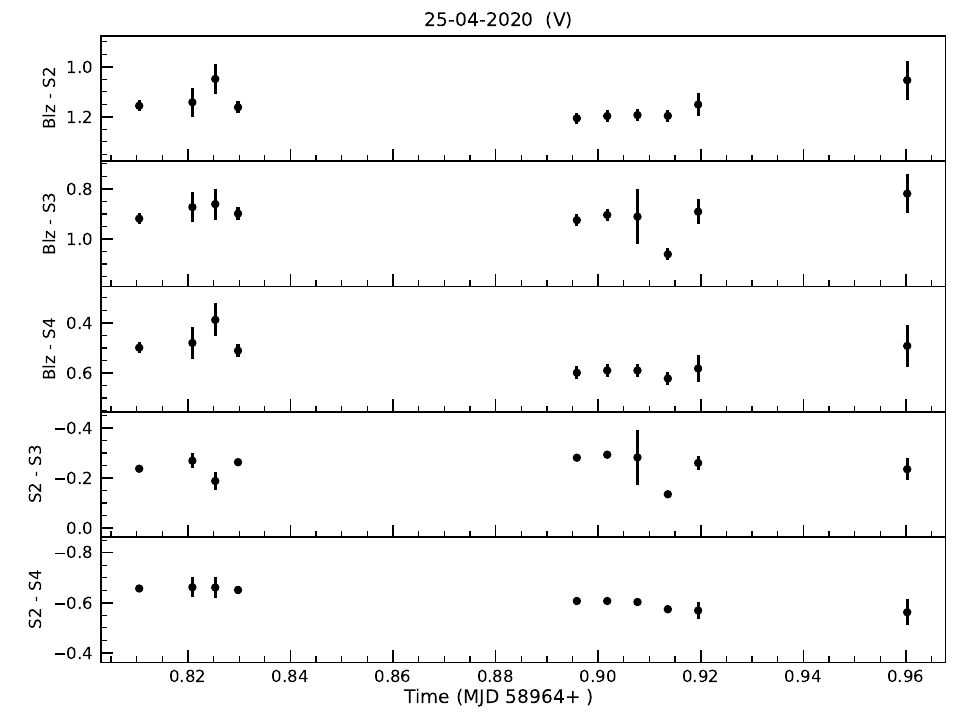}
\includegraphics[height=6cm,width=6cm]{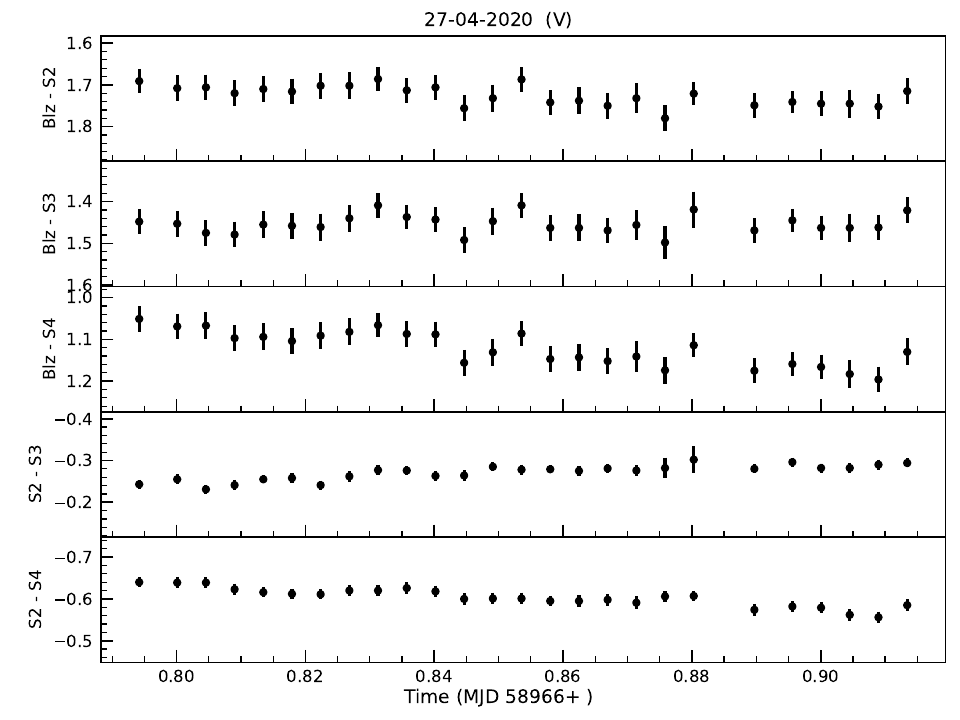}
\includegraphics[height=6cm,width=6cm]{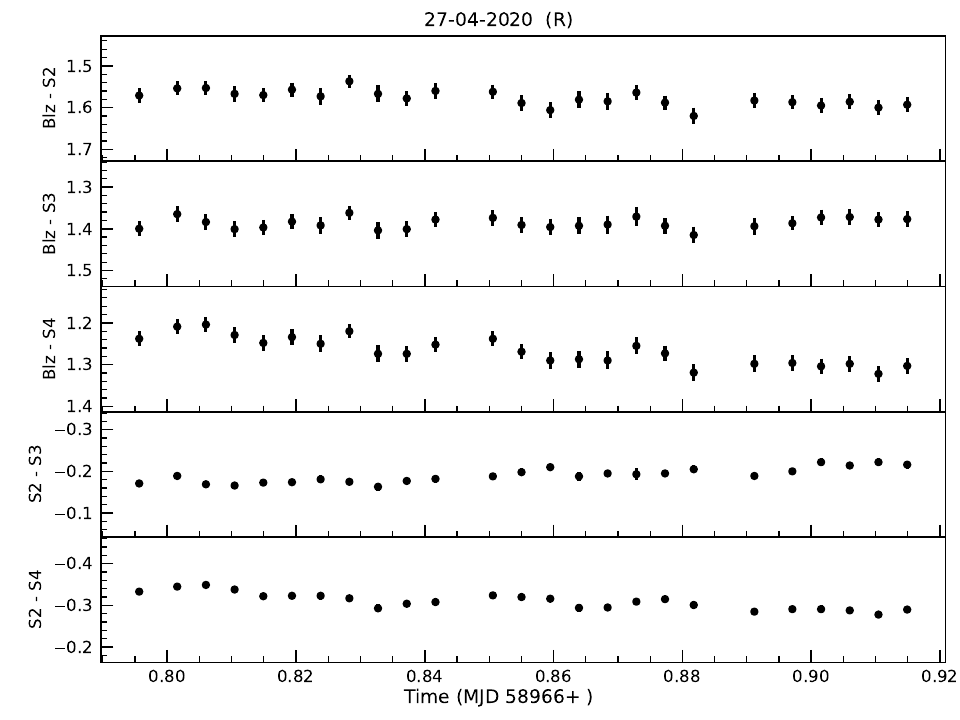}
}

\hbox{
\includegraphics[height=6cm,width=6cm]{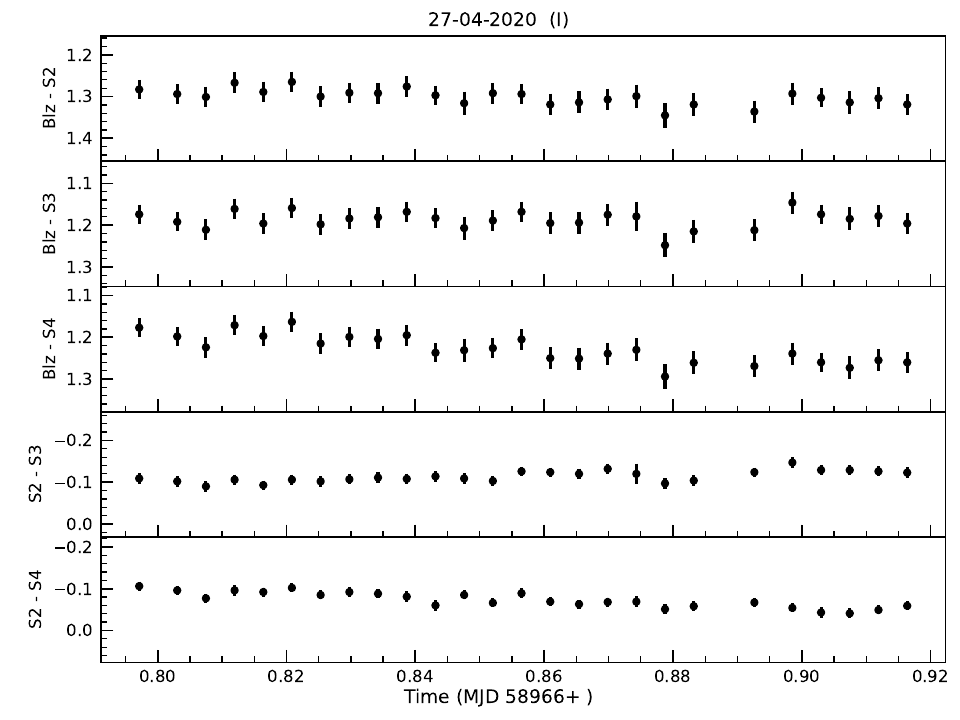}
\includegraphics[height=6cm,width=6cm]{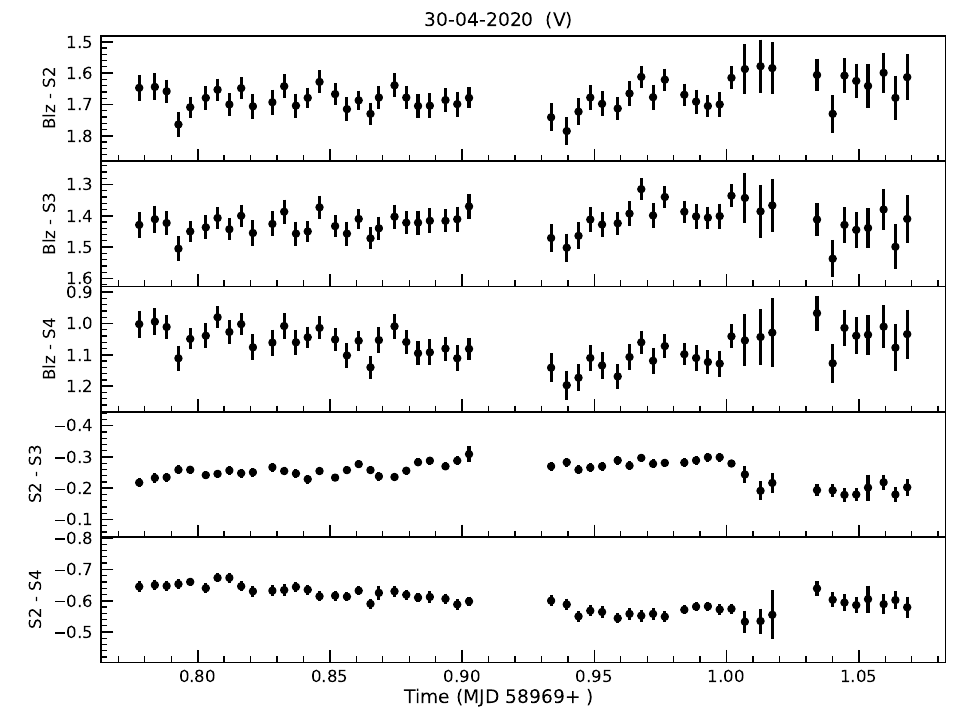}
\includegraphics[height=6cm,width=6cm]{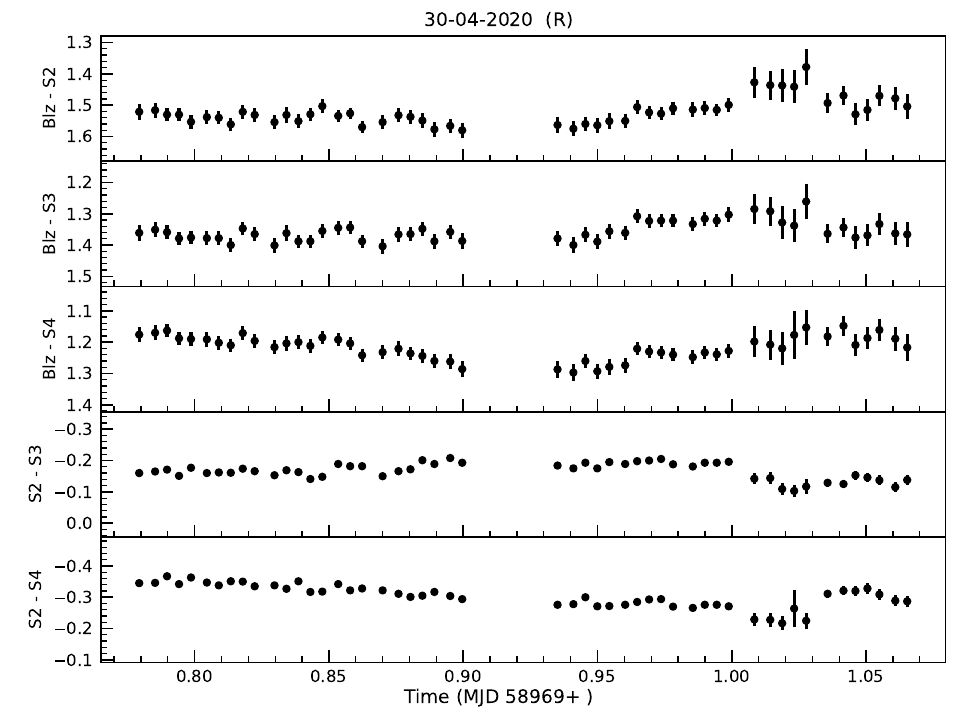}
}
\hbox{
\includegraphics[height=6cm,width=6cm]{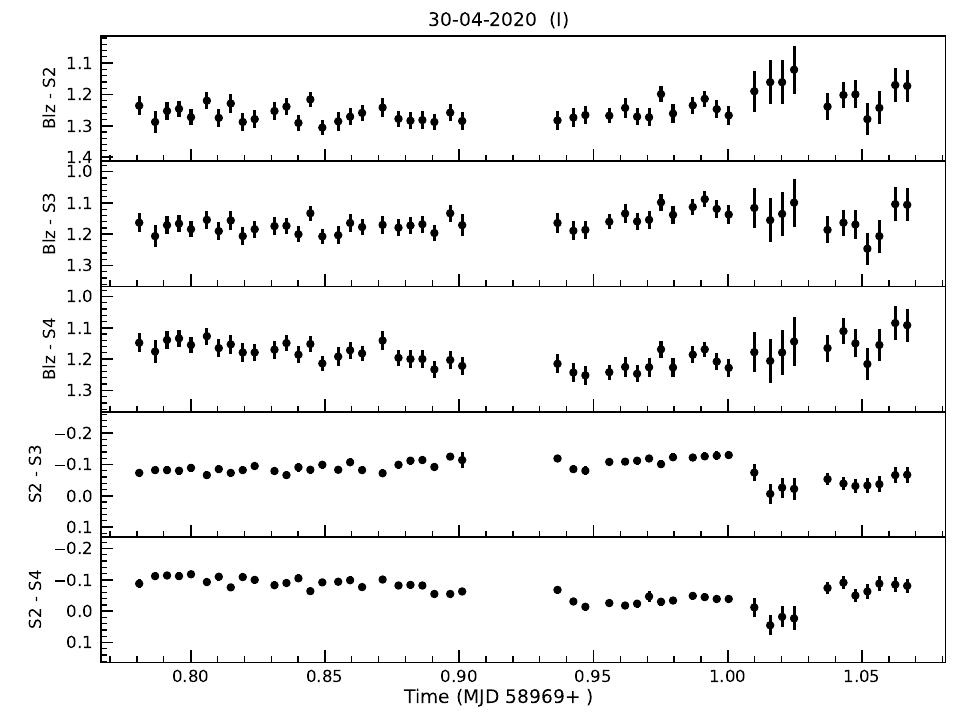}
\includegraphics[height=6cm,width=6cm]{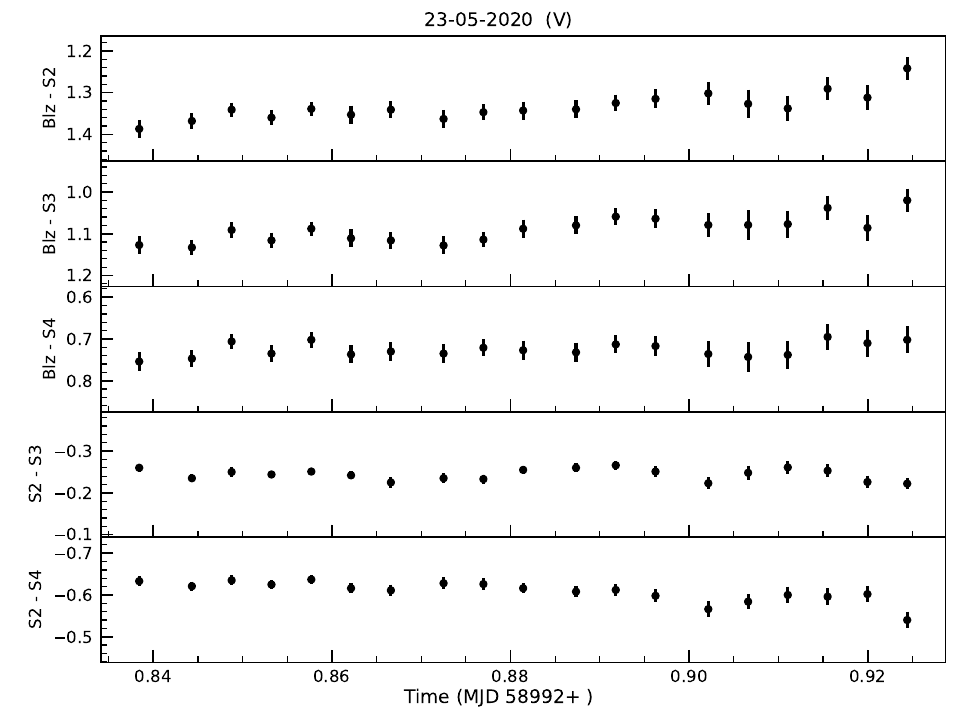}
\includegraphics[height=6cm,width=6cm]{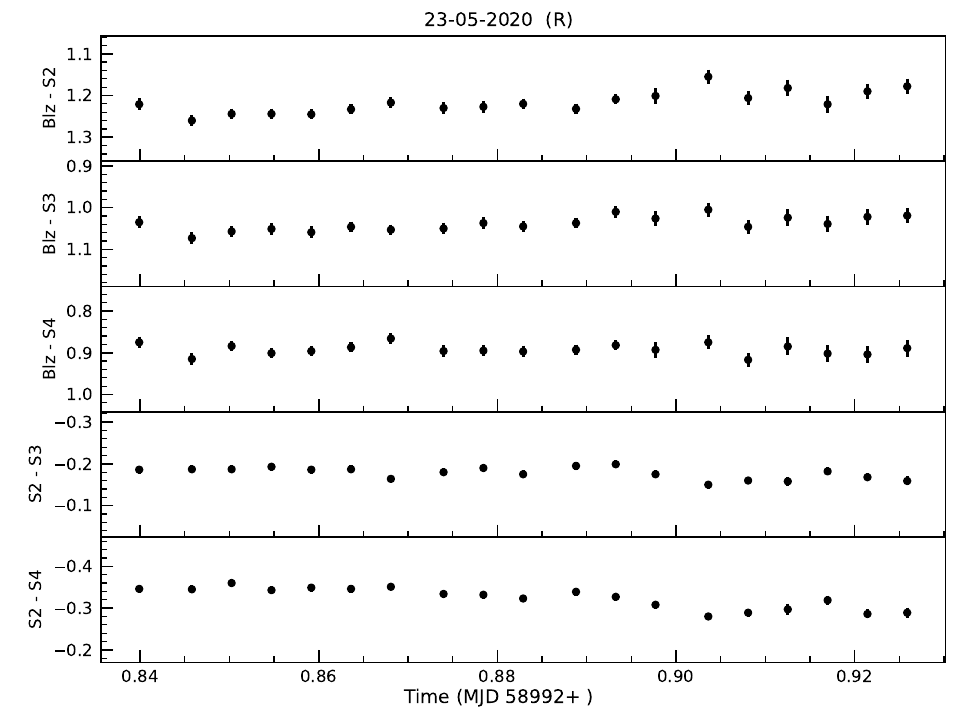}
}
}
\caption{Continued.}
\end{figure*}

\begin{figure*}[h]
 \ContinuedFloat
\vbox{
\hbox{
\includegraphics[height=6cm,width=6cm]{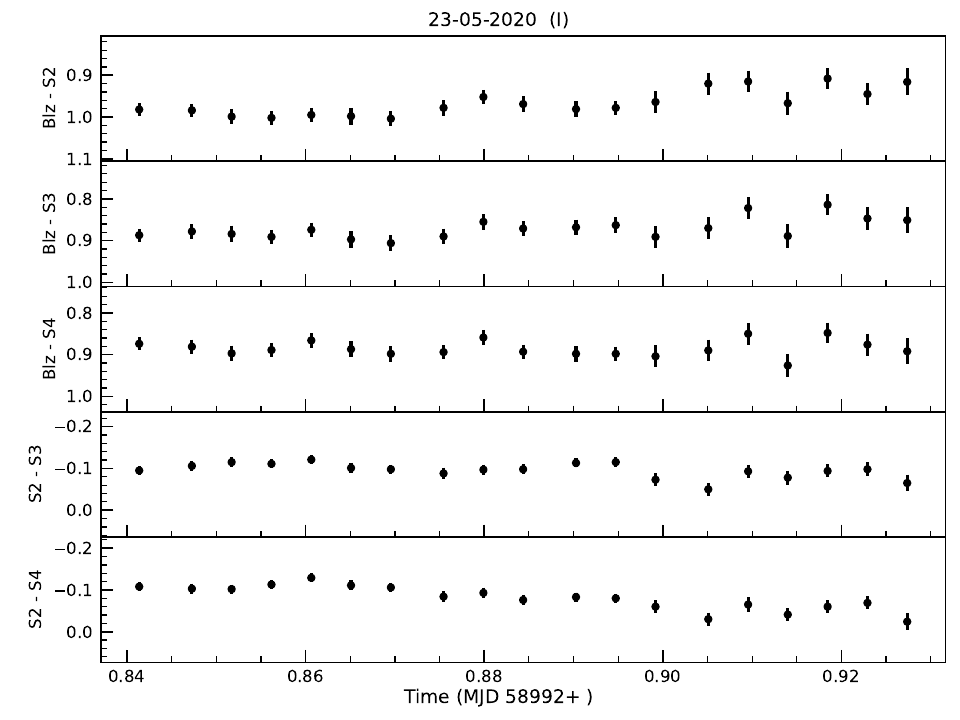}
\includegraphics[height=6cm,width=6cm]{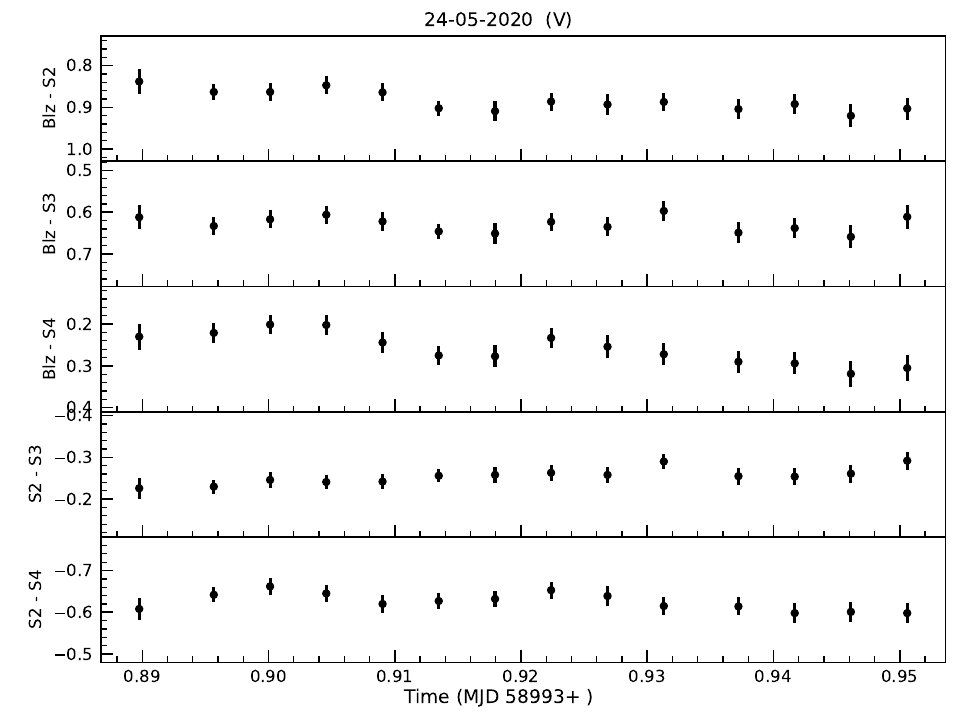}
\includegraphics[height=6cm,width=6cm]{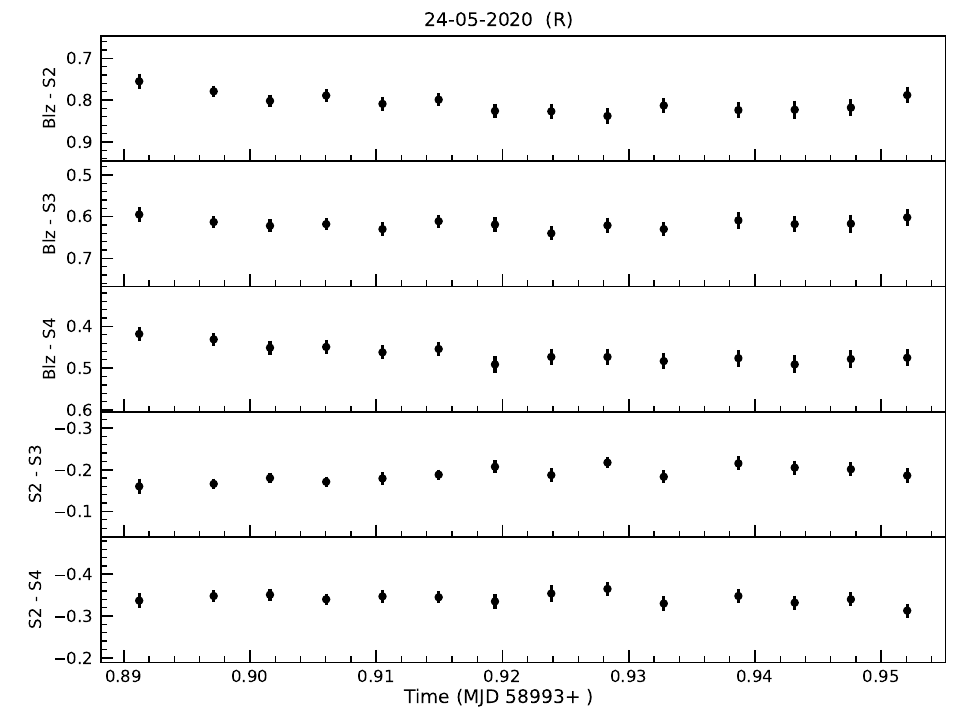}
}

\hbox{
\includegraphics[height=6cm,width=6cm]{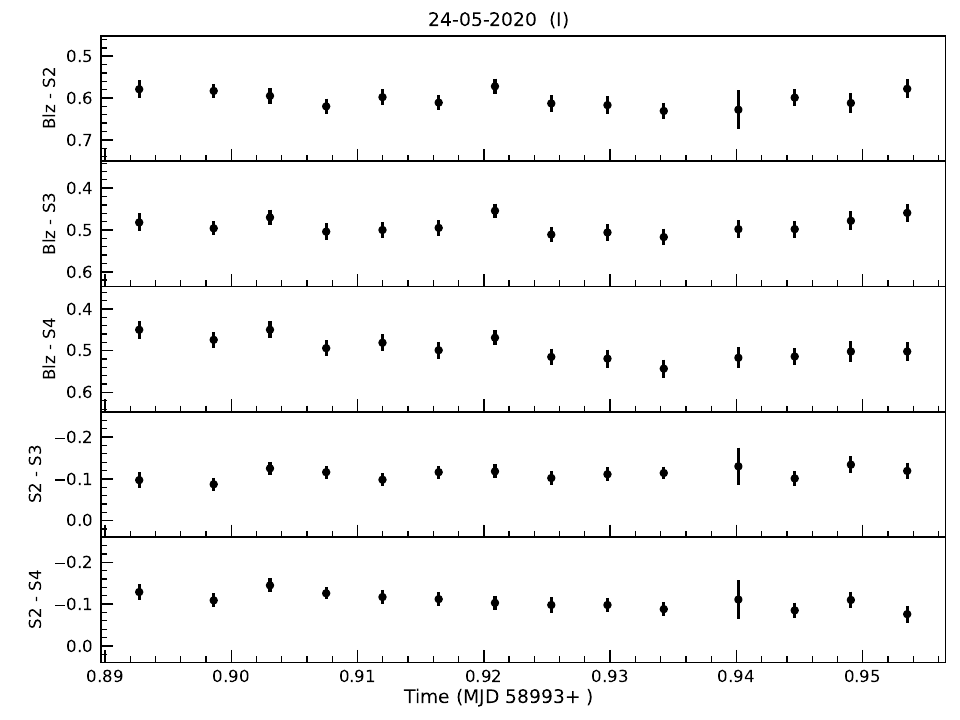}
\includegraphics[height=6cm,width=6cm]{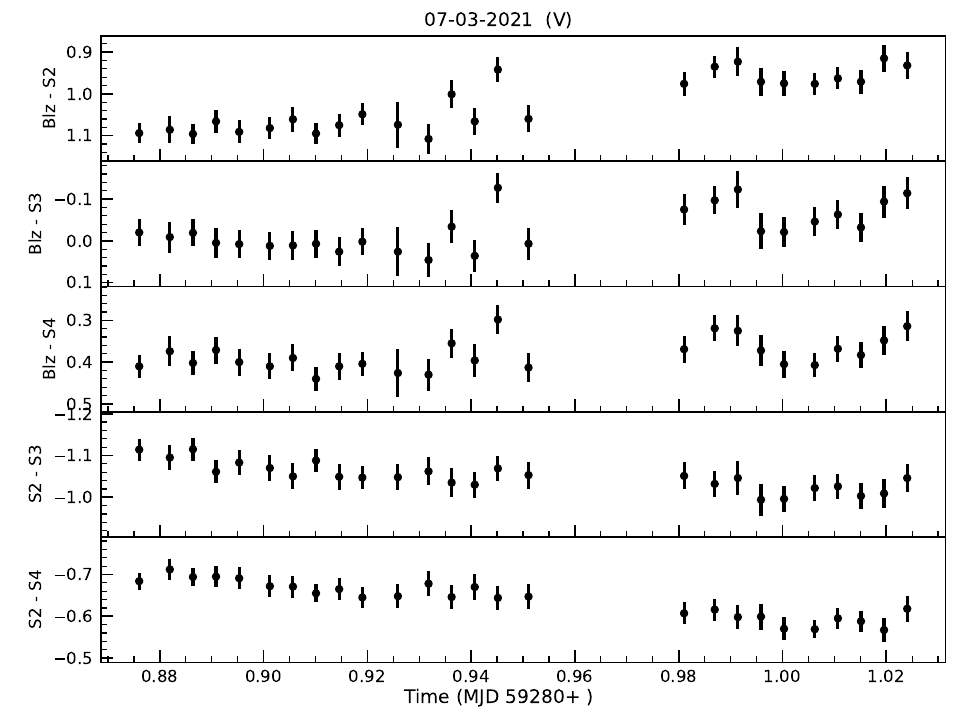}
\includegraphics[height=6cm,width=6cm]{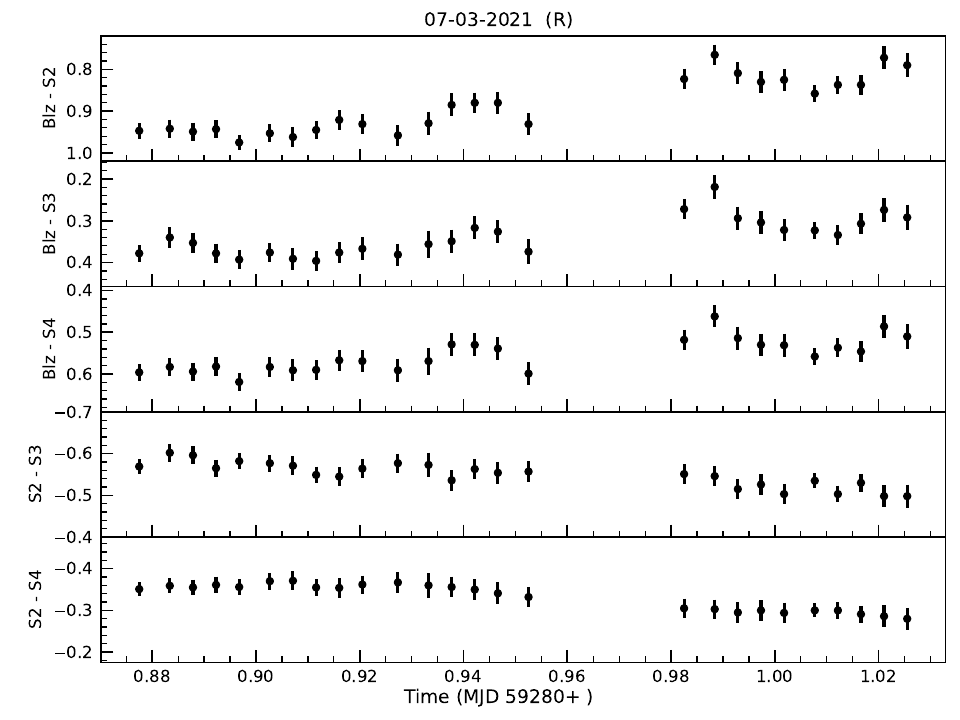}
}

\hbox{
\includegraphics[height=6cm,width=6cm]{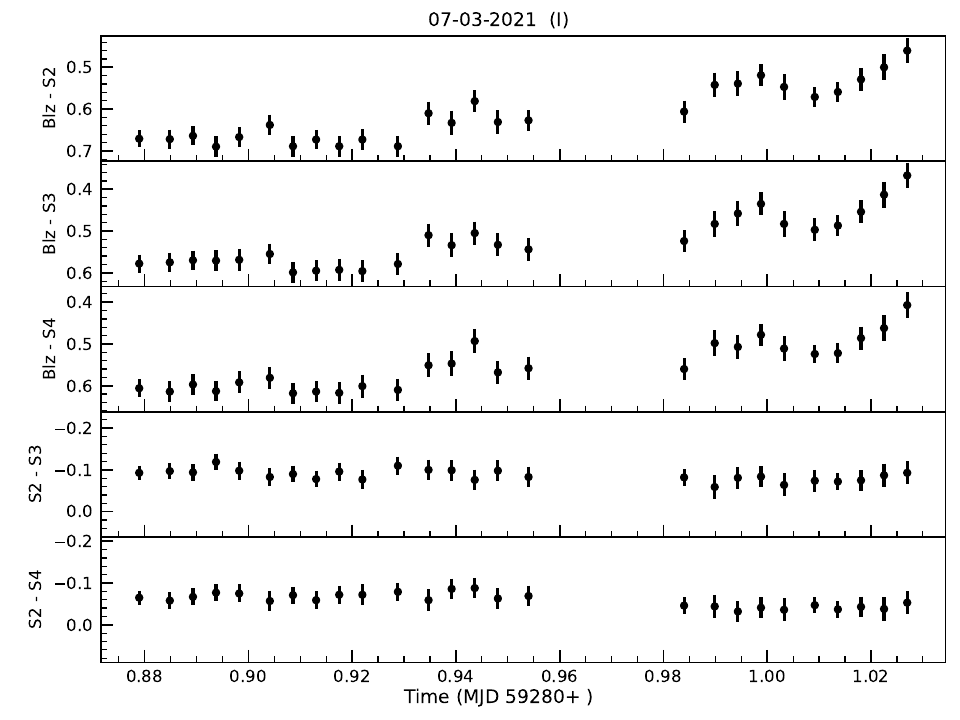}
\includegraphics[height=6cm,width=6cm]{20210308_flx_V.pdf}
\includegraphics[height=6cm,width=6cm]{20210308_flx_R.pdf}
}
\hbox{
\includegraphics[height=6cm,width=6cm]{20210308_flx_I.pdf}
\includegraphics[height=6cm,width=6cm]{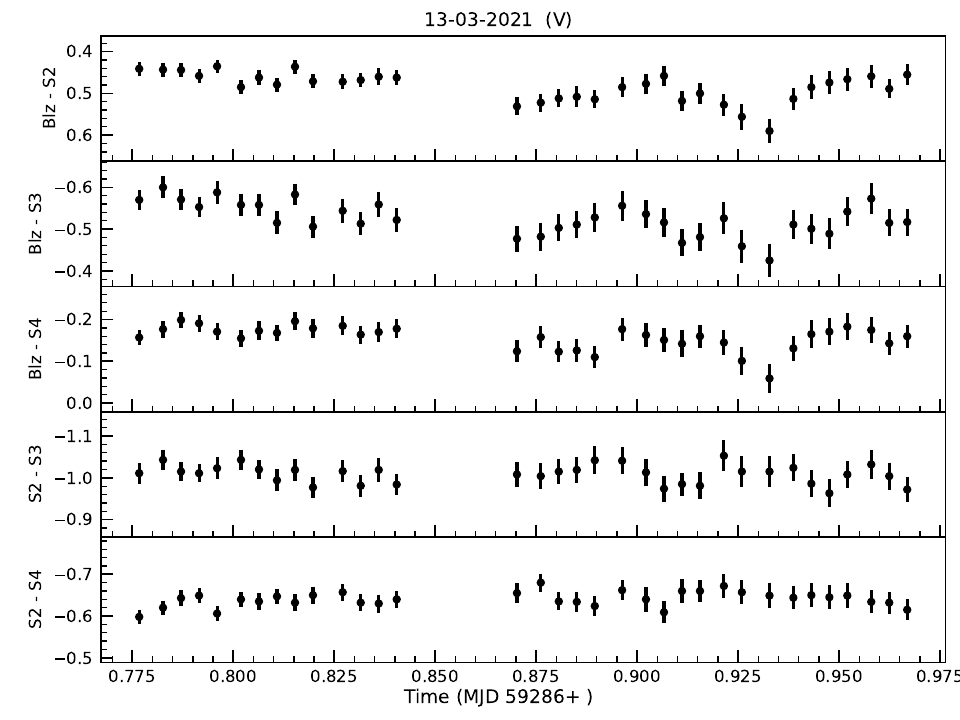}
\includegraphics[height=6cm,width=6cm]{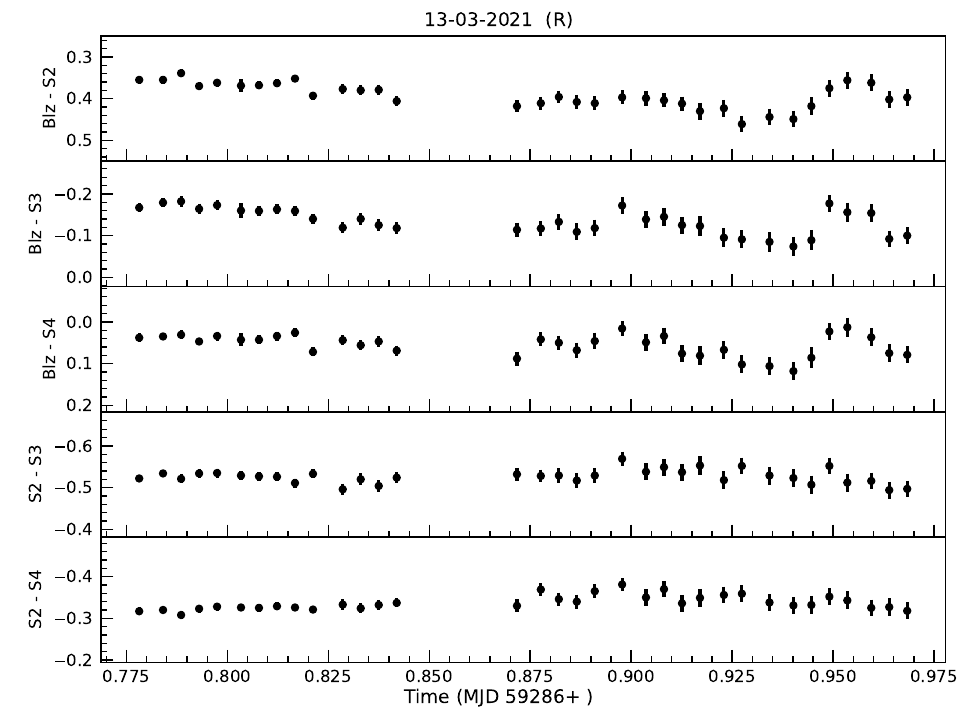}
}
}
\caption{Continued.}
\end{figure*}

\begin{figure*}[h]
 \ContinuedFloat
\vbox{
\hbox{
\includegraphics[height=6cm,width=6cm]{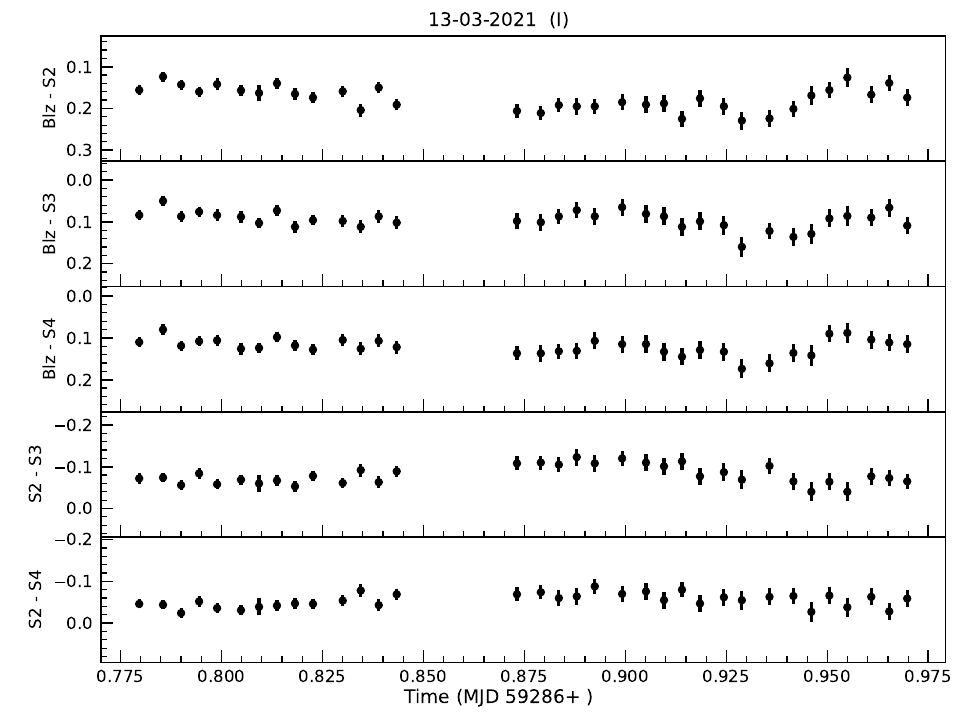}
\includegraphics[height=6cm,width=6cm]{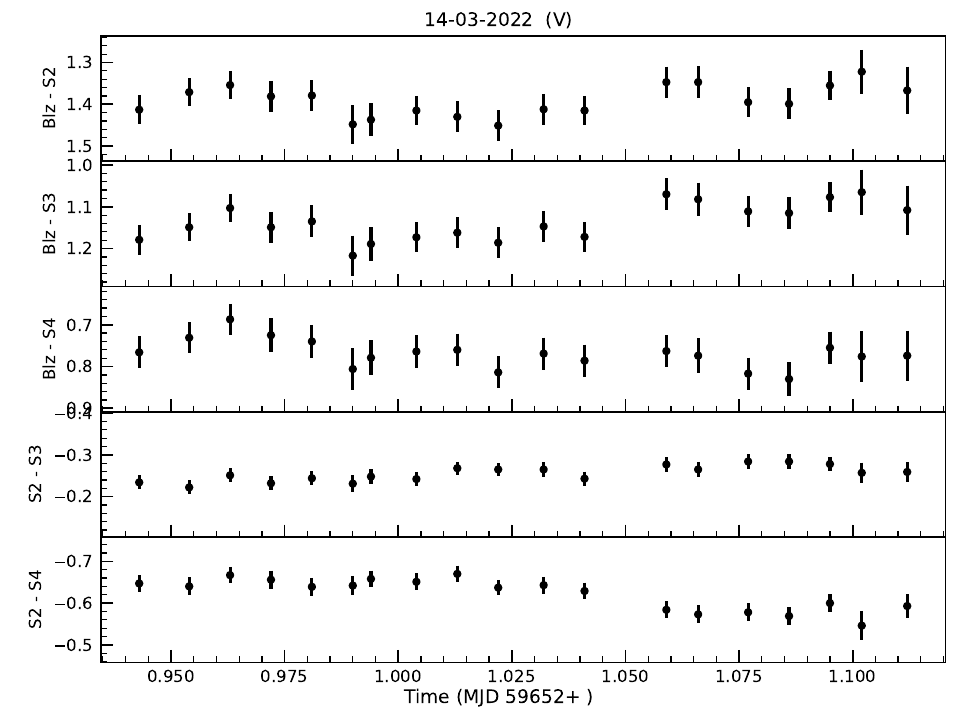}
\includegraphics[height=6cm,width=6cm]{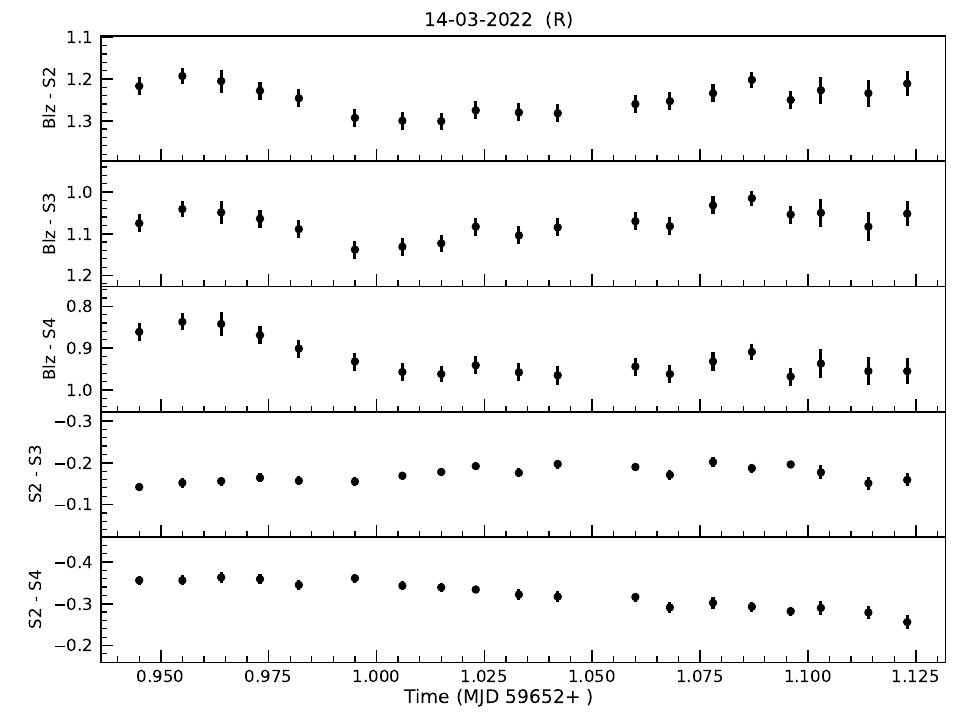}
}

\hbox{
\includegraphics[height=6cm,width=6cm]{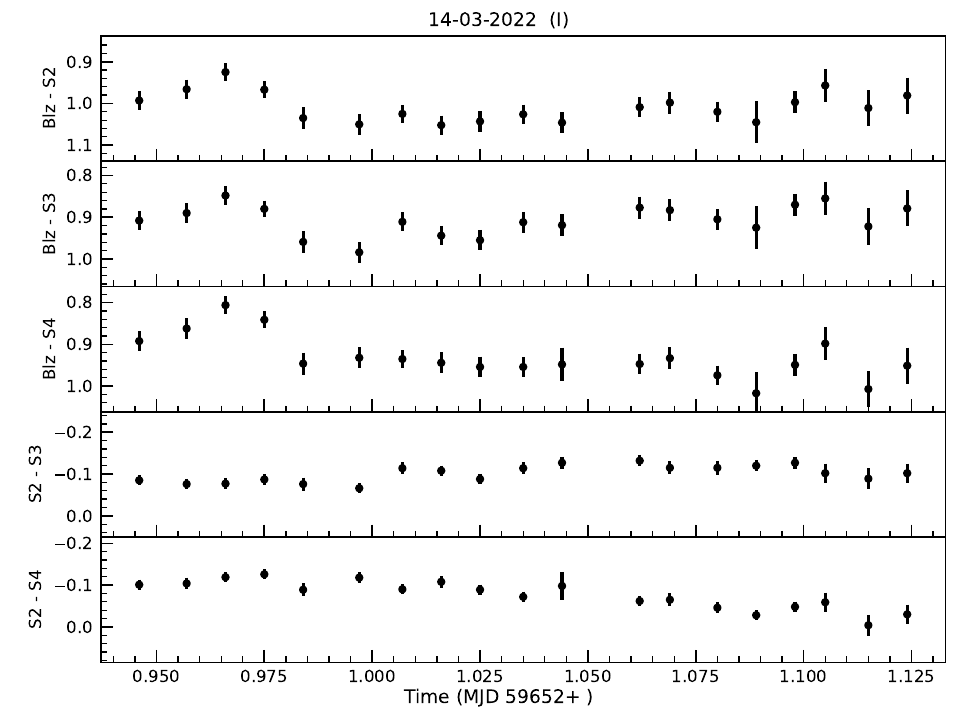}
\includegraphics[height=6cm,width=6cm]{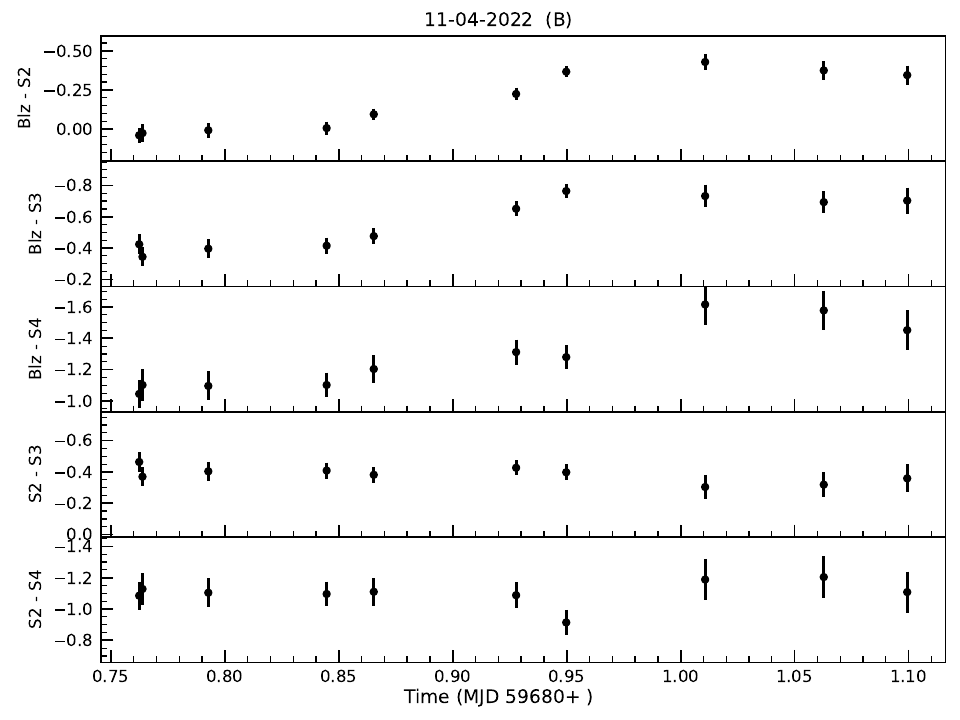}
\includegraphics[height=6cm,width=6cm]{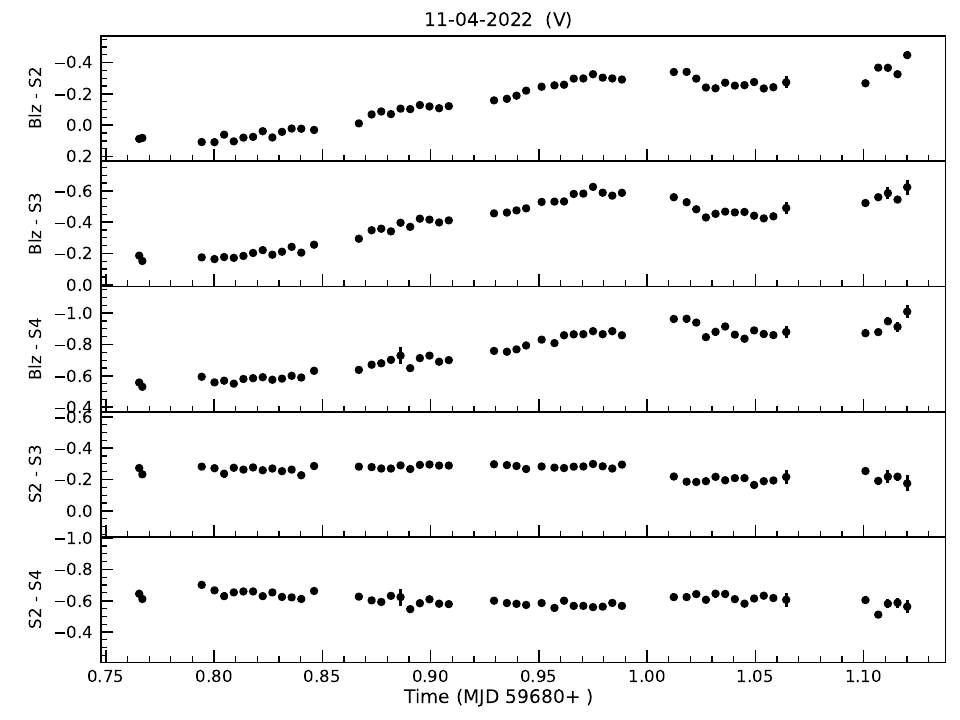}
}

\hbox{
\includegraphics[height=6cm,width=6cm]{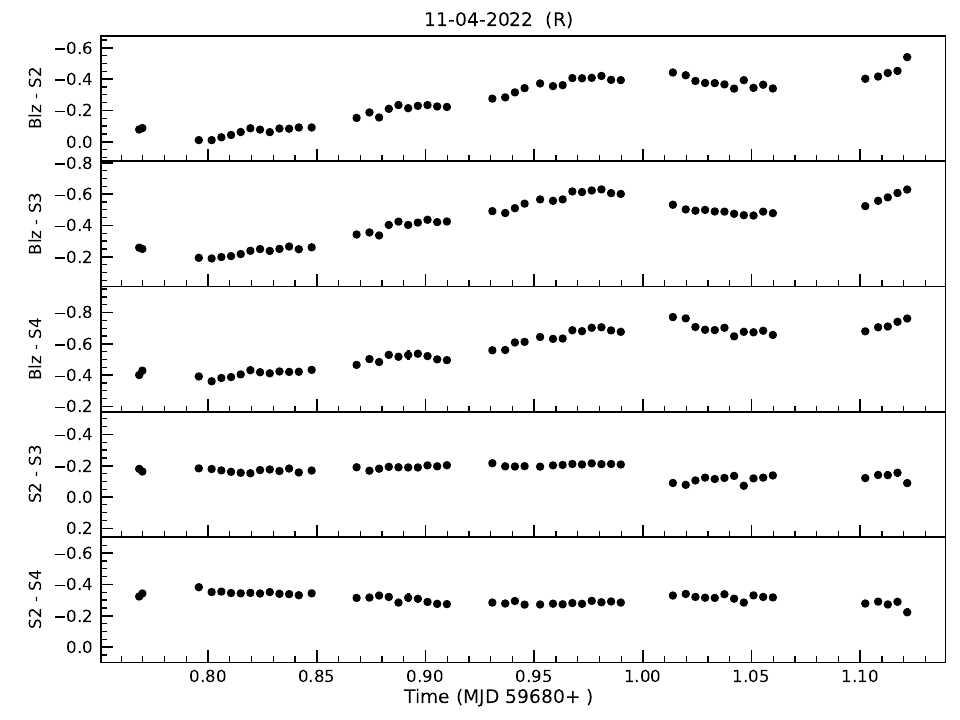}
\includegraphics[height=6cm,width=6cm]{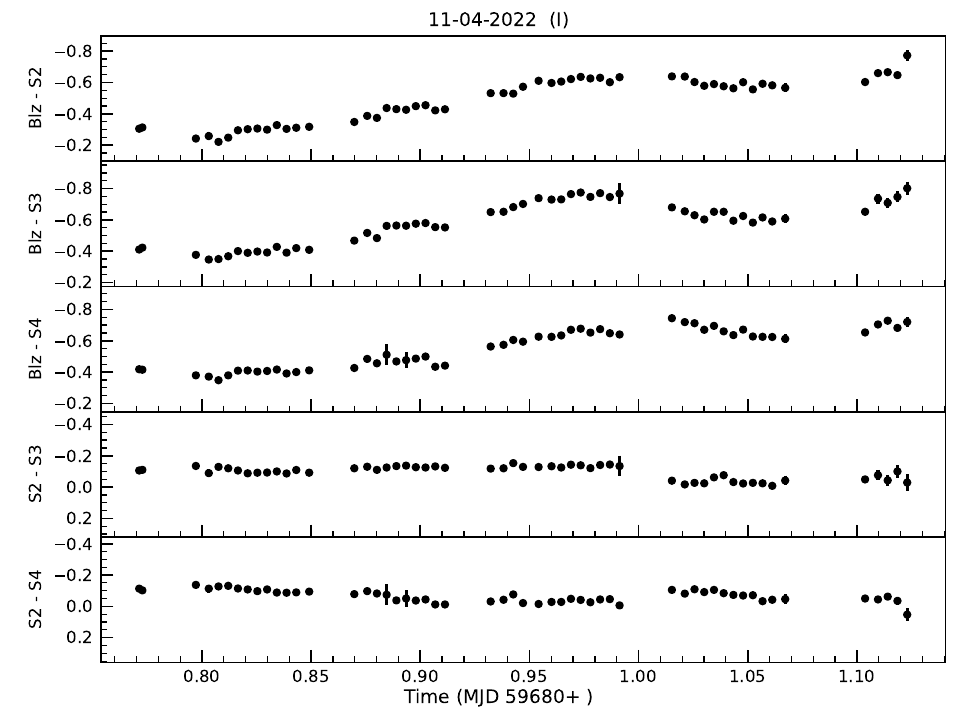}
\includegraphics[height=6cm,width=6cm]{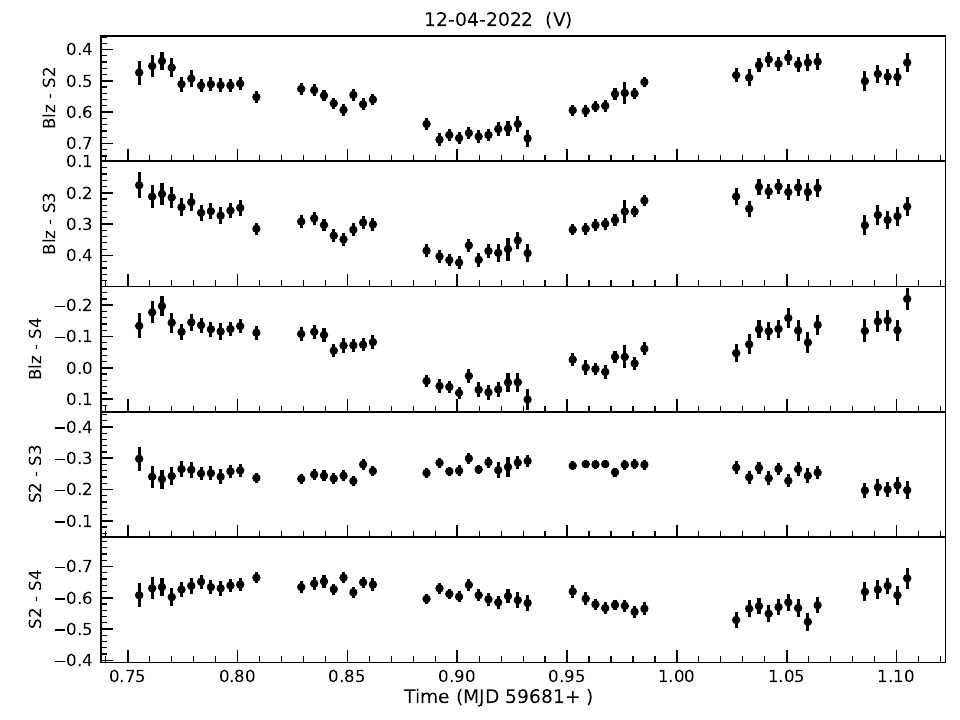}
}
\hbox{
\includegraphics[height=6cm,width=6cm]{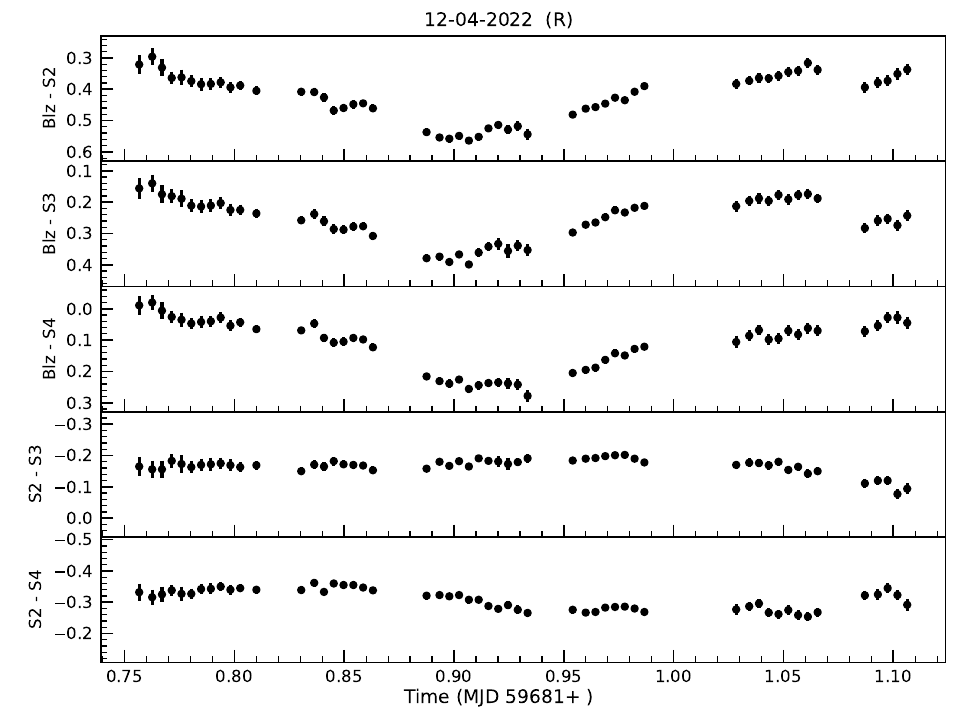}
\includegraphics[height=6cm,width=6cm]{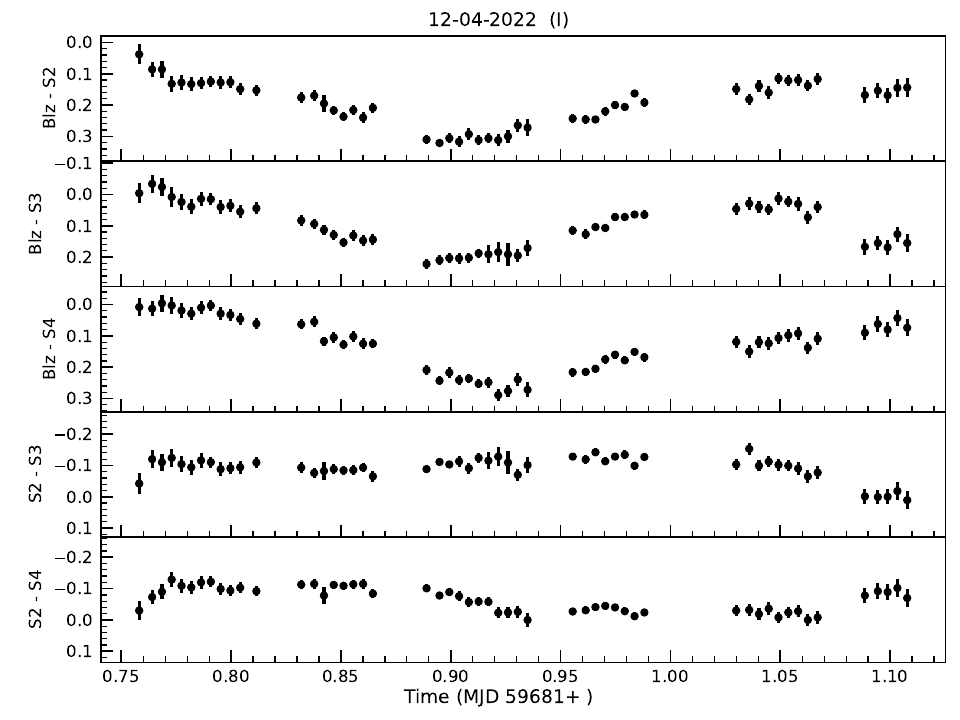}
\includegraphics[height=6cm,width=6cm]{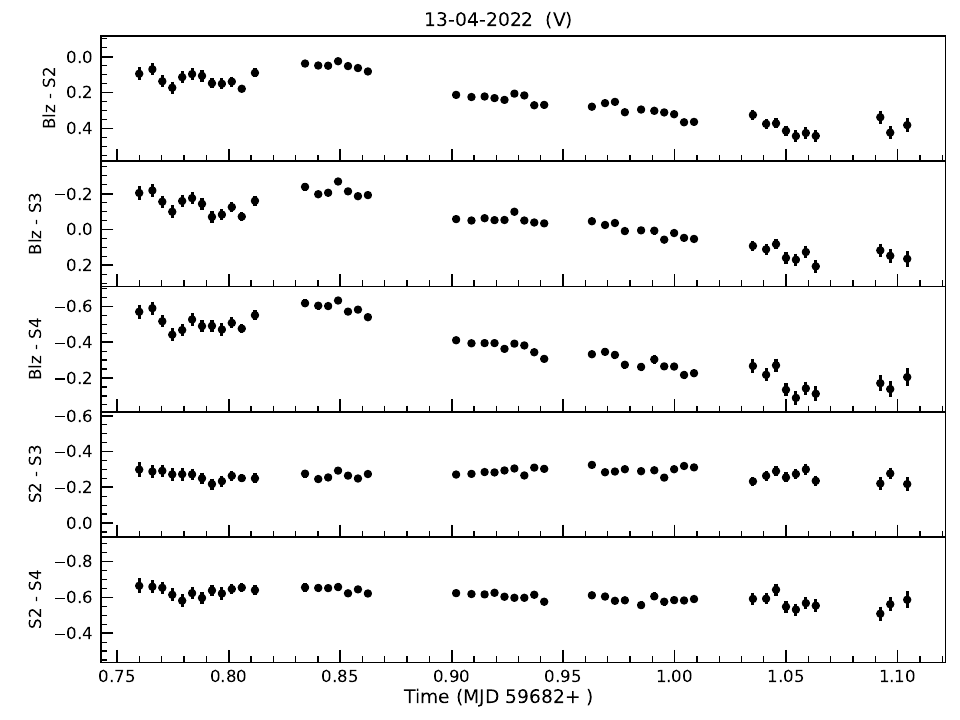}
}
}
\caption{Continued.}
\end{figure*}

\begin{figure*}[h]
 \ContinuedFloat
\vbox{
\hbox{
\includegraphics[height=6cm,width=6cm]{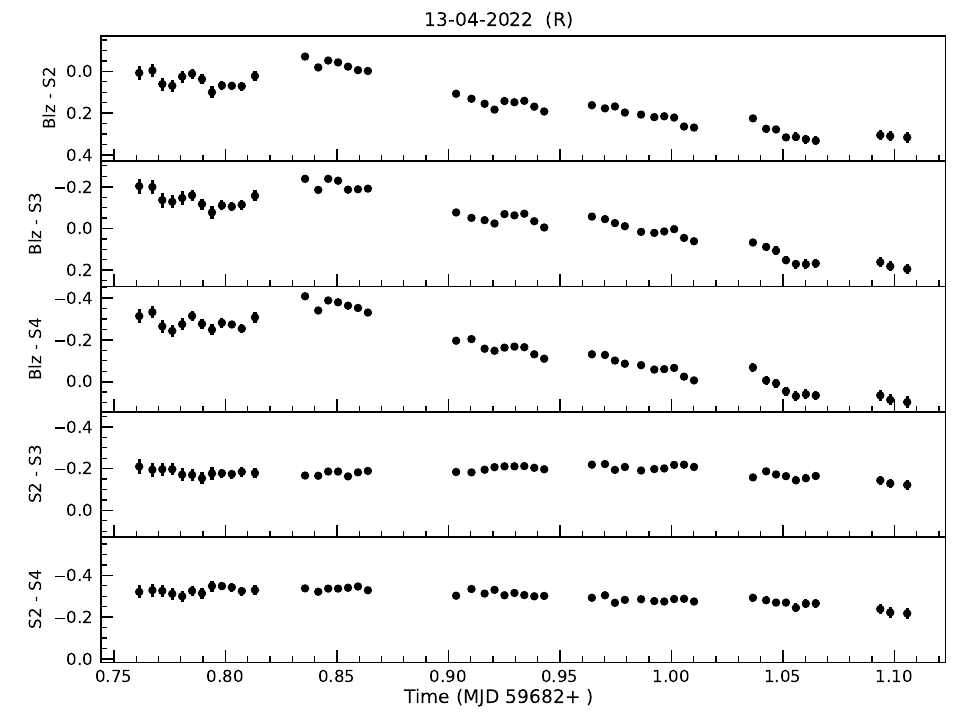}
\includegraphics[height=6cm,width=6cm]{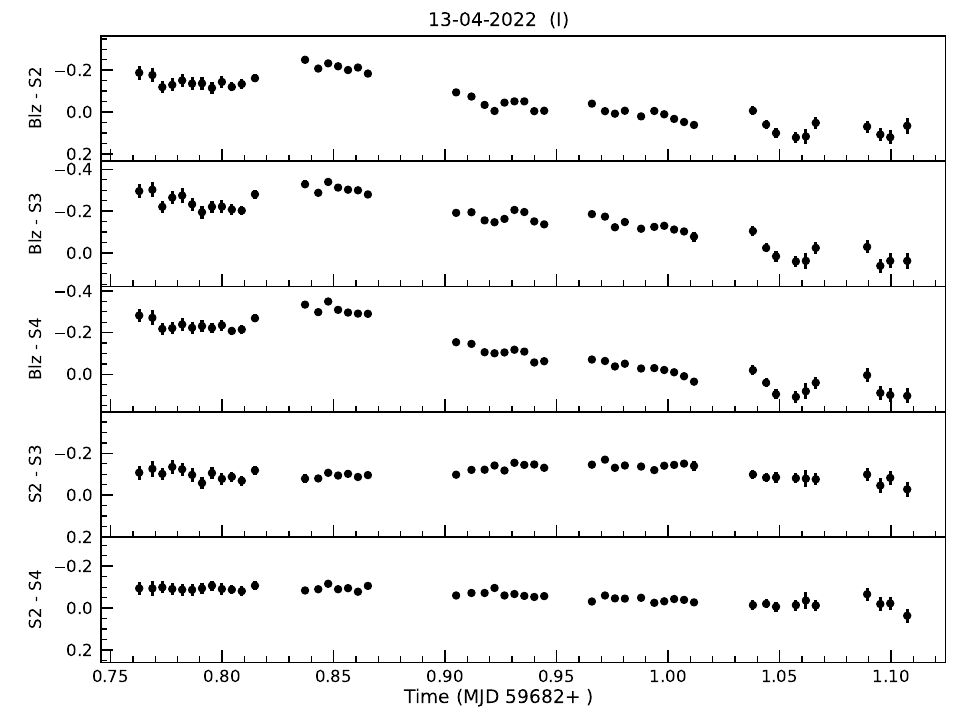}
\includegraphics[height=6cm,width=6cm]{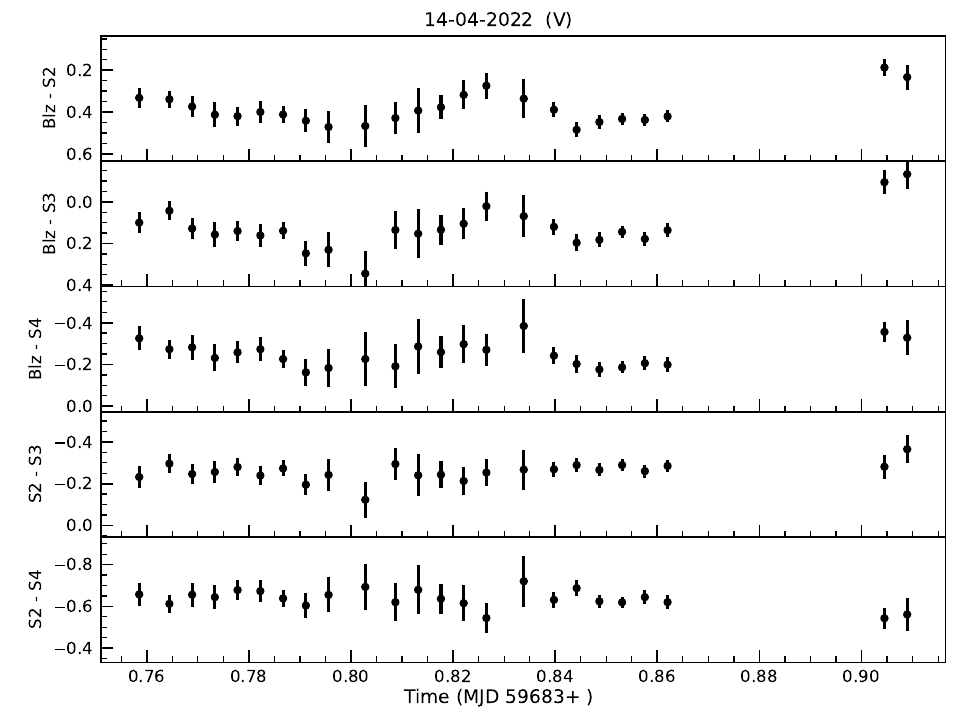}
}

\hbox{
\includegraphics[height=6cm,width=6cm]{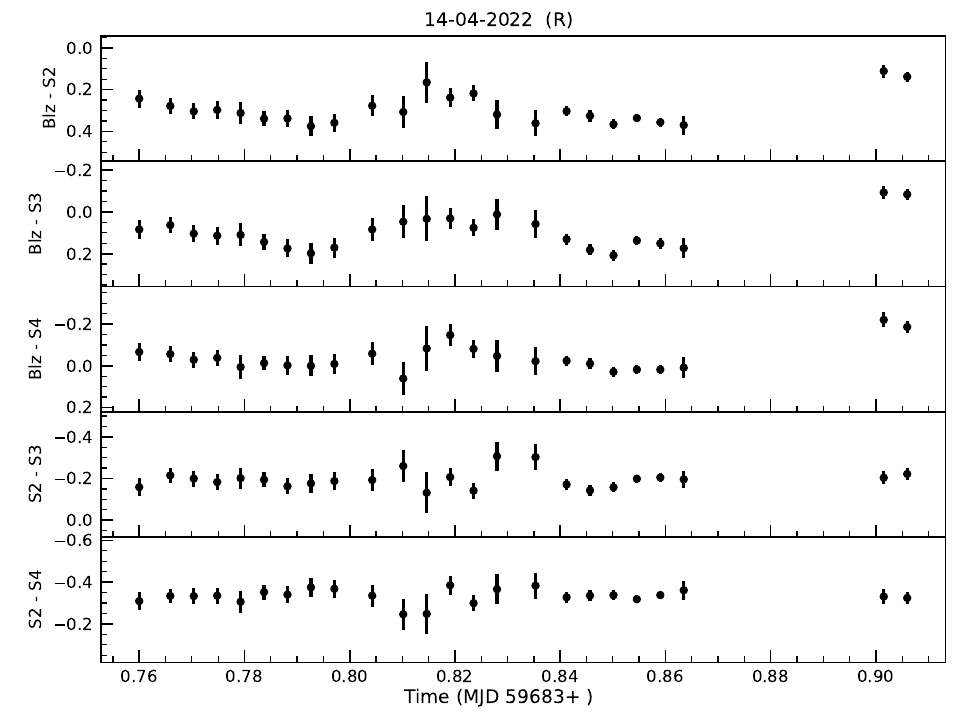}
\includegraphics[height=6cm,width=6cm]{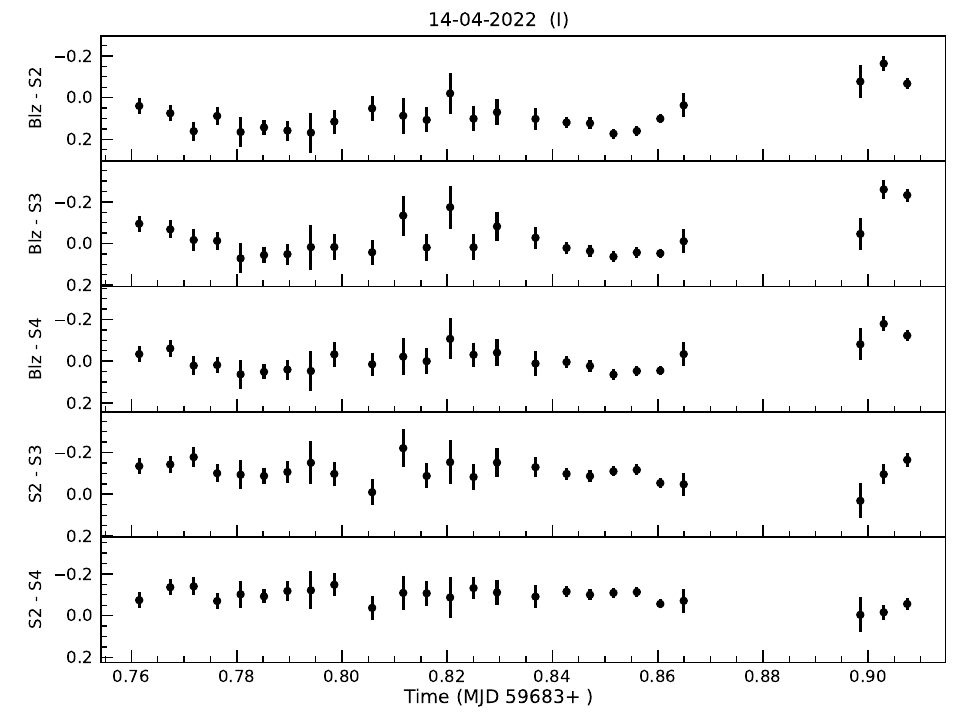}
\includegraphics[height=6cm,width=6cm]{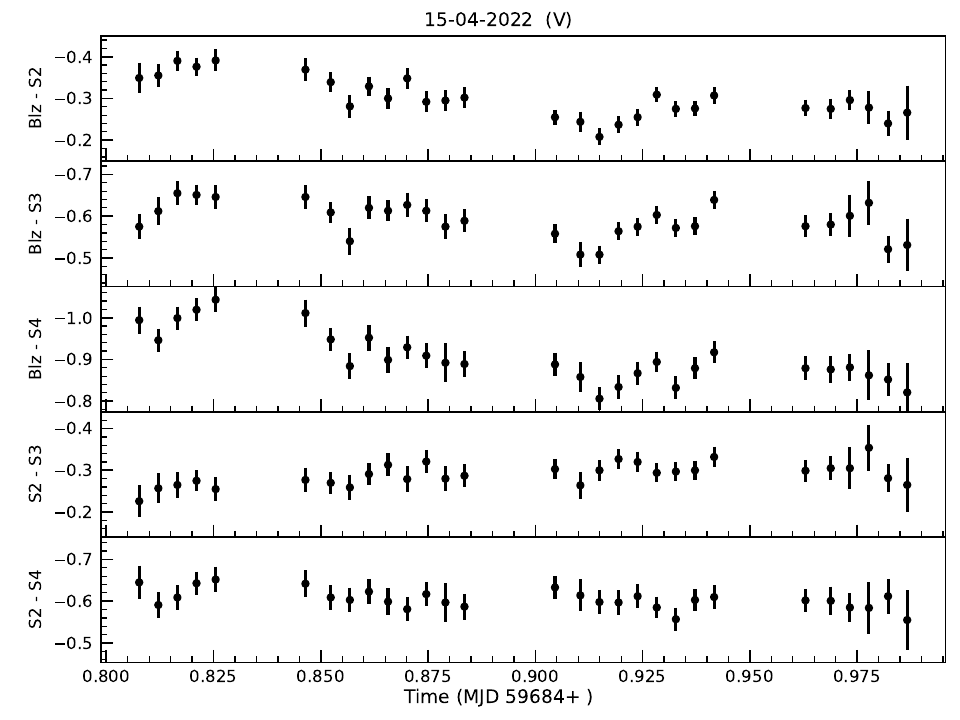}
}

\hbox{
\includegraphics[height=6cm,width=6cm]{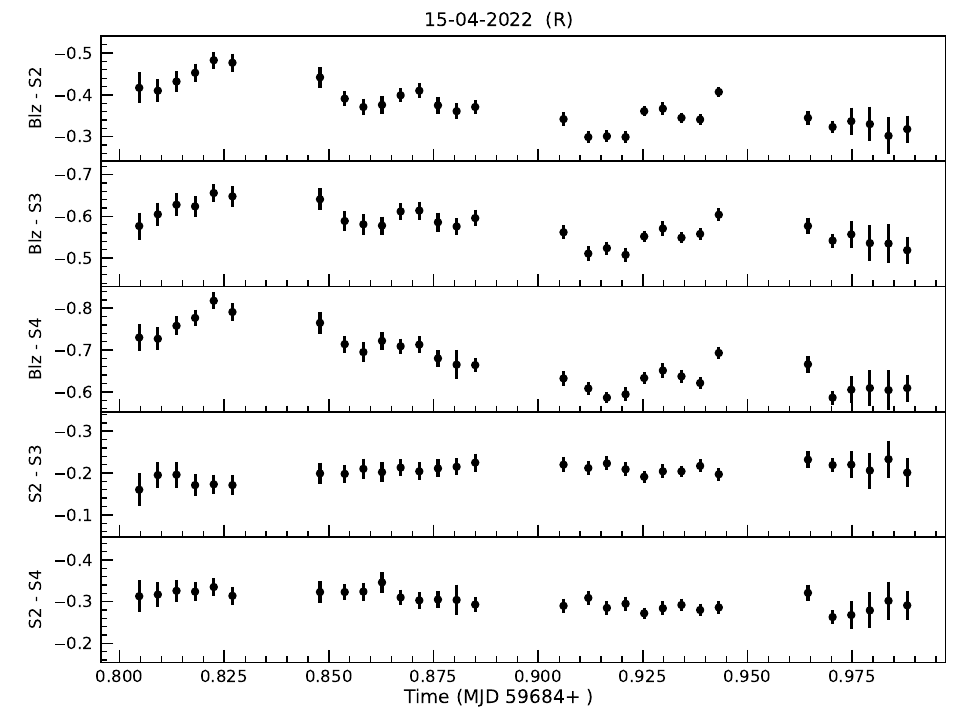}
\includegraphics[height=6cm,width=6cm]{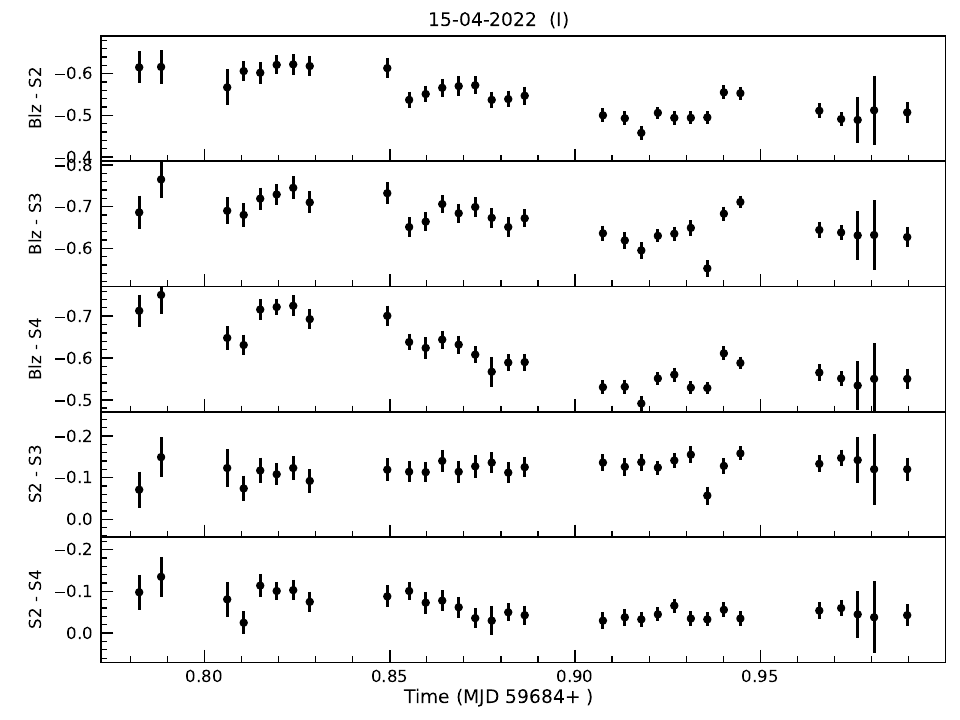}
\includegraphics[height=6cm,width=6cm]{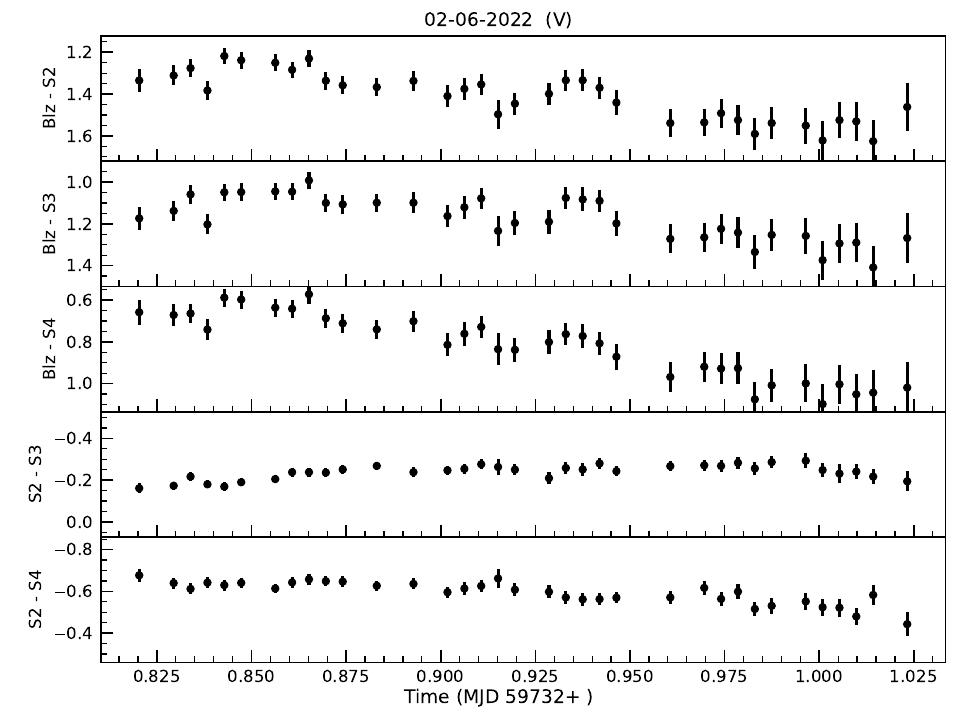}
}
\hbox{
\includegraphics[height=6cm,width=6cm]{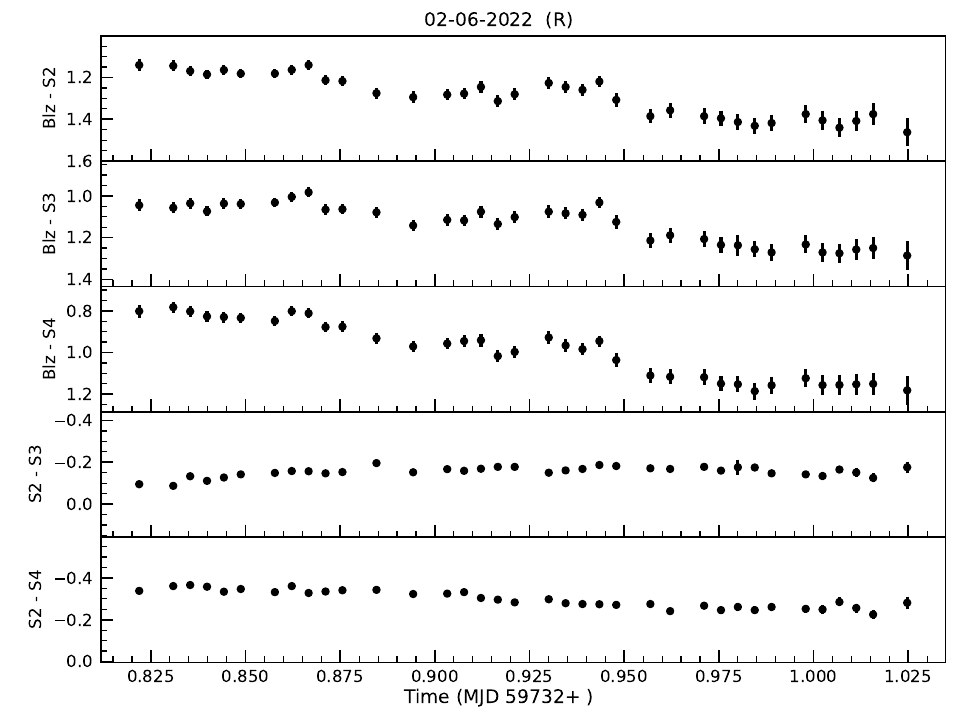}
\includegraphics[height=6cm,width=6cm]{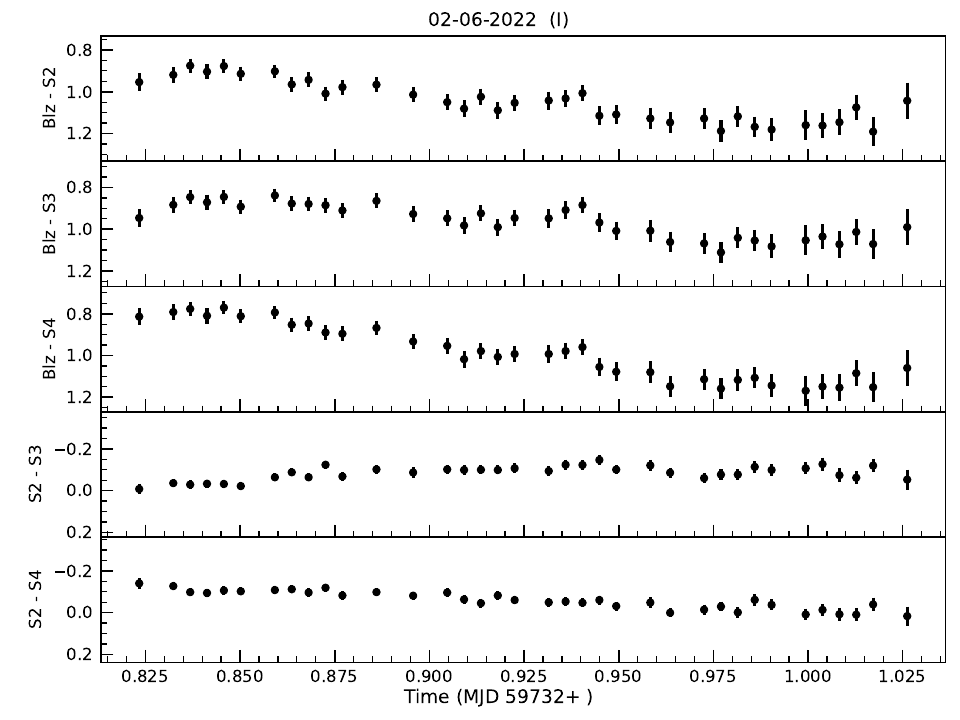}
\includegraphics[height=6cm,width=6cm]{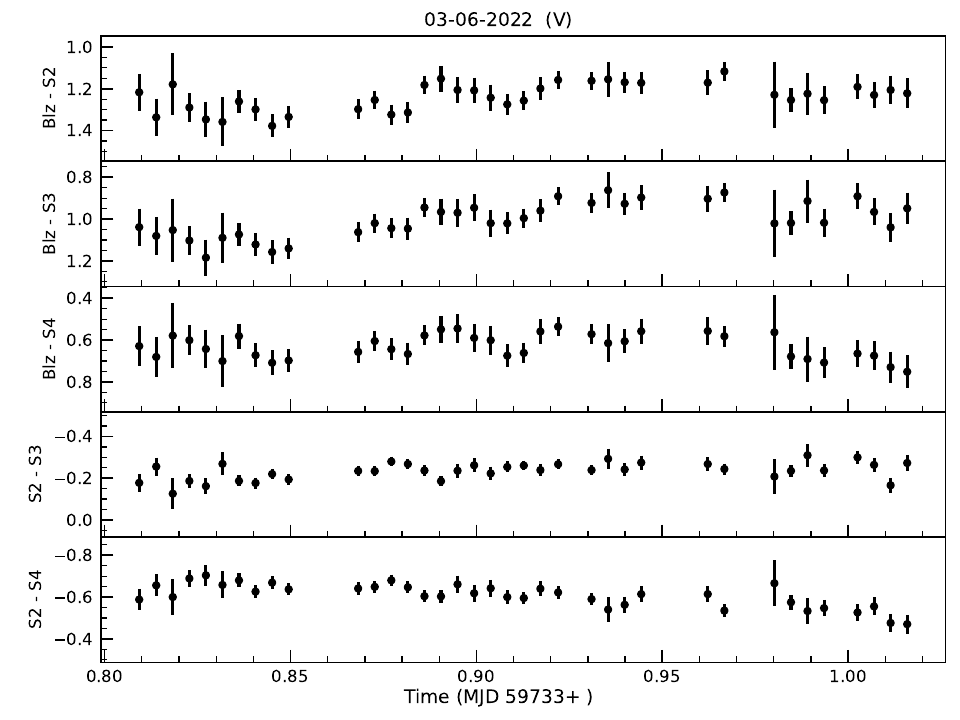}
}
}
\caption{Continued.}
\end{figure*}

\begin{figure*}[h]
 \ContinuedFloat
\vbox{
\hbox{
\includegraphics[height=6cm,width=6cm]{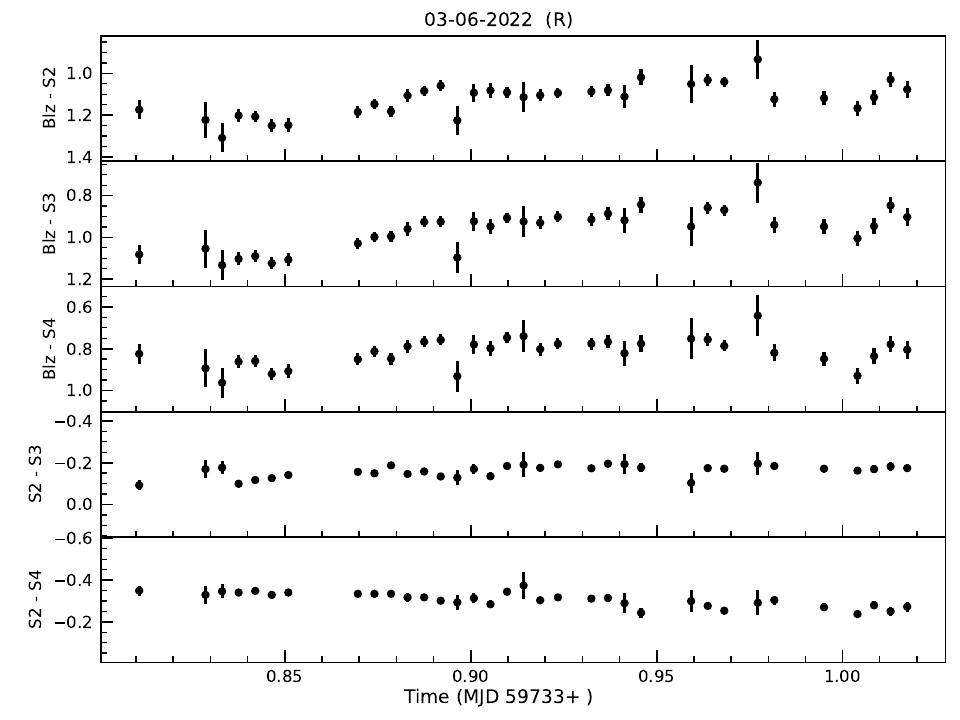}
\includegraphics[height=6cm,width=6cm]{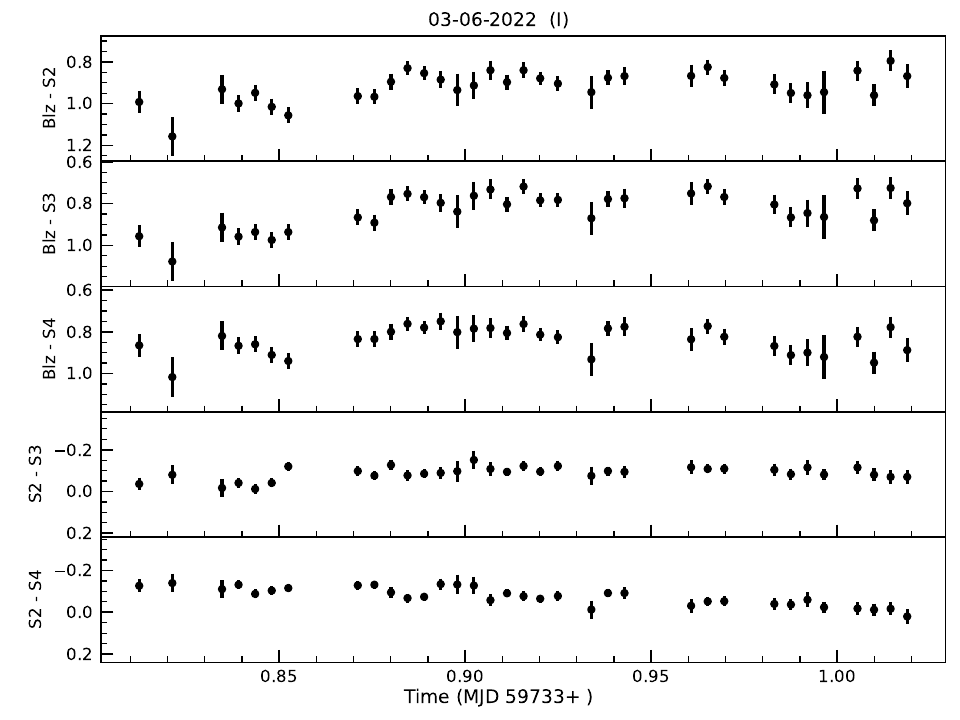}
\includegraphics[height=6cm,width=6cm]{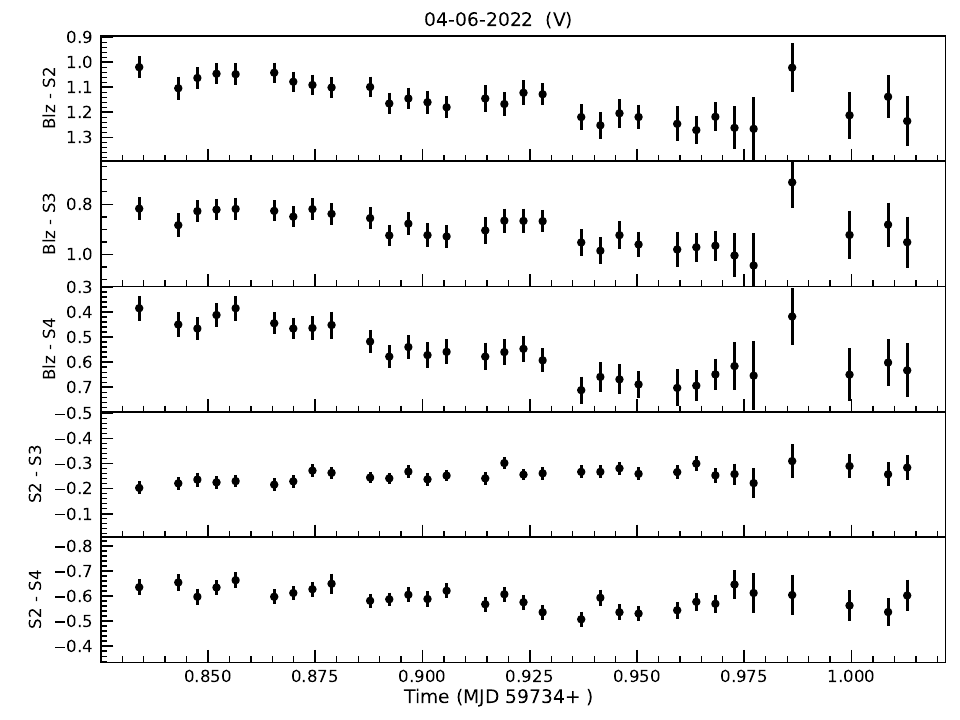}
}

\hbox{
\includegraphics[height=6cm,width=6cm]{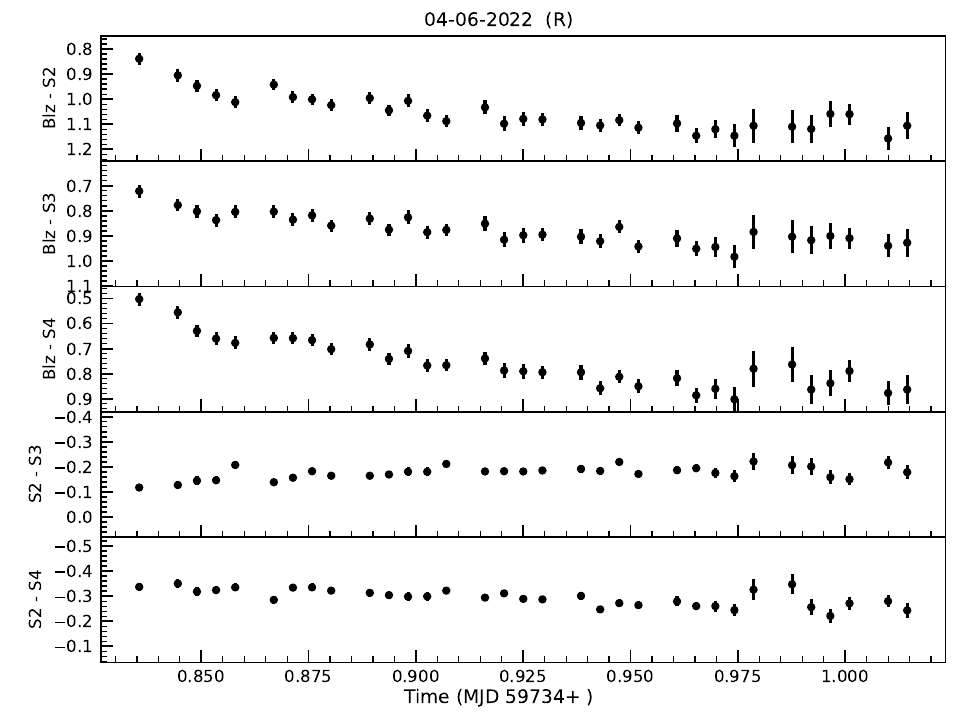}
\includegraphics[height=6cm,width=6cm]{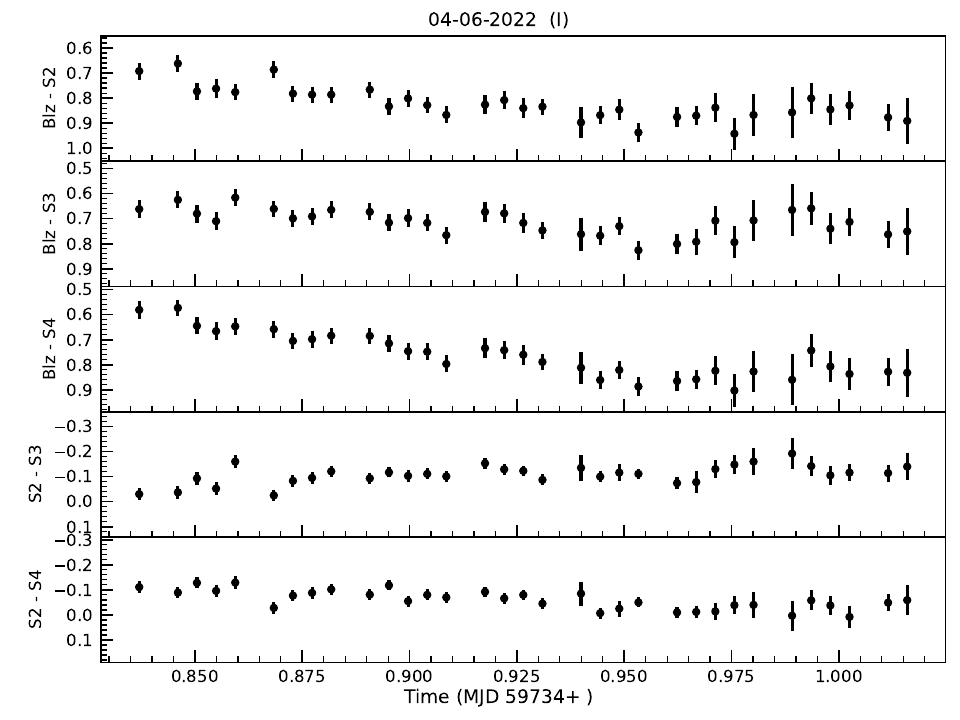}
\includegraphics[height=6cm,width=6cm]{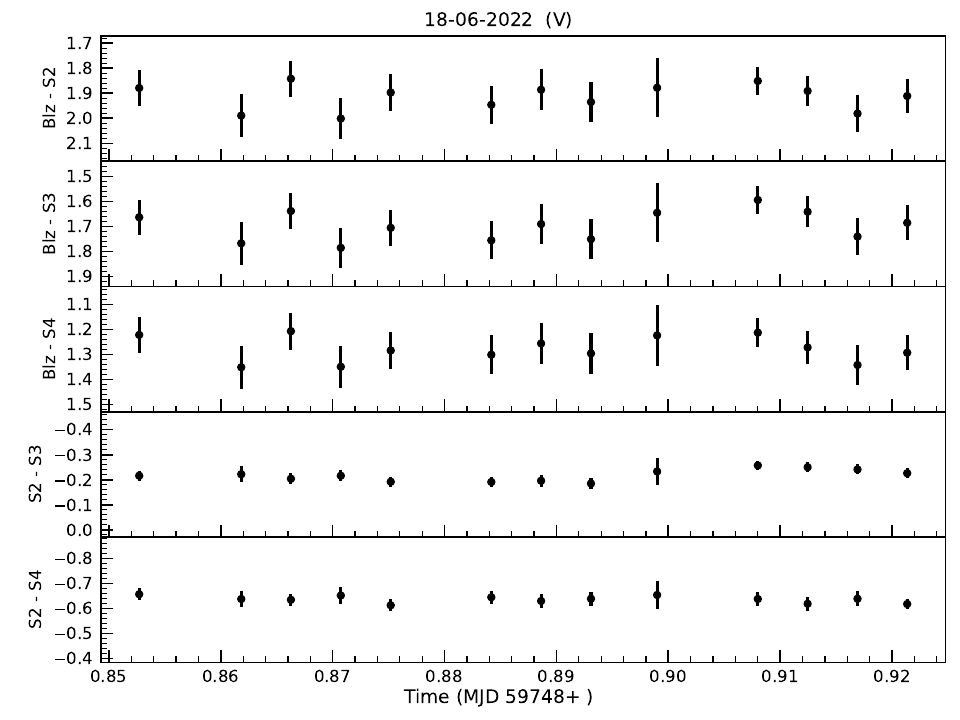}
}

\hbox{
\includegraphics[height=6cm,width=6cm]{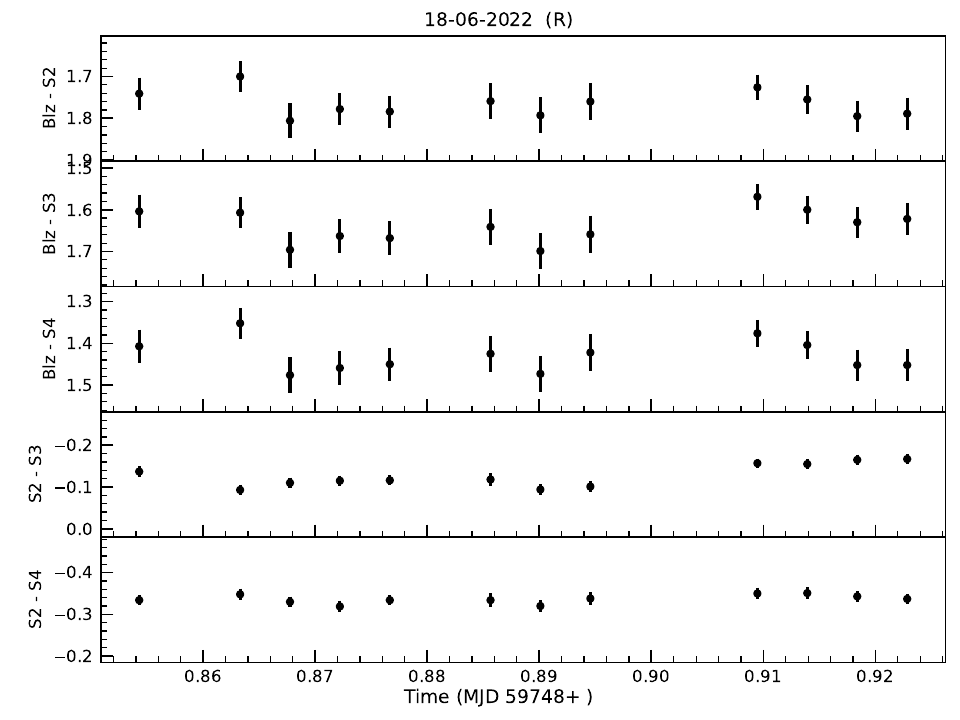}
\includegraphics[height=6cm,width=6cm]{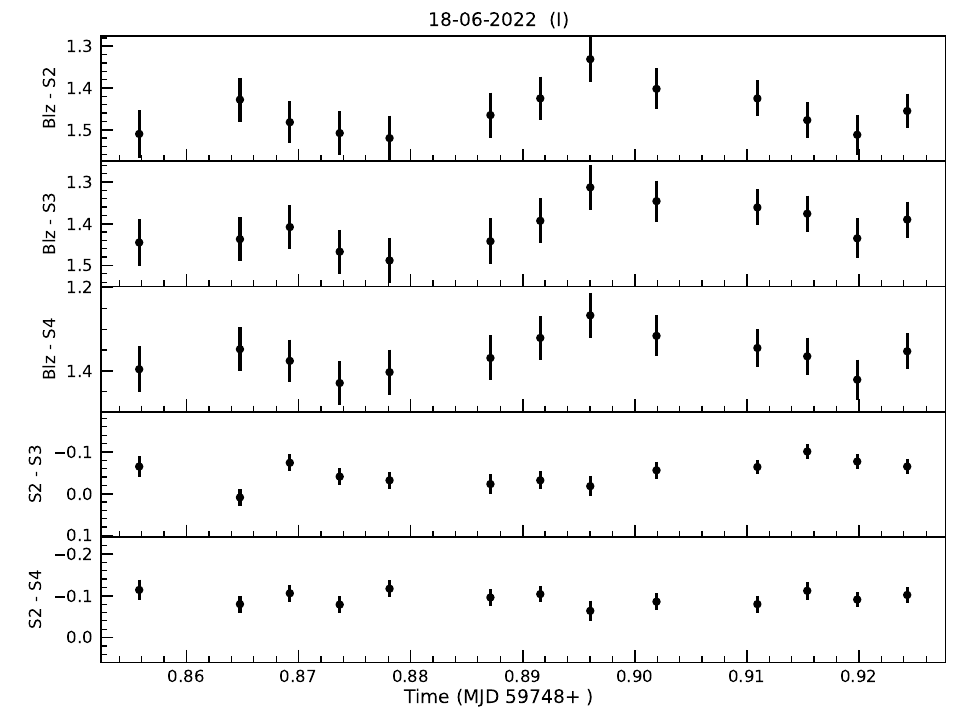}
\includegraphics[height=6cm,width=6cm]{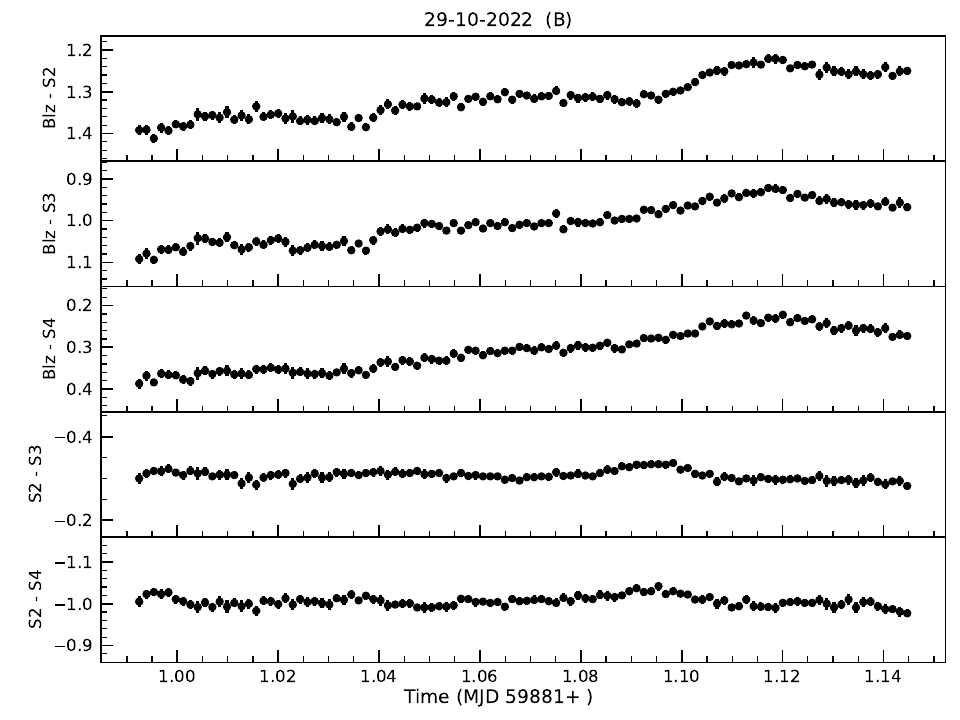}
}
\hbox{
\includegraphics[height=6cm,width=6cm]{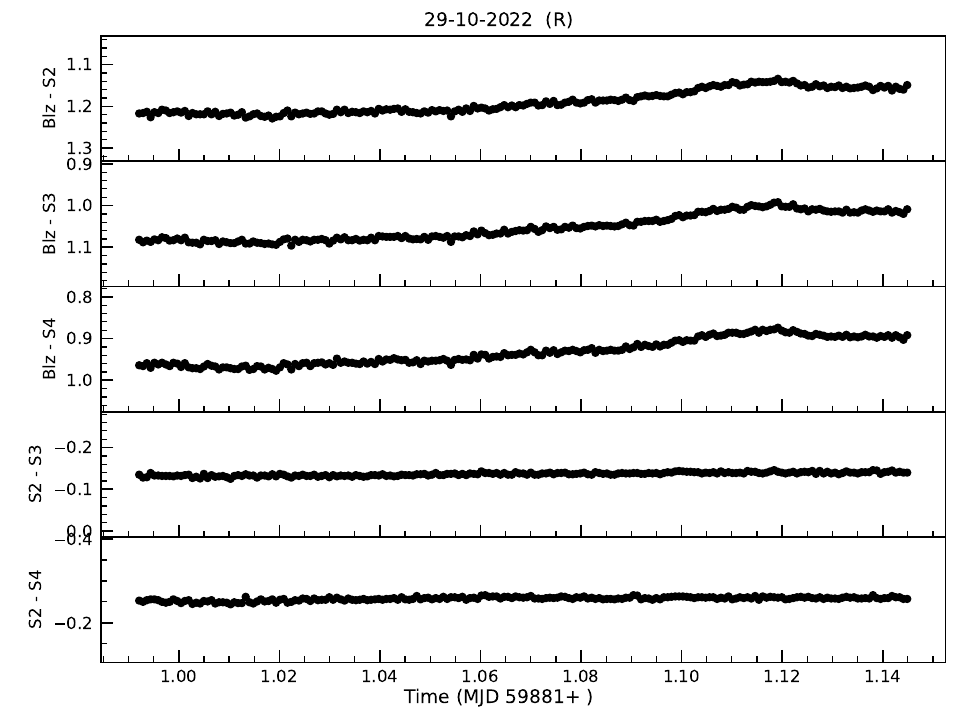}
\includegraphics[height=6cm,width=6cm]{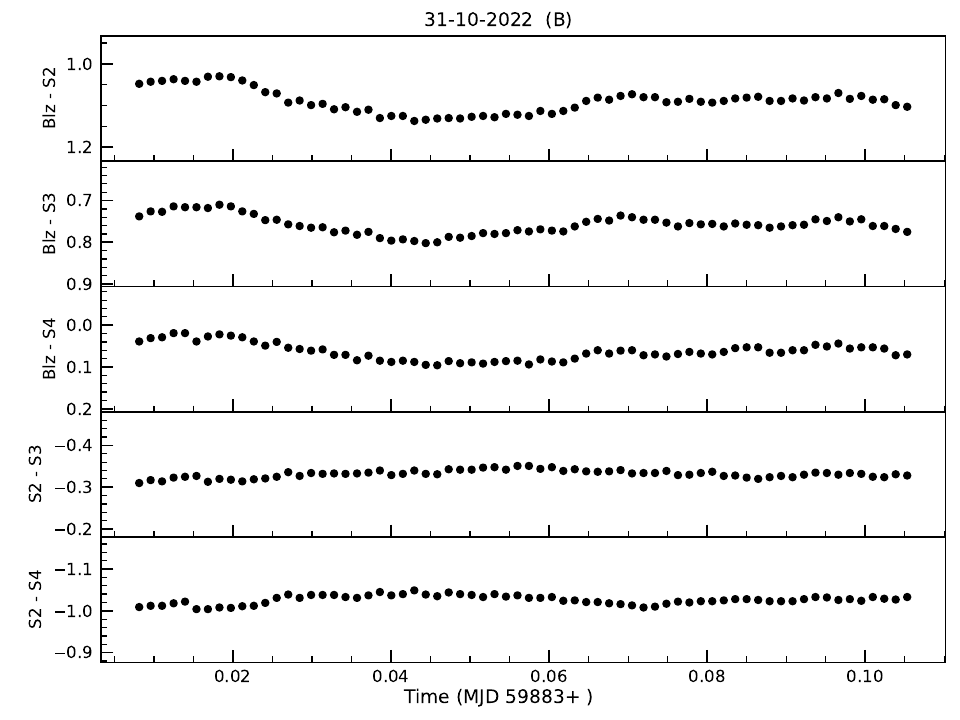}
\includegraphics[height=6cm,width=6cm]{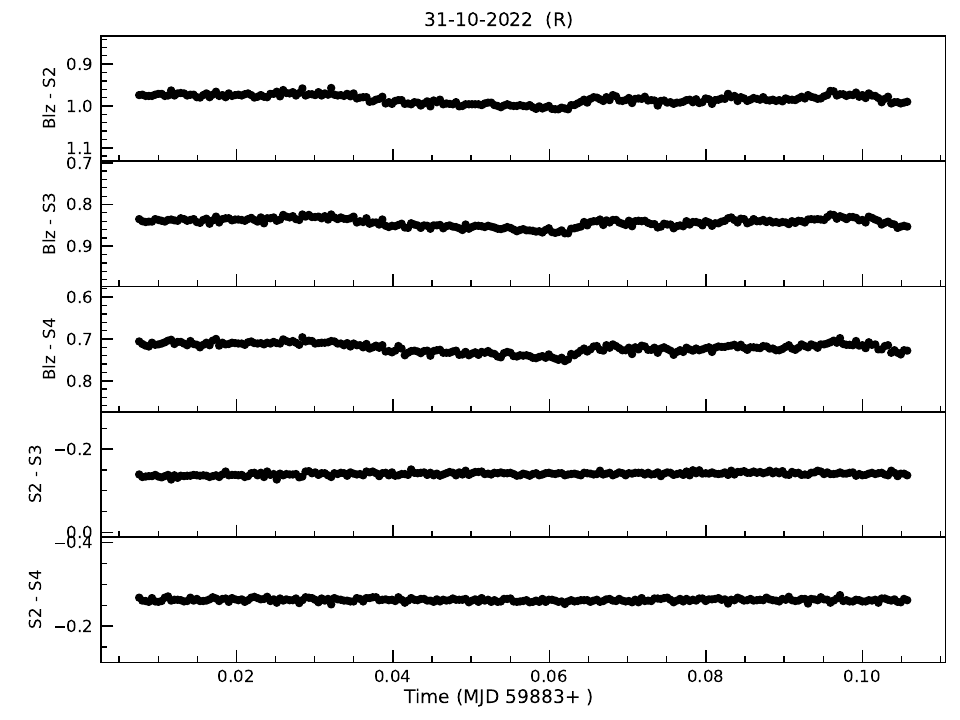}
}
}
\caption{Continued.}
\end{figure*}

\begin{figure*}[h]
 \ContinuedFloat
\vbox{
\hbox{
\includegraphics[height=6cm,width=6cm]{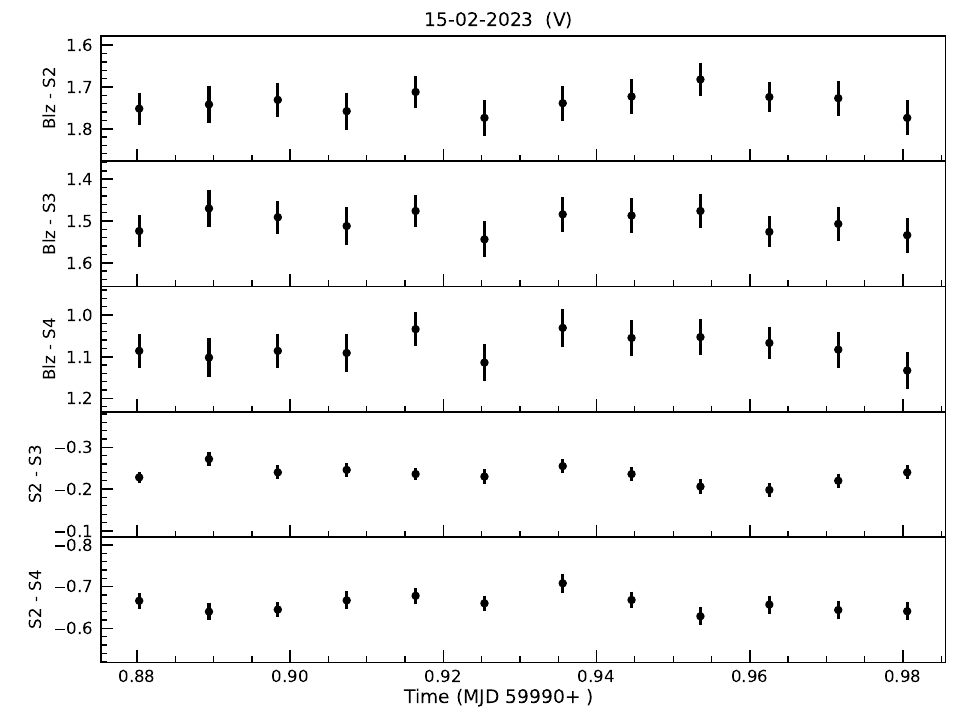}
\includegraphics[height=6cm,width=6cm]{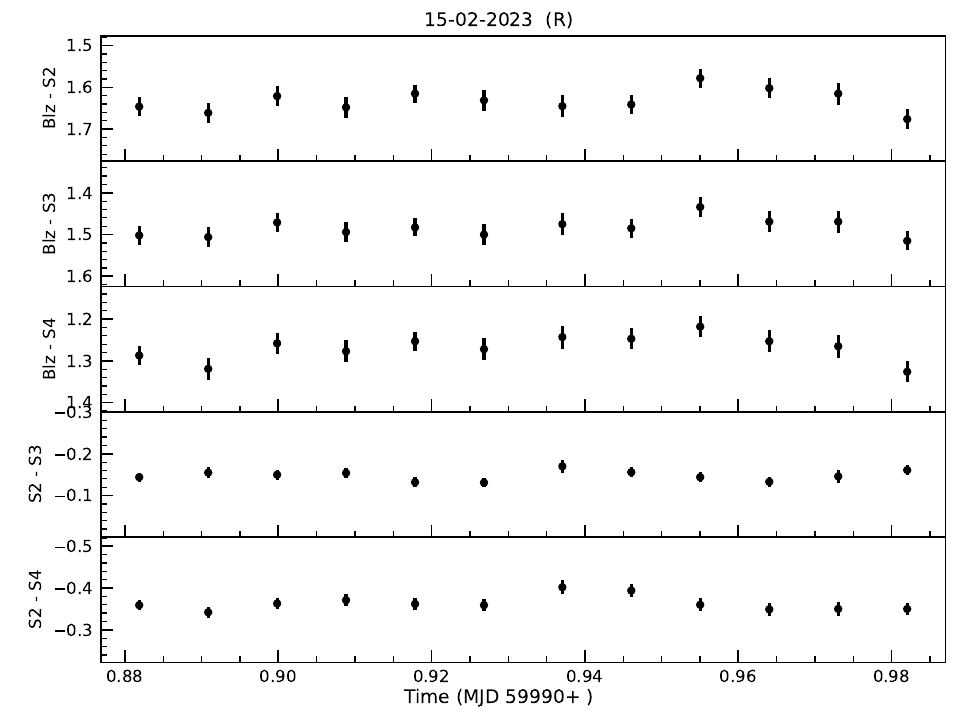}
\includegraphics[height=6cm,width=6cm]{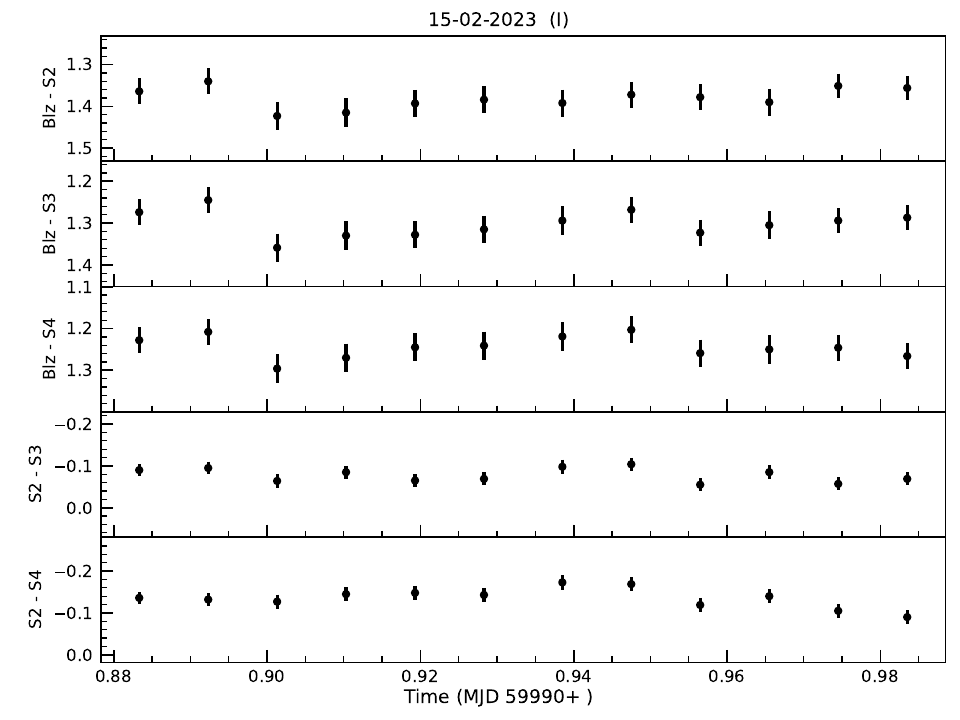}
}

\hbox{
\includegraphics[height=6cm,width=6cm]{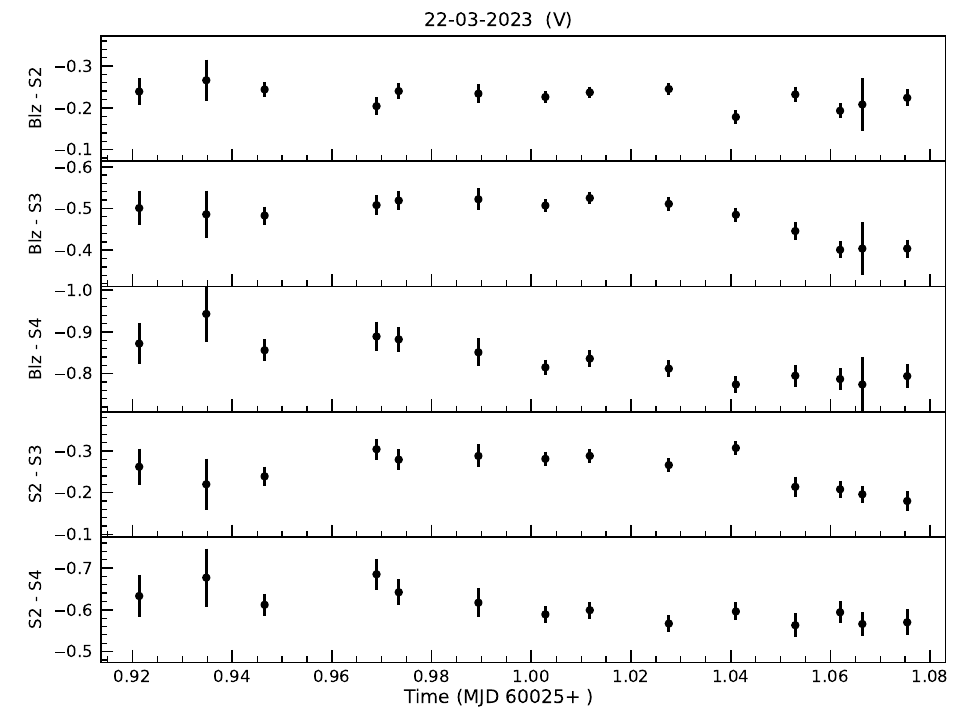}
\includegraphics[height=6cm,width=6cm]{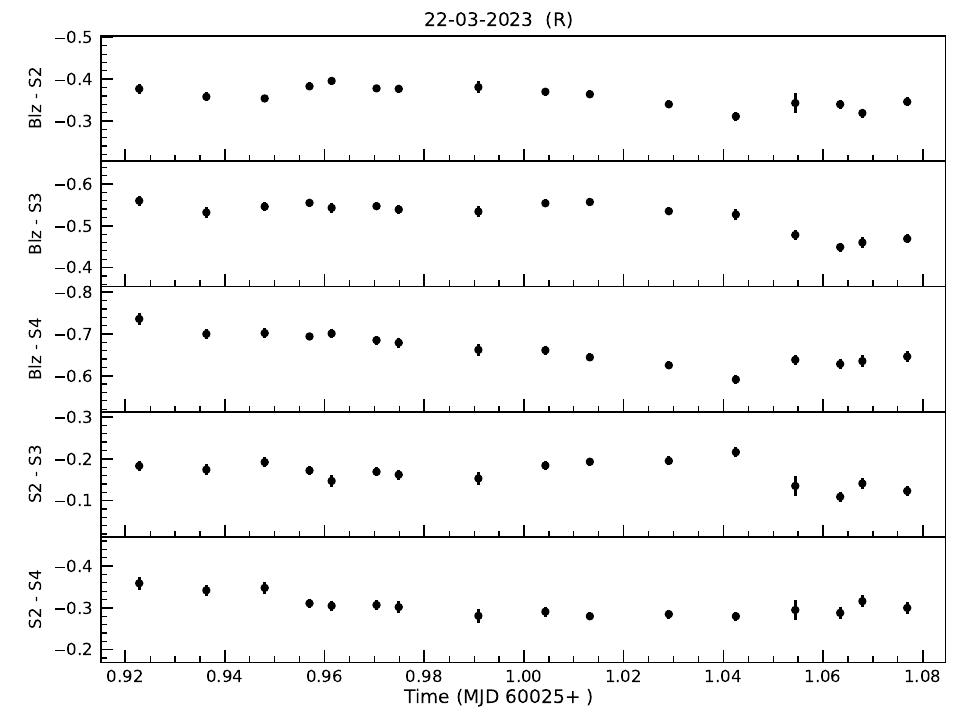}
\includegraphics[height=6cm,width=6cm]{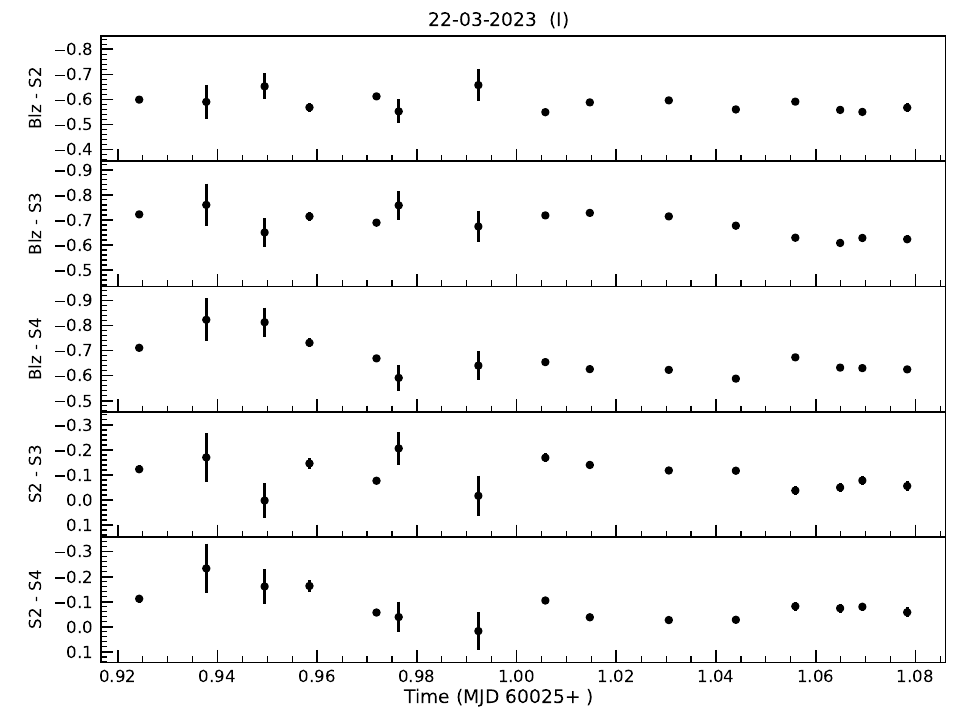}
}

\hbox{
\includegraphics[height=6cm,width=6cm]{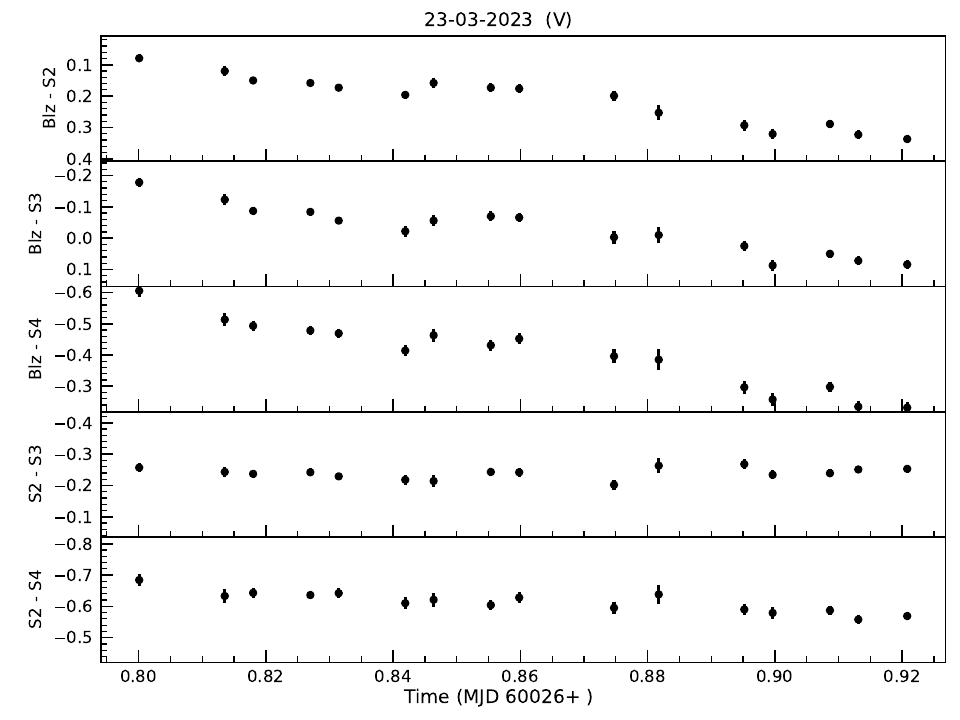}
\includegraphics[height=6cm,width=6cm]{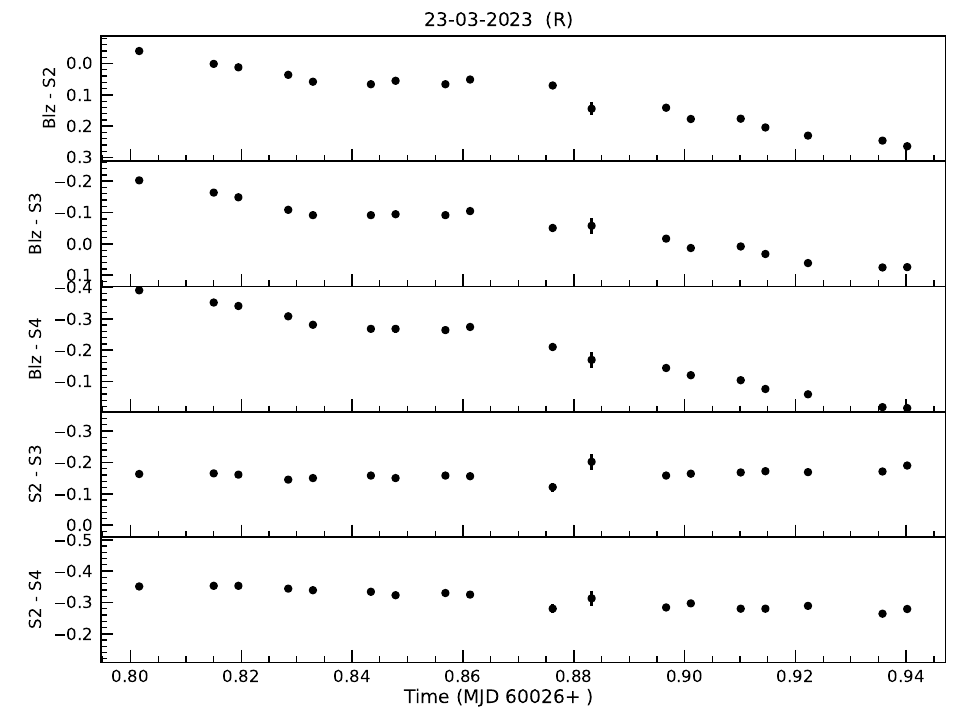}
\includegraphics[height=6cm,width=6cm]{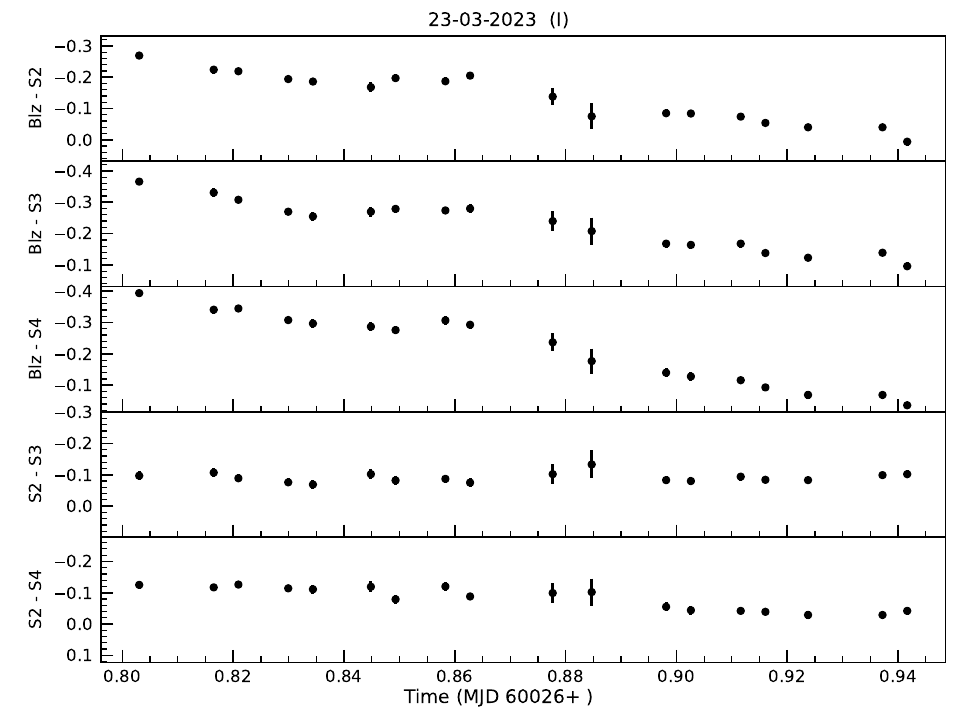}
}
\hbox{
\includegraphics[height=6cm,width=6cm]{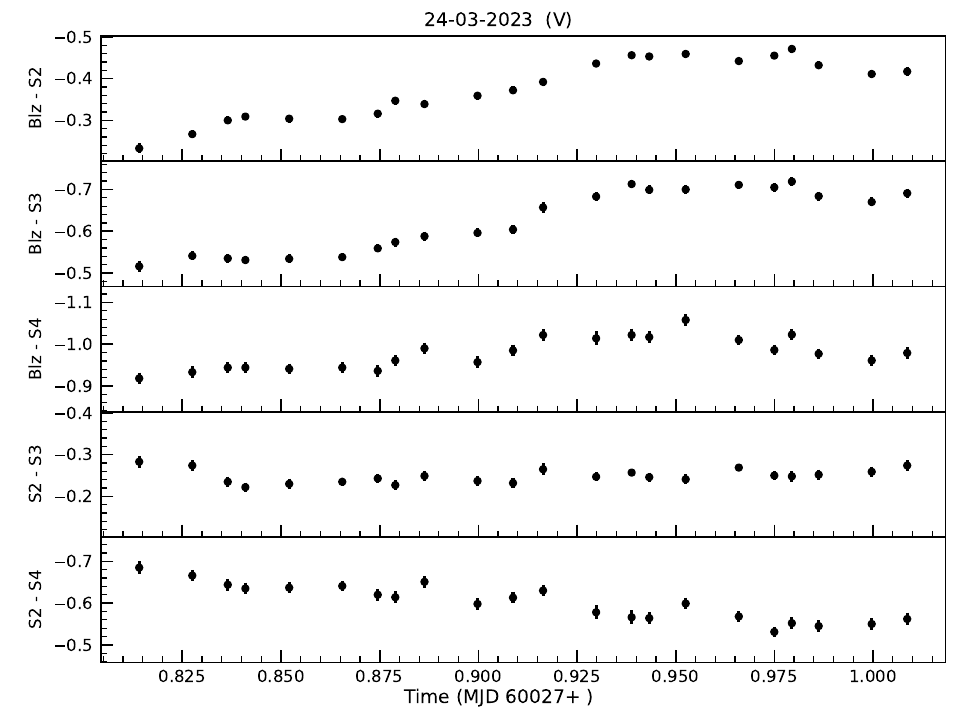}
\includegraphics[height=6cm,width=6cm]{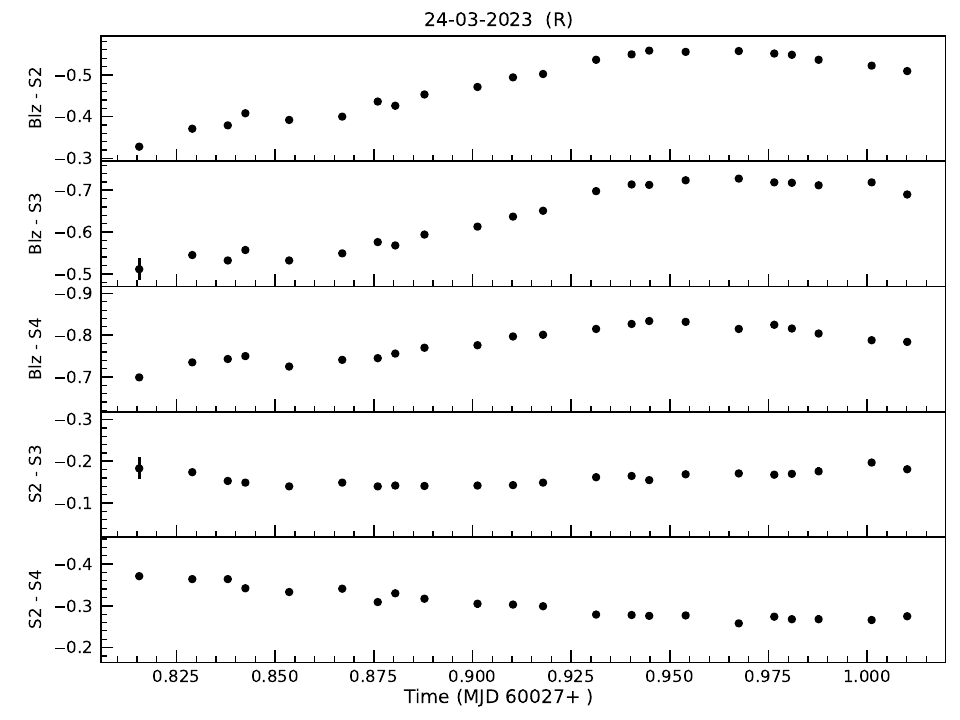}
\includegraphics[height=6cm,width=6cm]{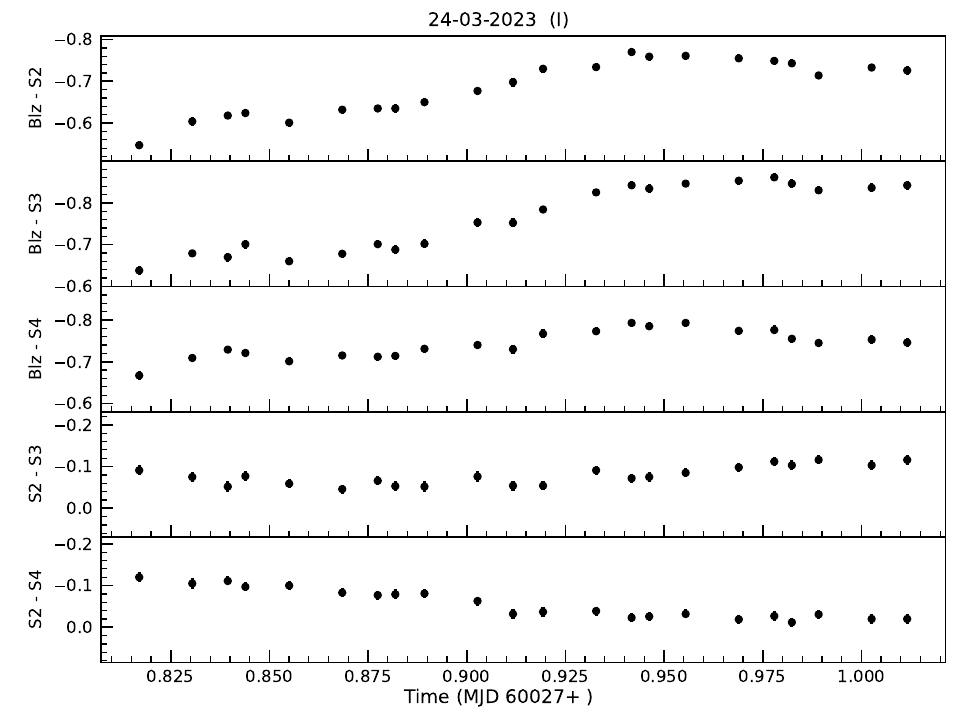}
}
}
\caption{Continued.}
\end{figure*}

\begin{figure*}[h]
 \ContinuedFloat
\vbox{
\hbox{
\includegraphics[height=6cm,width=6cm]{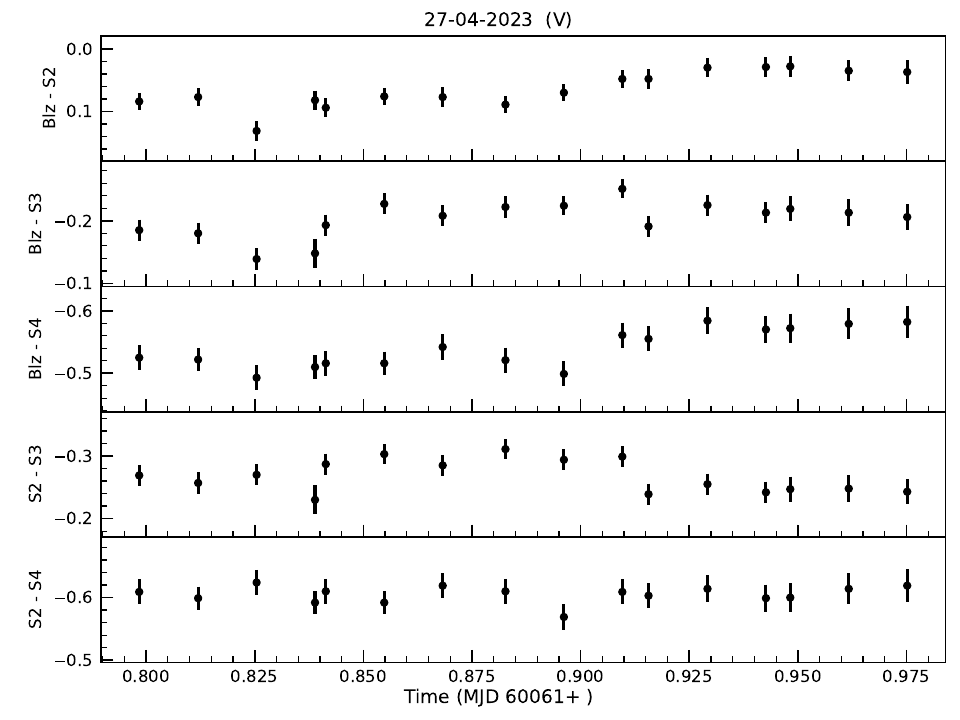}
\includegraphics[height=6cm,width=6cm]{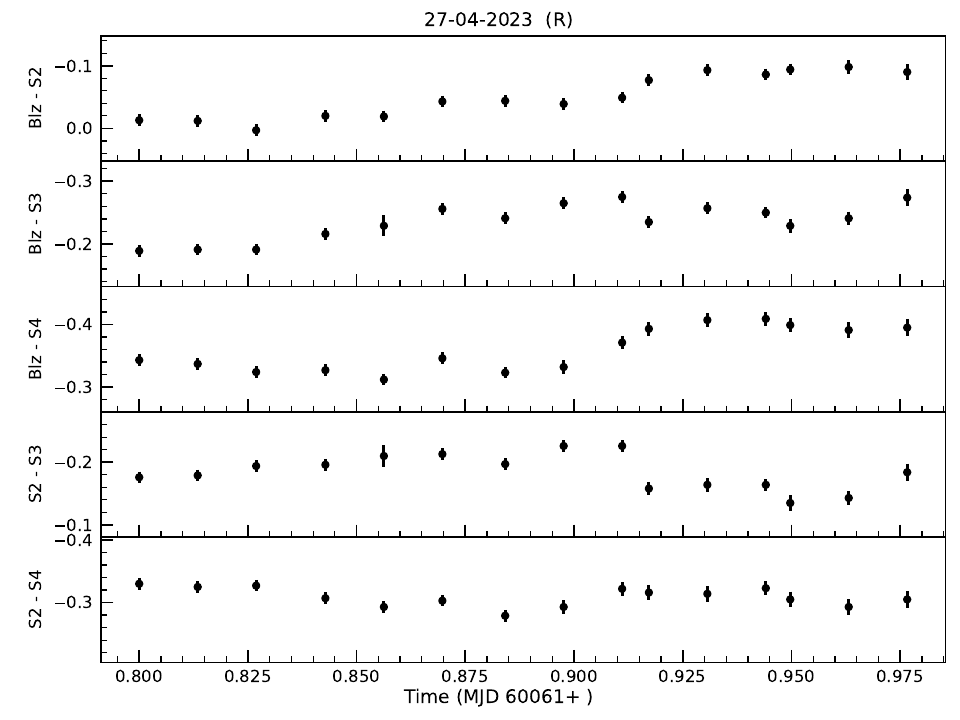}
\includegraphics[height=6cm,width=6cm]{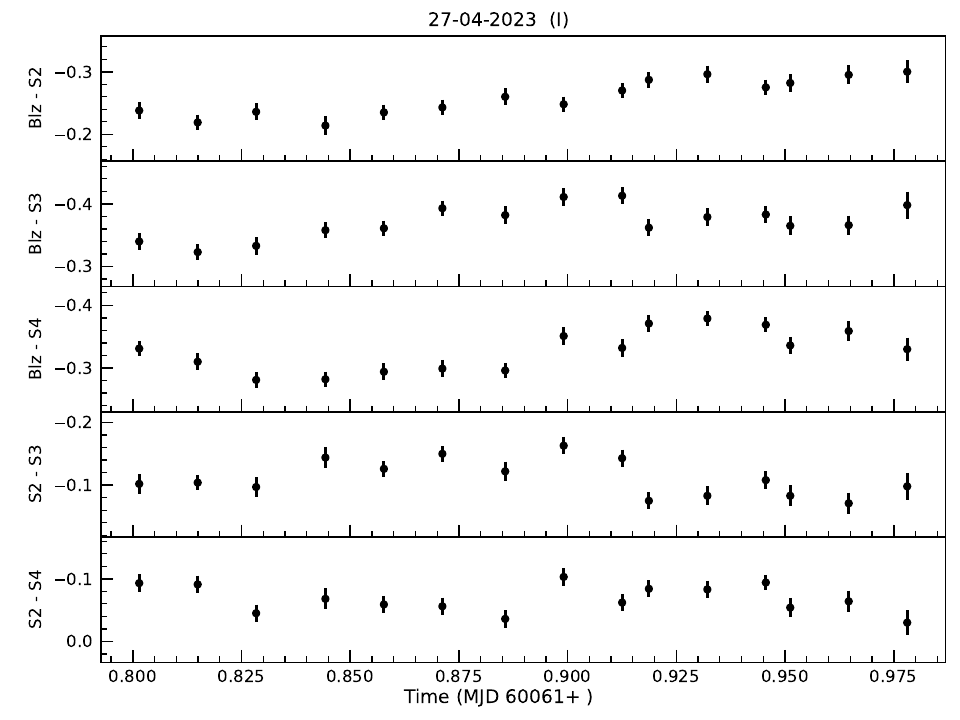}
}

\hbox{
\includegraphics[height=6cm,width=6cm]{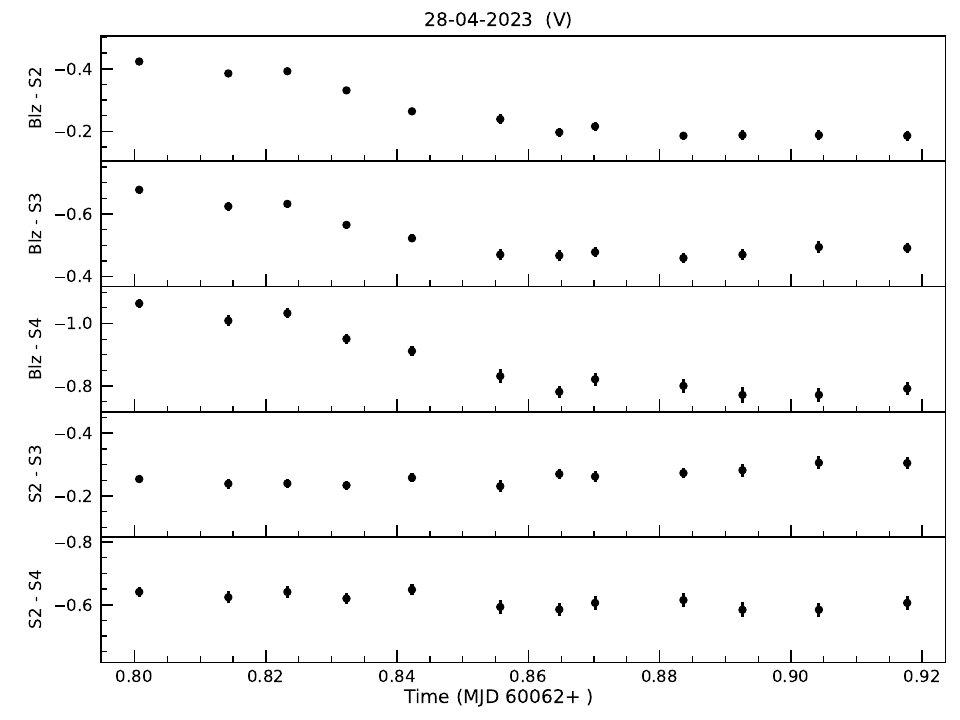}
\includegraphics[height=6cm,width=6cm]{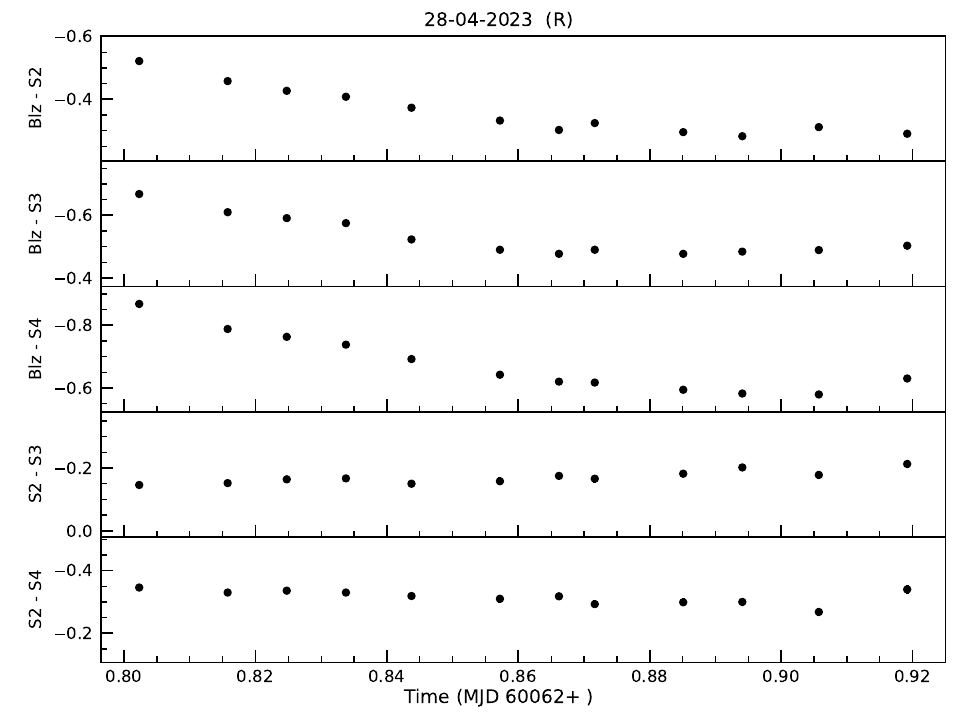}
\includegraphics[height=6cm,width=6cm]{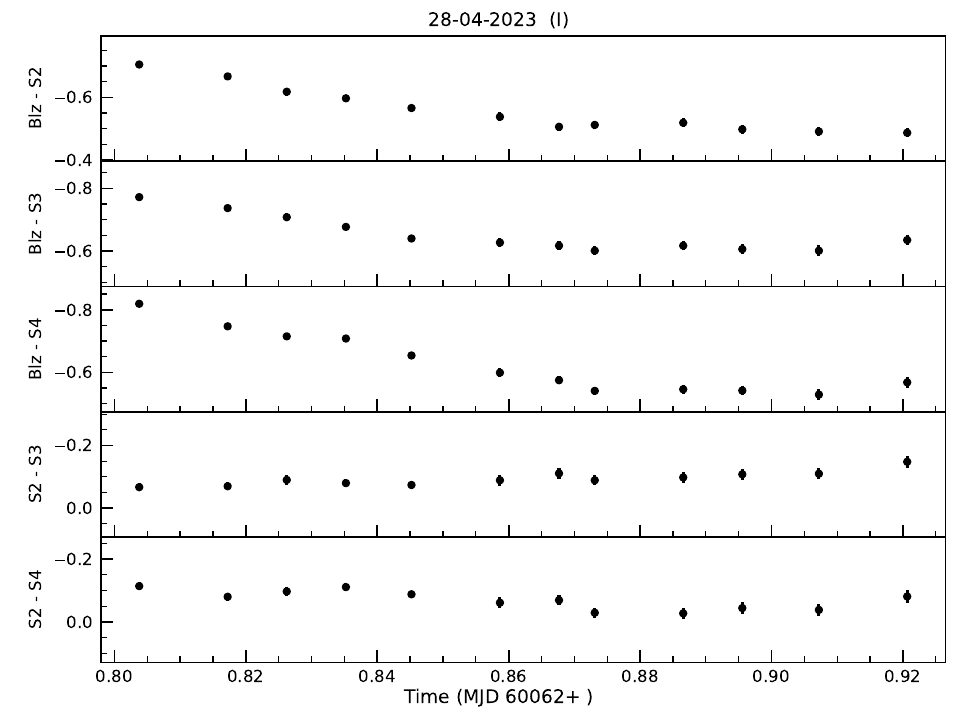}
}

}
\caption{Continued.}
\end{figure*}

\section{Intraday optical and polarization light curves.}

\begin{figure*}
\vbox{
\hbox{
\includegraphics[height=6cm,width=6cm]{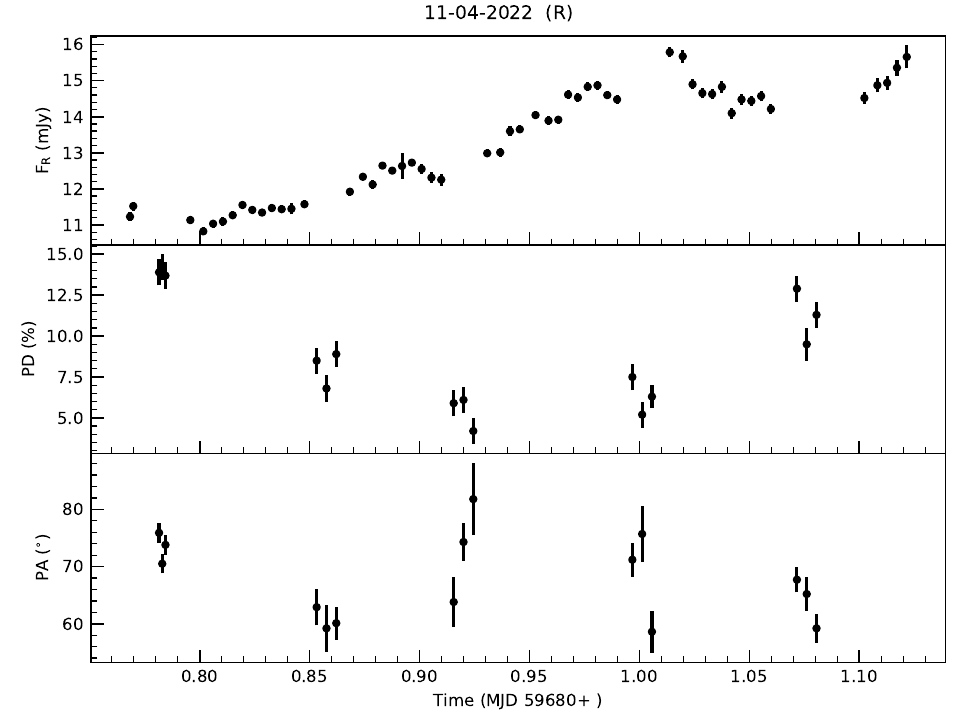}
\includegraphics[height=6cm,width=6cm]{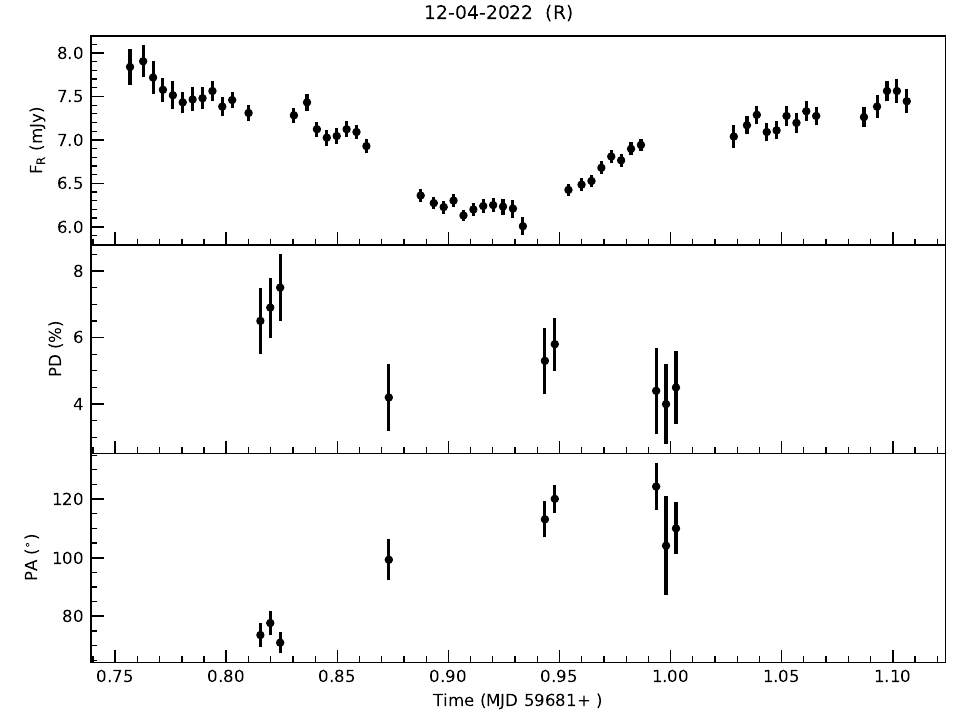}
\includegraphics[height=6cm,width=6cm]{20220413_flx_R_op_pol.pdf}
}

\hbox{
\includegraphics[height=6cm,width=6cm]{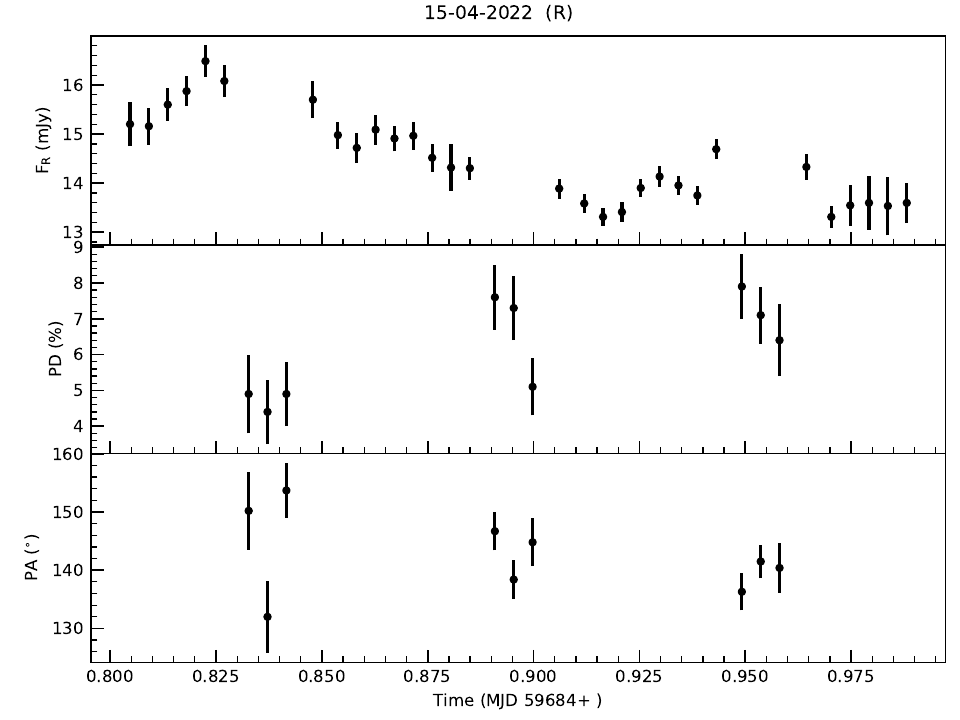}
\includegraphics[height=6cm,width=6cm]{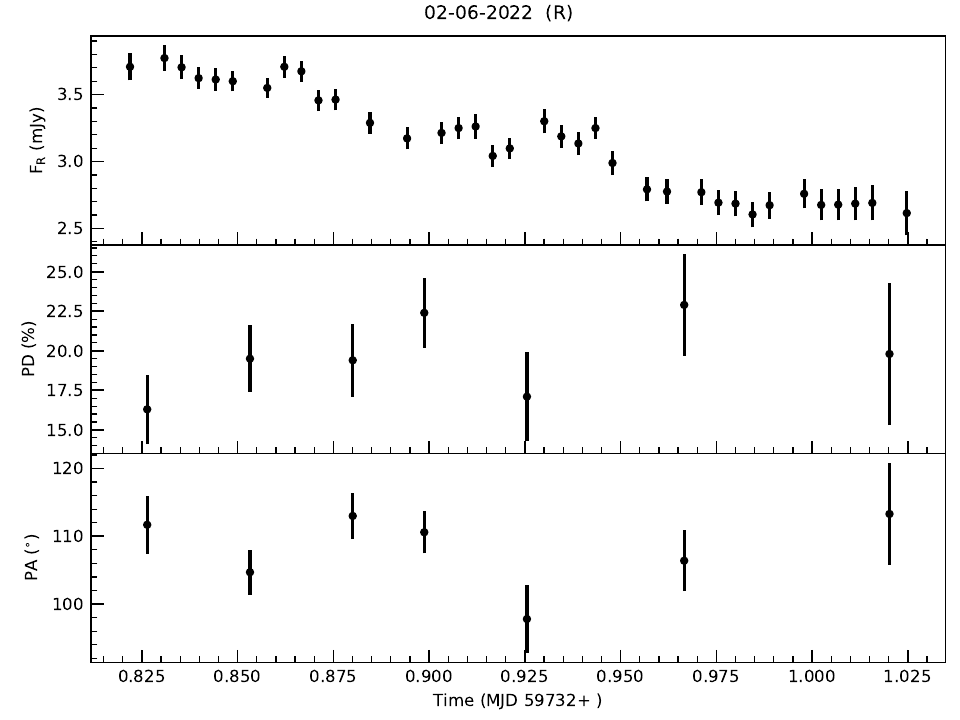}
\includegraphics[height=6cm,width=6cm]{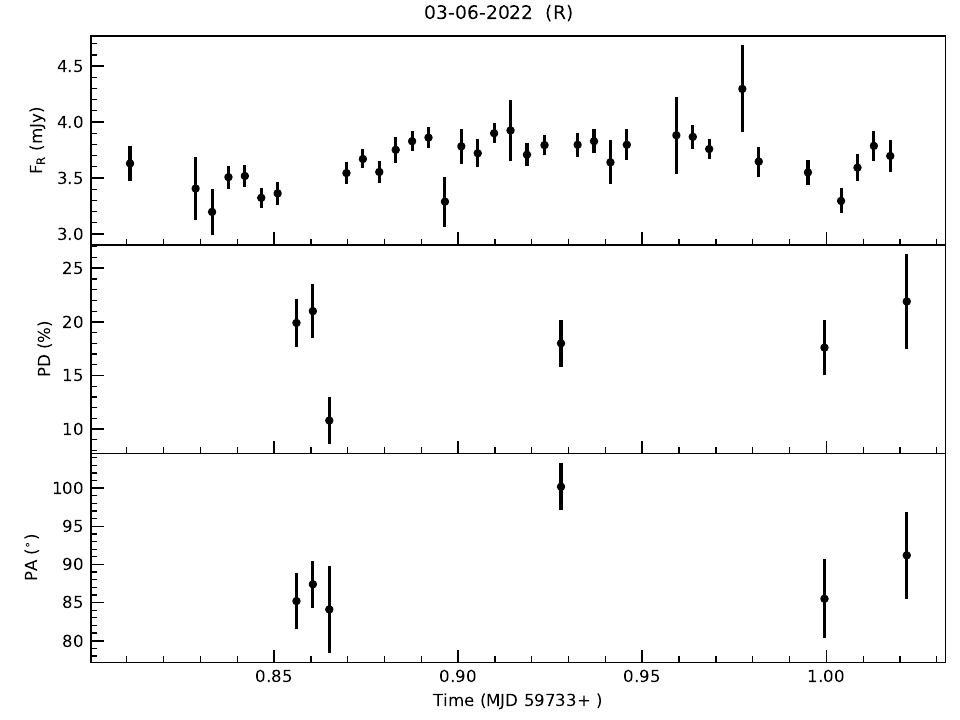}
}

\hbox{
\includegraphics[height=6cm,width=6cm]{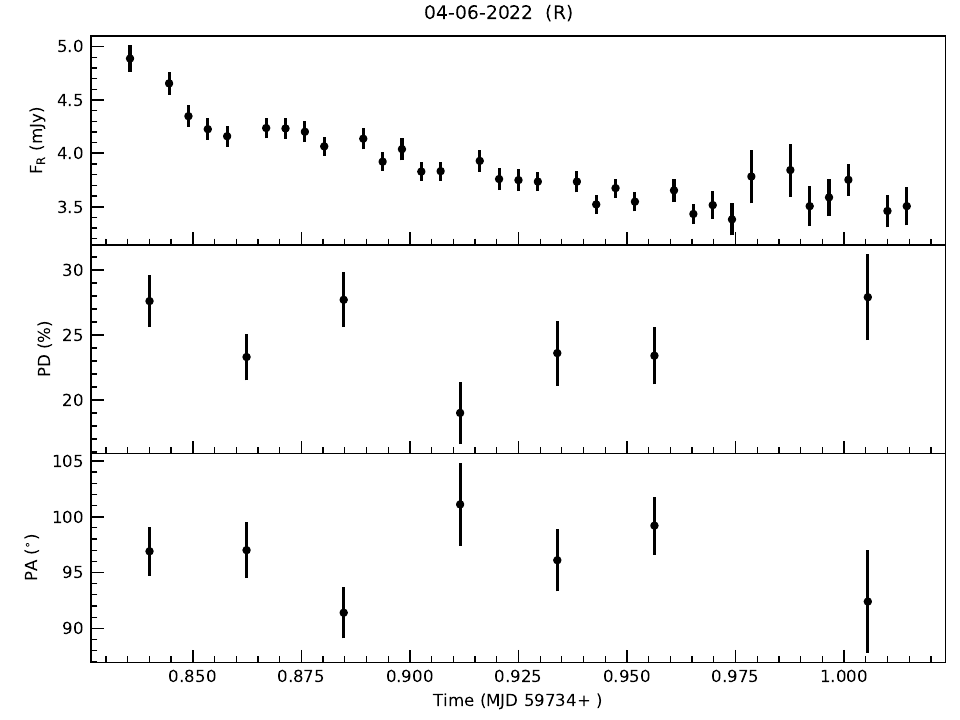}
\includegraphics[height=6cm,width=6cm]{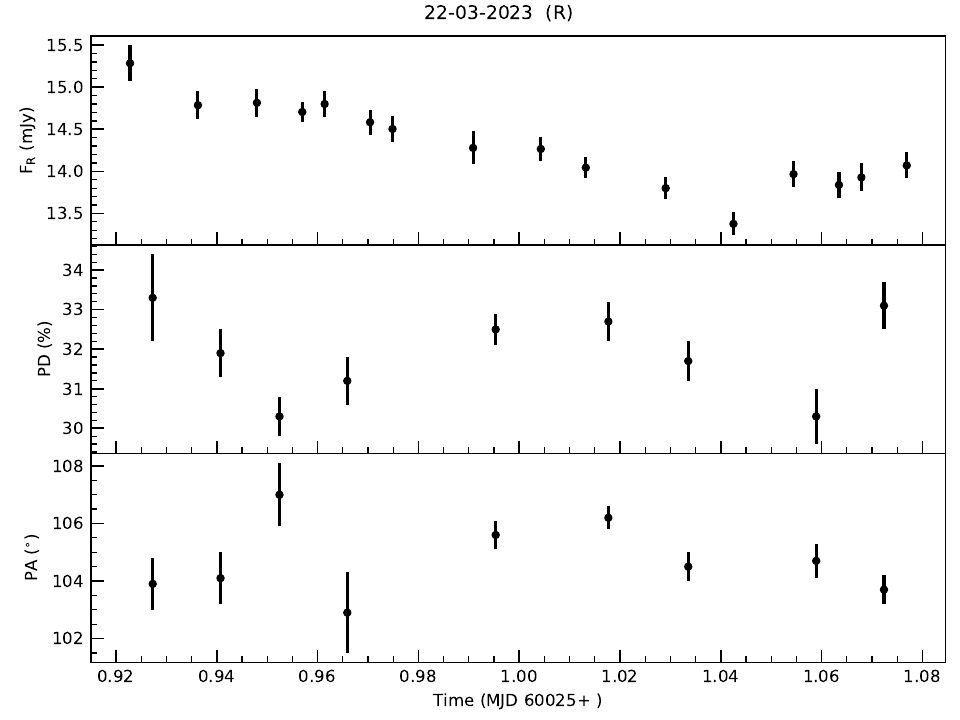}
\includegraphics[height=6cm,width=6cm]{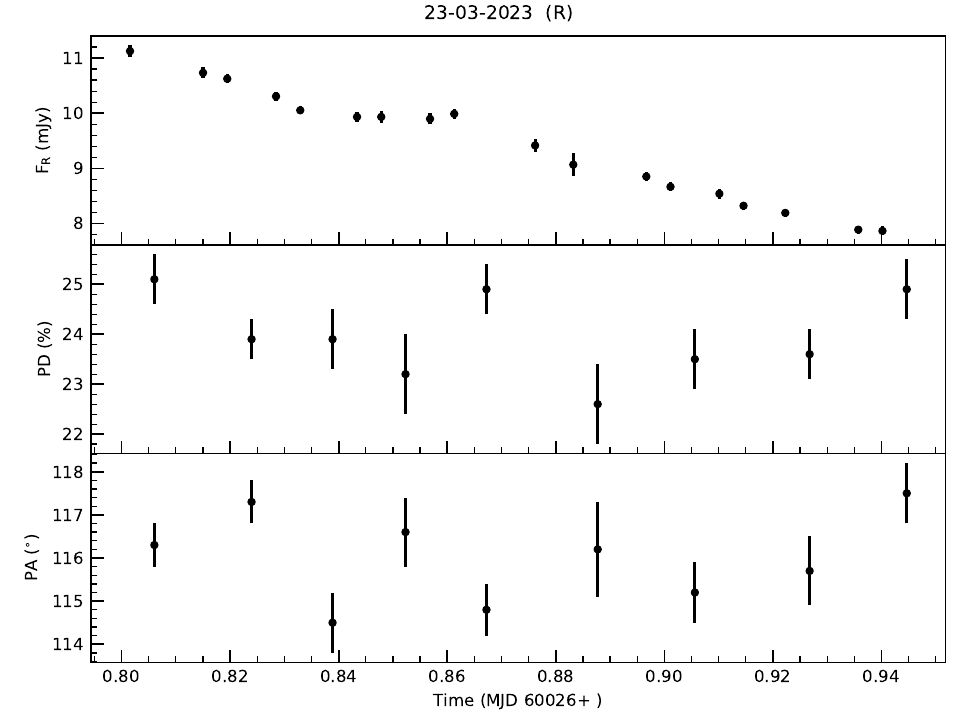}
}
\hbox{
\includegraphics[height=6cm,width=6cm]{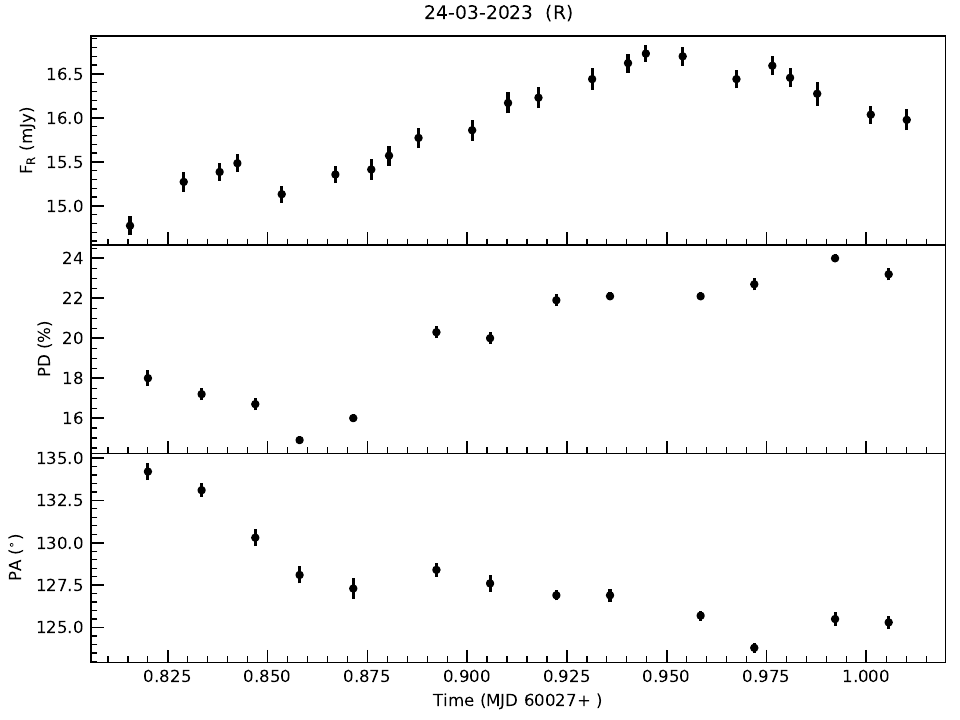}
\includegraphics[height=6cm,width=6cm]{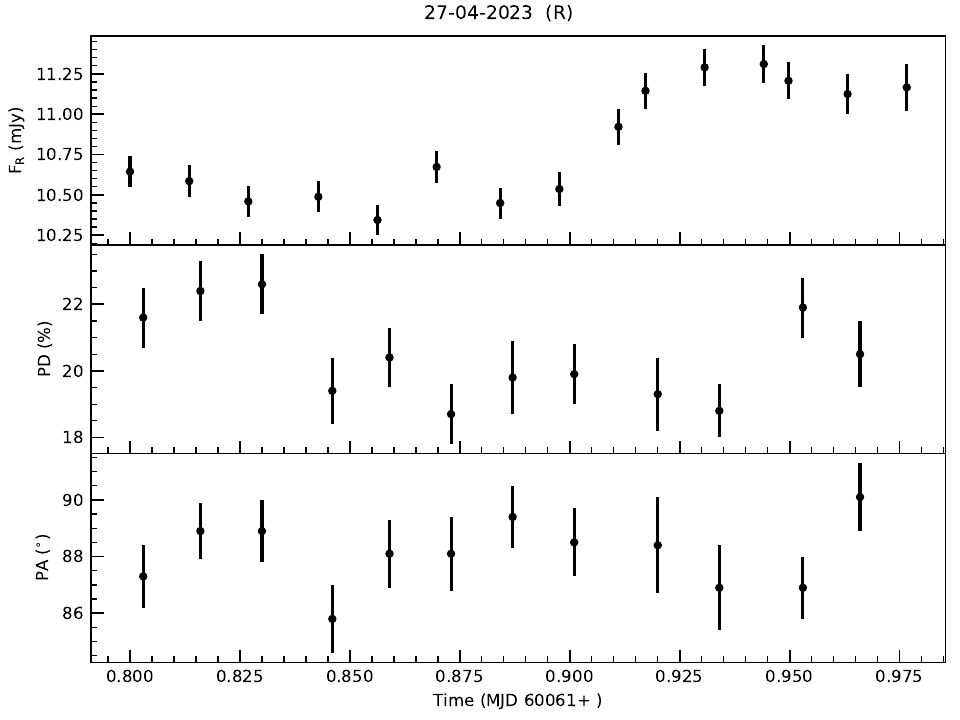}
\includegraphics[height=6cm,width=6cm]{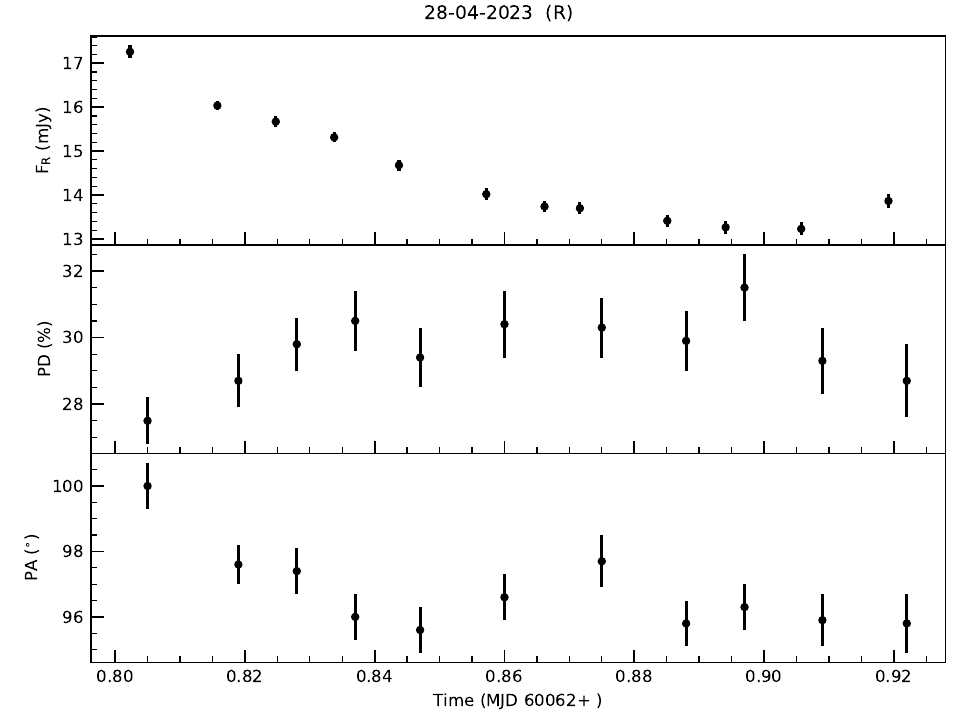}
}
}
\caption{\label{fig:idv_op_pol}Indtraday R-band light curves (in mJy) along with polarization measurements. The observation date and the name of the filter used are given at the top of each plot.}
\end{figure*}

\end{appendix}

\end{document}